%% file: main.tex
\documentclass[11pt,a4paper]{article}

\usepackage[a4paper, left=1in, right=1in, top=1in, bottom=1in]{geometry}
\usepackage{setspace}
\AtBeginDocument{%
  \abovedisplayskip=10pt
  \belowdisplayskip=10pt
  \abovedisplayshortskip=6pt
  \belowdisplayshortskip=6pt
}

\usepackage{graphicx}
\usepackage{multirow}
\usepackage{amsmath}
\usepackage{amssymb}
\usepackage{amsfonts}
\usepackage{stmaryrd}
\usepackage{amscd}
\usepackage{physics}
\usepackage{rotating}
\usepackage[parfill]{parskip}
\usepackage{textcomp}
\usepackage{changepage}
\usepackage[colorlinks=true, linkcolor=blue, urlcolor=blue, citecolor=blue, breaklinks=true]{hyperref}
\usepackage{longtable}
\usepackage[numbers,sort&compress]{natbib}
\usepackage[flushleft]{threeparttable}
\usepackage{threeparttablex}
\usepackage{bbding}
\usepackage{booktabs}
\usepackage{tabularx}
\usepackage{tcolorbox}
\usepackage{paracol}
\usepackage{adjustbox}
\usepackage{siunitx}
\usepackage{cleveref}
\crefformat{footnote}{#2\footnotemark[#1]#3}
\usepackage{sidecap}

\usepackage{color}
\usepackage[normalem]{ulem}
\usepackage[dvipsnames]{xcolor}
\usepackage{tikz}
\usetikzlibrary{positioning,arrows.meta}
\usepackage{wrapfig}
\usepackage{subcaption}
\usepackage{enumitem}

\newlist{compactdesc}{description}{1}
\setlist[compactdesc]{%
  labelindent=0pt,
  leftmargin=0pt,
  labelsep=0.5em,
  itemsep=0pt,
  topsep=0pt,
  partopsep=0pt,
  parsep=0pt,
  style=nextline
}

\usepackage{listings}
\usepackage{xcolor}
\usepackage{newfloat}
\usepackage{fancyvrb}
\DeclareFloatingEnvironment[fileext=lol,name=Listing]{listing}

\definecolor{codegreen}{rgb}{0,0.6,0}
\definecolor{codegray}{rgb}{0.5,0.5,0.5}
\definecolor{codepurple}{rgb}{0.58,0,0.82}
\definecolor{codeblue}{rgb}{0,0,0.8}
\definecolor{backcolour}{rgb}{0.95,0.95,0.95}

\lstdefinelanguage{Rust}{
    keywords={as, break, const, continue, crate, else, enum, extern, false, fn, for, if, impl, in, let, loop, match, mod, move, mut, pub, ref, return, self, Self, static, struct, super, trait, true, type, unsafe, use, where, while, async, await, dyn},
    keywordstyle=\color{codeblue}\bfseries,
    ndkeywords={bool, char, i8, i16, i32, i64, i128, isize, u8, u16, u32, u64, u128, usize, f32, f64, str, String, Vec, Option, Result, Some, None, Ok, Err, Box, Rc, Arc, Cell, RefCell},
    ndkeywordstyle=\color{codepurple},
    sensitive=true,
    comment=[l]{//},
    morecomment=[s]{/*}{*/},
    morestring=[b]",
    morestring=[b]',
}

\lstdefinelanguage{Coq}{
    keywords={Abort, About, Add, All, Arguments, Axiom, Back, Bind, Canonical, Cd, Check, Class, Close, Coercion, CoFixpoint, CoInductive, Combined, Compute, Context, Corollary, Create, Declare, Defined, Definition, Delimit, Derive, Deterministic, Drop, End, Eval, Example, Existential, Export, Fact, Field, File, Fixpoint, Focus, Functional, Generalizable, Global, Goal, Grab, Guarded, Hint, Hypotheses, Hypothesis, Identity, If, Implicit, Import, Include, Inductive, Infix, Instance, Interactive, Intro, Lazy, Lemma, Let, Load, Local, Locate, Ltac, Module, Morphism, Next, NoInline, Notation, Obligation, Obligations, Opaque, Open, Optimize, Parameter, Parameters, Polymorphic, Preterm, Print, Program, Projections, Proof, Prop, Property, Pwd, Qed, Quit, Rec, Record, Register, Require, Restart, Return, Rewrite, Ring, Scheme, Search, Searchpattern, Section, Separate, Set, Show, Solve, Strategy, Structure, SubClass, Tactic, Term, Theorem, Time, Transparent, Type, Typeclasses, Types, Undelimit, Undo, Unfocus, Unset, Variable, Variables, Variant, View, Visibility, apply, intro, intros, auto, simpl, reflexivity, rewrite, destruct, induction, case, exact, assumption, trivial, unfold, split, left, right, exists, constructor, discriminate, inversion, subst, clear, assert, pose, specialize, generalize, remember, set},
    keywordstyle=\color{codeblue}\bfseries,
    sensitive=true,
    comment=[l]{(*},
    morecomment=[s]{(*}{*)},
    morestring=[b]",
}

\lstdefinelanguage{diff}{
    morecomment=[f][\color{codeblue}]{@@},
    morecomment=[f][\color{red}]{-},
    morecomment=[f][\color{codegreen}]{+},
    morecomment=[f][\color{codepurple}]{!},
}

\newcommand{\aref}[1]{\hyperref[#1]{Appendix~\ref{#1}}}

\usepackage{caption}
\usepackage{multicol}
\usepackage{bm}
\usepackage{ifpdf}
\usepackage{float}
\usepackage{pdflscape}
\usepackage[table,xcdraw]{xcolor}
\usepackage{pifont}
\newcommand{\cmark}{\ding{51}}
\newcommand{\xmark}{\ding{55}}



\newcommand{\sublistingcaption}[1]{%
  \caption{#1}%
}

\title{\textbf{Refactoring and Equivalence in Rust}\\
\large Expanding the REM Toolchain with a Novel Approach to Automated Equivalence Proofs}

\author{
Matthew Britton\\
School of Computing, Australian National University\\
\texttt{matthew.britton@anu.edu.au}
\and
Sasha Pak\\
School of Computing, Australian National University\\
\texttt{sasha.pak@anu.edu.au}
\and
Alex Potanin\\
School of Computing, Australian National University\\
\texttt{alex.potanin@anu.edu.au}
}

\date{}

\newtcbox{\icodebox}{on line,
  boxrule=0pt,
  colback=gray!15,
  arc=2pt,
  boxsep=1pt,
  left=1pt, right=1pt, top=1pt, bottom=1pt,
}

\newcommand{\icodeverb}[1]{%
  \icodebox{%
    \raisebox{-.2ex}{\texttt{#1}}%
    \vphantom{Ay}
  }%
}

\begin{document}

\maketitle

\begin{abstract}
Refactoring tools are central to modern software development, with extract-function refactorings used heavily in day-to-day work. For Rust, however, the combination of ownership, borrowing, and advanced type system features makes automated extract-function refactoring particularly challenging. Existing tools, including the original Rusty Extraction Maestro (REM) prototype, either rely on slow, compiler-based analysis, support only a restricted fragment of the language, or provide little assurance beyond ``it still compiles''.

This paper presents REM2.0, a new extract-function and verification toolchain for Rust. REM2.0 works on top of \texttt{rust-analyzer} as a persistent daemon, reusing its analysis infrastructure to provide low-latency extract-function refactorings with a VSCode front-end. On top of this extraction engine, it adds a repairer that automatically adjusts lifetimes and function signatures when extraction exposes borrow-checker issues, and an optional verification pipeline that connects to \texttt{CHARON} and \texttt{AENEAS} to generate Coq proofs of equivalence between original and extracted functions for a supported subset of Rust.

The architecture is evaluated on three benchmark suites. First, on the original REM artefact (being forty cases drawn from real projects like gitoxide), REM2.0 achieves 100\% compatibility: it matches all original successes of the prototype and fixes several previous failures. It also reduces extraction latency from an average of around a thousand milliseconds to low single-digit milliseconds inside the daemon and well under a quarter of a second as perceived in the editor. Second, on a new corpus of forty feature-focused extractions from twenty highly starred GitHub repositories, REM2.0 successfully handles the majority of examples involving \texttt{async}/\texttt{await}, \texttt{const fn}, non-local control flow, generics, and higher-ranked trait bounds, with remaining failures concentrated around dynamic trait objects and complex generic bounds. Third, on a set of twenty verification benchmarks, the CHARON/AENEAS pipeline is able to construct end-to-end equivalence proofs for those cases that fall within its current subset, at a significantly higher performance cost than extraction but with correspondingly stronger guarantees.

Overall, the results show that a Rust Analyzer based design can provide fast, feature-rich extract-function refactoring for real Rust programs, while an opt-in verification pipeline can deliver machine-checked behaviour preservation for safety-critical refactorings. REM2.0 is therefore able to offer a highly practical path toward routinely verified refactoring in the Rust ecosystem.
\end{abstract}

\input{introduction}
\input{chapter1}
\input{chapter2}
\input{chapter3}
\input{chapter4}
\input{conclusions}

\appendix
\input{appendix_main}

\bibliographystyle{unsrtnat}
\bibliography{bibfile}

\end{document}

%% file: introduction.tex
\section{Introduction}
\label{sec:introduction}
Rust is a modern systems language that combines low-level control with strong safety guarantees~\cite{the_rust_language}. This is very different to previous norms, where a programmer had to make the choice between low-level control (with languages such as C/C++) and memory safety (with garbage collected languages such as Java). These guarantees, however, make automated refactoring—particularly ``Extract Method''—far more complex than in garbage-collected languages. Prior work such as \textit{Adventure of a Lifetime}\cite{AdventureOfALifetime} and the Rusty Extraction Maestro (REM) prototype\cite{BorrowingWithoutSorrowing} demonstrated that automated extraction in Rust is possible, but only within a narrow fragment of the language and with heavy dependence on unstable tooling.

This project revisits the problem from first principles, redesigning REM into a standalone, Rust-Analyzer (RA) driven toolchain capable of handling real-world Rust features such as \texttt{async}/\texttt{await}, generics, trait objects, and const contexts, whilst also bringing the execution time down by a factor of five. We also introduce an optional verification pipeline to check that extractions preserve behaviour. Together, these contributions aim to make safe, semantics-preserving extract-method refactoring practical for everyday Rust development rather than a research prototype.

\vspace*{-5mm}
\subsection{Context and Motivation}
\label{sec:motivation}
Modern software is rarely written from scratch. Instead, engineers spend much of their time extending, adapting, and repairing existing codebases. As systems grow, crates gain more responsibilities, functions become longer, and dependencies become more complex. Without a deliberate effort to manage this complexity, technical debt rapidly accumulates and future changes become increasingly costly. Refactoring, the task of restructuring a program's internals without altering its observable behaviour---is one of the main tools developers use to keep codebases maintainable.

Empirical evidence suggests that refactoring is not an occasional clean-up, but a routine part of development practice. A large-scale survey of 1{,}183 developers by JetBrains found that most respondents refactor at least weekly, and many reported spending an hour or more in a single refactoring session~\cite{OneThousandOneStories-SoftwareRefactoring}. This indicates that refactoring is both widespread and potentially time-consuming when performed manually, and that tool support plays an important role in enabling small, safe transformations to be applied regularly.

Within the broad scope of refactorings, \emph{Extract Method} is one of the most common operations. Given a contiguous block of code (for example, the body of a loop), Extract Method moves that block into a new function and replaces the original code with a call to that function. This improves readability by assigning a descriptive name to a piece of logic and shortening long functions, and, by isolating the codes behaviour, we enable reuse and targeted unit testing. In garbage-collected languages such as Java or C\#, performing Extract Method is relatively straightforward: the tool identifies the free variables in the selected fragment, turns them into parameters or return values, and relies on the runtime to manage object lifetimes and memory.

Rust, however, presents additional challenges. As a modern systems programming language, Rust provides C-like control over memory and performance while garunteeing of memory safety through its ownership and borrowing model~\cite{the_rust_language}. Rather than relying on garbage collection, the compiler enforces a static discipline: each value has a unique owner; borrows are restricted in number and kind; and lifetimes are tracked to ensure that references never outlive the data they point to. This eliminates broad classes of bugs, such as use-after-free and many data races, but also tightly couples control-flow structure, data ownership, and aliasing patterns.

As a consequence, na\"ively reusing Extract Method algorithms from garbage-collected languages does not work in Rust. Moving code into a new function can change which scope owns a value or how many mutable and immutable references to that value are simultaneously in play. A transformation that appears harmless at the level of syntax may cause the borrow checker to reject the program, or may alter which parts of the program are permitted to mutate shared state. Even when behaviour is preserved, the transformed code may require explicit lifetime parameters or more complex type signatures, making it harder for developers to read and maintain.

Rust is increasingly used for performance-critical and safety-critical infrastructure. In these domains, developers want both the maintainability benefits of refactoring and the safety guarantees that motivate the use of Rust. If refactoring tools frequently break compilation, produce unreadable code, or fail to support common language features such as \icodeverb{async/await}, closures, and generics, developers are less likely to rely on them and may instead resort to manual editing or just ignoring the problem.

Prior work has shown that non-trivial Extract Method refactoring in Rust is possible. In particular, \emph{Adventure of a Lifetime} (AoL) and the associated Rusty Extraction Maestro (REM) prototype introduced a Rust-specific refactoring pipeline that explicitly reasons about ownership, borrowing, and non-local control flow~\cite{AdventureOfALifetime}. However, the original REM implementation is tightly coupled to an outdated IntelliJ IDEA-based Rust frontend and heavily relies on highly unstable \texttt{rustc} internals, which at best result in behaviour changes from one version of \texttt{rustc} to the next. It also provides no guarantees of semantic preservation beyond the informal assurance of passing tests. This project is motivated by the gap between the clear need for robust refactoring support in real-world Rust development and the current lack of comprehensive, lifetime-aware tools that integrate into modern Rust workflows.

\vspace*{-5mm}
\subsection{Problem Statement}
\label{sec:problem_statememt}
\vspace{-2.5mm}
Current IDE support for Extract Method in Rust is adequate for simple cases, but limited when code involves realistic borrowing patterns, complex generics, macros, or asynchronous constructs. Tools such as RA and the IntelliJ-based plugins generally succeed on small, local fragments, but tend either to refuse more complex refactorings or to produce code that no longer compiles. They also do not attempt to repair lifetimes or ownership, and they provide no guarantees that the refactoring has preserved behaviour.

The Rusty Extraction Maestro (REM) prototype from \emph{Adventure of a Lifetime}~\cite{AdventureOfALifetime} addressed some of these shortcomings by introducing Rust-specific analyses and, in particular, an automated lifetime and ownership repair loop. After applying an Extract Method transformation, REM iteratively adjusts the extracted function's signature in response to compiler feedback until the borrow checker accepts it. To the best of our knowledge, this form of automated lifetime repair is not available in any other refactoring tool for Rust. However, REM remains a research prototype: it is tied to an obsolete IntelliJ-based frontend and fragile \texttt{rustc} internals, it incurs substantial latency due to repeated \texttt{cargo check} invocations, and its support for advanced language features such as \icodeverb{async}/\icodeverb{await}, closures, complex generics, macros, and partial moves of struct fields is non-existent.

A further limitation, shared by both REM and IDE-native refactorings, is the lack of strong guarantees that transformations preserve program behaviour. REM performs complex structural changes, including modifications to function boundaries and ownership patterns, and may introduce auxiliary types or explicit lifetime parameters. Test suites can provide some assurance that the refactored program behaves as intended, but they are necessarily incomplete and only indicate that an error exists, not where it was introduced. For a refactoring tool intended to operate on non-trivial Rust code, there is a clear need for an updated design that combines REM's advanced lifetime-repair capabilities with modern toolchain support and a lightweight verification pipeline that offers stronger evidence of behaviour preservation.

\vspace*{-5mm}
\subsection{Project Aims}
\label{sec:project_aims}
\vspace{-2.5mm}
This project aims to turn the original REM prototype into a practical tool for modern Rust development. The first goal is to re-engineer REM as a standalone pipeline that no longer relies on outdated IntelliJ-based infrastructure or fragile \texttt{rustc} internals, and to extend its Extract Method capabilities to better handle modern Rust language features such as \icodeverb{async}/\icodeverb{await}, closures, macros, and more complex uses of generics.

The second goal is to connect our Extract Method refactorings to a equivalence checking pathway that can provide stronger evidence of semantic preservation than tests alone. Specifically, the project aims to integrate REM with existing verification-oriented toolchains (in particular CHARON and AENEAS) so that refactorings can be translated into a form suitable for logical reasoning about equivalence between the original and refactored code.

The third goal is to expose all of this new functionality through an interface that fits naturally into existing workflows. This includes a command-line interface suitable for scripting and batch evaluation, as well as a Visual Studio Code extension that allows developers to invoke REM directly from the editor. A longer-term ambition, which is beyond the scope of this project, is to explore integration with RA itself, for example by reusing its intermediate representations more directly or embedding a subset of REM's analysis into the language server.

\vspace*{-5mm}
\subsection{Key Contributions}
\label{sec:key_contributions}
\vspace{-2.5mm}
To achieve these aims, this project makes the following contributions:

\textbf{A re-engineered REM core for modern toolchains.}
The original REM implementation has been refactored into a standalone Extract Method pipeline that decouples the analysis and transformation logic from the obsolete IntelliJ-based frontend and fragile \texttt{rustc} internals. The new daemon based approach is readily compatible with existing IDEs and workflows. \\
\textbf{Integration with Rust Analyzer for fast, in-memory analysis.}
The project introduces a new extraction engine that builds on RA's semantic information and in-memory workspace model. This gives REM2.0 a far deeper semantic understanding of code than previous AST-based approaches. \\
\textbf{Extended extraction coverage for modern Rust features.}
The Extract Method pipeline has been extended to support a broad range of real-world Rust constructs including selected uses of \texttt{async}/\texttt{await}, generics, non-local control flow, const evaluation, and higher-ranked trait bounds. \\
\textbf{Editor/IDE integration via \texttt{rem-server} and VSCode.}
The core implementation is exposed as \texttt{rem-server}, a long-running JSON-RPC process suitable for integration with arbitrary IDEs. A lightweight CLI and prototype VSCode extension demonstrate practical usage today.\footnote{REM-VSCode is available on the VSCode marketplace, \url{https://marketplace.visualstudio.com/items?itemName=MatthewBritton.remvscode}} \\
\textbf{A prototype equivalence checking pipeline using CHARON and AENEAS.}
REM2.0 leverages recent advances in program verification to translate Rust into proof-oriented functional languages such as Coq, enabling automated behavioural equivalence checking for Extract Method refactorings. \\
\textbf{An empirical evaluation on many real-world Rust projects.}
We evaluate the REM pipeline across a suite of widely used Rust projects, measuring coverage, performance, and failure modes. This demonstrates that the new architecture preserves REM's strengths—particularly automated lifetime and ownership repair—while significantly improving practical applicability and language coverage.

\vspace*{-5mm}
\subsection{Method Overview and Scope}
\label{sec:method-overview-scope}
\vspace{-2.5mm}
At a high level, this project constructs a multi-stage Extract Method pipeline for Rust and evaluates it on real-world codebases. The core extraction and analysis stages reuse RA's in-memory workspace and semantic information to identify candidate fragments and generate initial refactorings. A dedicated repairer then processes the extracted functions fixing lifetimes, and ownership patterns until the transformed code is valid Rust. For a selected subset of examples, the original and refactored programs are translated via CHARON and AENEAS into verification-oriented intermediate representations, enabling reasoning about semantic preservation. The resulting toolchain is exposed through a JSON-RPC server (\texttt{rem-server})\footnote{https://crates.io/crates/rem-server}, a lightweight CLI (\texttt{rem-cli})\footnote{https://crates.io/crates/rem-command-line}, and a Visual Studio Code extension, and is evaluated on a suite of open-source Rust crates. The scope of this work is limited to Extract Method refactorings; other refactoring operators are not considered. Support for macros and very complex generic patterns remains partial, and the verification pipeline is applied only to representative case studies rather than entire codebases. Full integration into RA itself is treated as a long-term direction rather than an immediate goal.

%% file: chapter1.tex
\section{Background}
\label{sec:background}

Rust's ownership and borrowing discipline makes it possible to write
memory-safe, data-race–free systems code without a garbage collector, but it
also raises the bar for both automated refactoring and formal reasoning about
program behaviour~\cite{the_rust_language,automated_refactoring_of_rust_programs}.
This section surveys the background needed for the rest of this report. We begin
by reviewing refactoring and, in particular, the Extract Method transformation.
We then introduce Rust's ownership, borrowing, lifetimes, and control-flow
features, and explain how they interact with Extract Method. Building on this,
we describe prior work on automated refactoring for Rust, with a focus on
\emph{Adventure of a Lifetime} and the Rusty Extraction Maestro (REM)
prototype~\cite{AdventureOfALifetime,BorrowingWithoutSorrowing}. Finally, we
outline relevant work on verification and formal semantics of Rust, and briefly
discuss existing tooling and IDE integration. Together, these sections position
the contributions of this project within the current state of the art in making
Rust code easier to evolve and reason about.

\vspace*{-5mm}
\subsection{Overview of Refactoring}
\label{sec:overview_refactoring}
\vspace{-2mm}
Refactoring is the process of restructuring existing code to improve its internal
structure without changing its externally observable behaviour. It is a key
practice for improving code maintainability and controlling technical debt in
large software projects~\cite{OneThousandOneStories-SoftwareRefactoring}. Modern
IDEs such as Eclipse and IntelliJ IDEA provide a wide range of automated
refactorings, allowing developers to apply semantics-preserving transformations
quickly and, in principle, safely~\cite{AdventureOfALifetime,Formal_Specifiation_JAVA}.

As discussed in Section~\ref{sec:motivation}, empirical studies show that
refactoring is a routine part of professional development, but automated
refactoring tools are not always used to their full potential. Developers often
prefer manual edits, either because they are unfamiliar with the available
refactorings or because they are unsure whether the tool will preserve
behaviour. This motivates research into refactoring tools that are both more
reliable (through stronger guarantees) and more transparent in how they operate.

In this report we focus on the \emph{Extract Method} refactoring, already
introduced in Section~\ref{sec:motivation}. In the refactoring space,
Extract Method is often treated as a basic building block: a contiguous block of code
is moved into a newly created function, and the original code is replaced by a
call to that function. This is used to split long methods into smaller units, to
isolate cohesive behaviour, and as a stepping stone for more complex
refactorings such as Extract Class or Move Method~\cite{Formal_Specifiation_JAVA}.
In garbage-collected languages such as Java or C\#, most implementations are
largely syntactic: they identify the free variables within the selected fragment,
turn them into parameters or return values, and rely on the runtime to manage
object lifetimes and memory.

\begin{paracol}{2}
\setlength{\columnsep}{2em}
\sloppy
\switchcolumn[0]

The Java example in Listing~\ref{lst:extract-method-java} illustrates this style
of transformation in a language where explicit memory management is not a
concern. A long method containing, for instance, input validation and
transformation logic can be refactored by selecting the validation block,
extracting it into a dedicated method, and replacing the original block with a
call. The resulting code is clearer, and the validation can be tested in
isolation without changing the behaviour of the program.

\switchcolumn

\begin{lstlisting}[
    basicstyle=\scriptsize\ttfamily,
    frame=lines,
    numbers=left,
    breaklines=true,
    language=Java
]
// Before:
public void doComplicatedWork(List<Integer> nums) {
    int sum = 0;
    for (int n : nums) { // Consider this as a 
        sum += n;        // Lengthy piece of 
    }                    // Business logic
    System.out.println("Sum is: " + sum);
}

// After:
public void doComplicatedWork(List<Integer> nums) {
    int sum = doSmallPartOfWork(nums); // Logic is now reusable
    System.out.println("Sum is: " + sum);
}
// This function can be unit tested on its own
private int doSmallPartOfWork(List<Integer> nums) {
    int sum = 0;         
    for (int n : nums) {
        sum += n;
    }
    return sum;
}
\end{lstlisting}
\captionsetup{type=listing}
\captionof{listing}{Example of Extract Method refactoring in Java}
\label{lst:extract-method-java}
\end{paracol}

In Rust, the high-level intent of Extract Method is the same, but the
transformation interacts with ownership, borrowing, and lifetimes in ways that
do not arise in typical garbage-collected languages. Understanding these
language features is therefore essential before discussing automated Extract
Method refactoring for Rust.

\vspace*{-5mm}
\subsection{Rust's Ownership Model \& Semantics}
\label{sec:rusts_ownership_model}

Rust enforces memory safety through a static ownership and borrowing discipline.
Each value has a unique owner; when the owner goes out of scope, the value is
dropped\footnote{We recommend reading this section of Rust By Example \cite{rust_by_example} to better understand lifetimes if this is your first encounter! \\ \url{https://doc.rust-lang.org/rust-by-example/scope/lifetime.html}}. Instead of relying on a garbage collector, Rust uses the type system
and a compile-time borrow checker to prevent use-after-free, data races, and
other common memory errors~\cite{the_rust_language,automated_refactoring_of_rust_programs,automatically_enforcing_rust_trait_properties}. Similar ownership disciplines have been explored in other languages, notably through generic ownership types for Java~\cite{PotaninNobleClarkeBiddle2006oopsla,ZibinPotaninLiAliErnst2010}.
While these guarantees are attractive for systems programming, they complicate
both manual and automated program transformations, including Extract Method.

\newpage
\vspace*{-2.5mm}
\subsubsection{Ownership and Moves}
\begin{paracol}{2}
\setlength{\columnsep}{2em}
At the core of Rust's model is the idea that each value has a single owner.
Assigning or passing a value typically moves ownership to a new variable or
function, leaving the previous binding invalid. Listing~\ref{lst:ownership-left} on the right
shows a simple example:

\switchcolumn

\begin{lstlisting}[
    basicstyle=\scriptsize\ttfamily,
    frame=lines,
    numbers=left,
    breaklines=true,
    language=Rust
]
fn main() {
    let s1 = String::from("Hello");
    let s2 = s1; // Ownership of the string data moves to s2
    // The following line would be invalid if uncommented:
    // println!("{}", s1);
    // ^ value borrowed here after move
    println!("{}", s2); // Prints "Hello"
}
\end{lstlisting}

\captionsetup{justification=centering}
\captionsetup{type=listing}
\captionof{listing}{Ownership and moves in Rust}
\label{lst:ownership-left}
\end{paracol}

In this example, the value is moved from \texttt{s1} to \texttt{s2}, making
\texttt{s1} unusable afterwards. An Extract Method transformation that passes
\texttt{s1} into a new function by value may therefore change which scope owns
the value or when it is dropped. Automated refactoring tools must account for
these moves to avoid introducing subtle ownership bugs or compilation failures.

\vspace*{-2.5mm}
\subsubsection{Borrowing}
Instead of transferring ownership, Rust encourages passing references. A value
can be borrowed immutably (using \icodeverb{\&T}) any number of times, or
mutably (using \icodeverb{\&mut T}) exactly once at a time. The borrow checker
enforces that mutable and immutable borrows do not conflict, preventing data
races and many classes of aliasing bugs~\cite{ZibinPotaninAliArtziKiezunErnst2007}. Listing~\ref{lst:borrowing-imm-mut}
illustrates the basic patterns.

\begin{figure}[H]
    \centering
    \begin{subfigure}[t]{0.47\textwidth}
        \centering
\begin{lstlisting}[
            basicstyle=\scriptsize\ttfamily,
            frame=lines,
            numbers=left,
            breaklines=true,
            language=Rust
        ]
fn print_message(message: &str) {
    println!("{}", message);
}

fn main() {
    let greeting = String::from("Hello, world!");
    print_message(&greeting); // Borrows greeting immutably
    println!("Still own greeting: {}", greeting);
}
\end{lstlisting}
        \captionsetup{justification=centering}
        \sublistingcaption{Immutable borrow: many readers allowed}
        \label{lst:borrow-immutable}
    \end{subfigure}
    \hfill
    \begin{subfigure}[t]{0.47\textwidth}
        \centering
\begin{lstlisting}[
            basicstyle=\scriptsize\ttfamily,
            frame=lines,
            numbers=left,
            breaklines=true,
            language=Rust
        ]
fn add_exclamation(message: &mut String) {
    message.push('!');
}

fn main() {
    let mut greeting = String::from("Hello");
    add_exclamation(&mut greeting); // Borrows greeting mutably
    println!("{}", greeting);       // Prints "Hello!"
}
\end{lstlisting}
        \captionsetup{justification=centering}
        \sublistingcaption{Mutable borrow: a single writer}
        \label{lst:borrow-mutable}
    \end{subfigure}
    \captionsetup{type=listing,justification=centering}
    \caption{Borrowing in Rust}
    \label{lst:borrowing-imm-mut}
\end{figure}

When a block of code is extracted into a new function, the tool must decide for
each captured variable whether to move it, borrow it immutably, or borrow it
mutably. Poor choices can cause the borrow checker to reject the refactored
program or force the tool to introduce unnecessary clones that change
performance characteristics. This is one of the core challenges for designing an Extract
Method algorithm in Rust.

\vspace*{-2.5mm}
\subsubsection{Lifetimes and Inference}
The Rust compiler tracks how long references remain valid using lifetime
parameters. Many simple cases do not require explicit annotations because the
compiler can infer relationships between input and output lifetimes. For
example, the function in Listing~\ref{lst:lifetime-inference} can be written
without explicit lifetime parameters:

\begin{figure}[h]
\begin{lstlisting}[frame=none, numbers=none, language=Rust]
fn echo(message: &str) -> &str { message }
\end{lstlisting}
    \captionof{listing}{\small How we as developers write Rust code}
    \captionsetup{justification=raggedright}
    \label{lst:lifetime-inference}
\end{figure}

\vspace{-2.5mm}
Conceptually, the compiler treats this as if the programmer had written
Listing~\ref{lst:lifetime-expanded}, introducing an explicit lifetime
\icodeverb{'a} shared by the argument and result:

\begin{figure}[h]
\begin{lstlisting}[frame=none, numbers=none, language=Rust]
fn echo<'a>(message: &'a str) -> &'a str { message }
\end{lstlisting}
    \captionof{listing}{\small How the Rust compiler ``sees'' Rust Code}
    \captionsetup{justification=raggedright}
    \label{lst:lifetime-expanded}
\end{figure}

\vspace{-2.5mm}
When code is extracted into a new function, however, the relationships between
references can become more complex, and the compiler may no longer be able to
infer lifetimes automatically. In such cases, explicit lifetime parameters are
required on the extracted function. Automated refactoring tools must therefore
be prepared to introduce, propagate, and most importantly simplify lifetime annotations as part
of the transformation.

\vspace*{-2.5mm}
\subsubsection{Non-local Control Flow}
Rust allows \icodeverb{return}, \icodeverb{break}, and \icodeverb{continue}
inside nested blocks and loops. Extracting a fragment that contains such
statements into a separate function changes the control-flow structure: an early
\icodeverb{return} from the original function becomes a \icodeverb{return} from
the extracted function instead. Any realistic Extract Method implementation for
Rust must either forbid such cases or explicitly encode and reconstruct
non-local control flow. This point will be revisited in the discussion of the
REM tool in Section~\ref{sec:rem_background}.

Overall, Rust's ownership, borrowing, lifetimes, and control-flow constructs
provide strong safety guarantees but tightly couple data flow, control flow, and
scope. This makes even seemingly local refactorings, such as Extract Method,
significantly more complex than in more permissive, garbage-collected
languages.

\vspace*{-5mm}
\subsection{Automated Refactoring Techniques for Rust}
\label{sec:automated_refactoring_techniques}

\vspace*{-2.5mm}
\subsubsection{Renaming and Simple Refactorings}
One of the first efforts to build a Rust refactoring tool was by G. Sam et al.
(2017), who created a proof-of-concept refactoring framework utilizing the Rust
compiler's internal APIs. The team partnered with Mozilla Research\footnote{Mozilla Research were the first major sponsors of the Rust project, and have been a driving force in its development over the last ten years. \cite{rust_wikipedia_2025}. \\ \url{https://blog.mozilla.org/en/mozilla/mozilla-welcomes-the-rust-foundation/}} to be among
the first to implement the Rust specific refactorings of \textit{Lifetime
Elision} and \textit{Lifetime Reification}. This allowed their program to
introuce explicit lifetime parameters in instances where the compiler was
implicitly inferring them - a refactoring that was brand new to Rust. The
challenges encountered illustrated how Rust's stricter scoping and shadowing
rules required careful handling of name conflicts during renaming (e.g. avoiding
situations where renaming a variable could unintentionally shadow another)
\cite{automated_refactoring_of_rust_programs}. Additionally, the 2017 study
concluded that many refactorings are possible with Rust's compiler
infrastructure, but ensuring \textit{behavioural preservation} (especially
around ownership transfers) requires additional static analyses not needed in
langauges without Rust's constraints. Their work on \textit{Lifetime Elision}
has since formed part of the REM toolchain, where the rich feedback from the
Rust compiler is leveraged to ensure that the final transformation is both
valid Rust and as legible as possible \cite{AdventureOfALifetime},
\cite{BorrowingWithoutSorrowing}.

\vspace*{-2.5mm}
\subsubsection{Extract Method (REM)}
\label{sec:rem_background}
A major advance in automated refactoring for Rust came with the Rusty Extraction
Maestro (REM). Costea et al.'s \emph{Adventure of a Lifetime: Extract Method
Refactoring for Rust}~\cite{AdventureOfALifetime} introduced the theoretical
framework, and Thy et al.'s \emph{Borrowing Without Sorrowing} provided a
practical implementation as an IntelliJ plugin~\cite{BorrowingWithoutSorrowing}.
REM tackles Extract Method in Rust by decomposing it into a sequence of
transformations, each designed to address a specific aspect of Rust's ownership,
borrowing, lifetimes, and control flow.

The process begins with a na\"ive hoisting step that moves the selected block of
code into a new function and replaces the original fragment with a call. This
initial transformation is rarely well-typed. REM then applies a series of
\emph{automated repairs} guided by specialised analyses called \emph{oracles}.
An ownership analysis determines, for each value crossing the new function
boundary, whether it should be passed by value, by shared reference
(\icodeverb{\&T}), or by mutable reference (\icodeverb{\&mut T}). Other passes
reify non-local control flow by encoding early \icodeverb{return},
\icodeverb{break}, and \icodeverb{continue} statements into an auxiliary enum,
allowing the caller to reconstruct the original behaviour by pattern matching on
the result. A final repair loop iteratively invokes the Rust compiler, uses the
borrow checker's error messages as feedback, and introduces or refines lifetime
parameters until the extracted function type-checks.

\begin{figure}[h]
    \centering
    \includegraphics[width=0.8\textwidth]{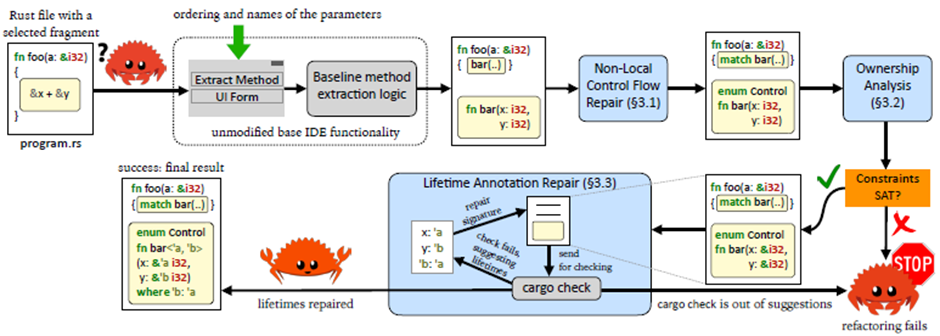}
    \caption{Overview of the REM extract-method pipeline (after Costea et al.~\cite{AdventureOfALifetime})}
    \label{fig:aol_process}
\end{figure}

Figure~\ref{fig:aol_process} (adapted from~\cite{AdventureOfALifetime}) shows
the overall REM pipeline. Each phase is deliberately simple, but their
composition yields a transformation that can handle (mostly) realistic Rust code. In an
evaluation on real-world projects, REM was able to perform extractions that IDE
tools could not handle, including cases involving multiple interacting borrows
and nested lifetimes~\cite{AdventureOfALifetime}. To the best of our knowledge,
REM's automated lifetime and ownership repair loop remains unique among
refactoring tools for Rust.

\vspace{-2.5mm}
Despite these strengths, the original REM prototype is tightly coupled to an
outdated IntelliJ-based Rust frontend and unstable \texttt{rustc} internals, as
discussed in Section~\ref{sec:introduction}. This limits its applicability in
modern toolchains. Our project builds on the conceptual design of REM,
particularly its lifetime and ownership repair capabilities, while re-engineering
the implementation around Rust Analyzer and a standalone daemon architecture to
make Extract Method refactoring practical for contemporary Rust development.

\vspace*{-2.5mm}
\subsubsection{Automated Fixes for Ownership Errors}
Another line of work relevant to Rust refactoring focuses on automatically repairing code that violates ownership or borrowing rules—transforming non-compiling programs into compiling ones without altering semantics. \textit{Rust-Lancet} (Yang et al., ICSE 2024) targets exactly this problem, aiming to fix ownership-rule violations while preserving behaviour. Given a Rust file that fails to compile, Rust-Lancet analyses its AST using \texttt{syn} and \texttt{quote}, applies targeted transformations, and repeatedly rechecks the program until the error is eliminated. Typical repairs include inserting a \icodeverb{clone()} to satisfy ownership constraints or reordering statements to avoid invalid borrows. Crucially, Rust-Lancet includes a behaviour-preservation check: it validates that each proposed fix does not change observable program output, using a specialised semantics for reasoning about programs that do not yet compile.

Across 160 real ownership violation cases, Rust-Lancet successfully repaired a large majority with zero false positives—never introducing an incorrect fix—outperforming both compiler suggestions and several LLM-based baselines. Although positioned as a bug-fixing tool rather than a refactoring engine, its pipeline of semantics-preserving AST rewrites overlaps strongly with automated refactoring. Their results highlight how Rust's strict ownership model can guide automated transformations: once the compiler accepts the repaired code and behaviour-preservation checks succeed, the tool can be confident the patch is correct \cite{RustLancet}.

\vspace*{-5mm}
\subsection{Verification and Formal Methods for Rust}
\label{sec:verification_formal_methods}
Rust's appeal for building reliable systems has driven a lot of active work and research into
formal semantics and verification. Traditional deductive verification and model
checking must be adapted to account for Rust's ownership, lifetimes, and
\texttt{unsafe} code, but Rust's strong static guarantees also make many
verification tasks much more manageable when compare to traditional approaches for C or C++.

\vspace*{-2.5mm}
\subsubsection{RustBelt and Type-System Soundness}
Jung et al.'s \textbf{RustBelt} gives the first machine-checked soundness proof
for a realistic subset of Rust's type system~\cite{RustBelt}. By embedding a
Rust-like calculus ($\lambda$Rust) in the Iris concurrent separation logic and
linking it to low-level memory models, RustBelt shows that well-typed safe Rust
code cannot exhibit undefined behaviour, even when using libraries that contain
\texttt{unsafe} internals. This foundational result allows later work to assume
that safe Rust is memory and thread-safe and focus verification effort either
on functional correctness or on carefully selected \texttt{unsafe}
fragments.

\vspace*{-2.5mm}
\subsubsection{Verification via Functional Translation}
A complementary approach is to translate Rust programs into pure functional
languages that work with existing proof assistants and verifiers. The CHARON\footnote{\url{https://github.com/AeneasVerif/charon/}}/AENEAS\footnote{\url{https://github.com/AeneasVerif/aeneas}}
toolchain follows this strategy. CHARON translates Rust into an intermediate
representation called \emph{Low-Level Borrow Calculus} (LLBC), which is related
to Rust's MIR but makes the ownership and borrowing structure explicit. AENEAS then
assigns a pure, value-based semantics to LLBC and translates it into a lambda
calculus suitable for verification in tools such as Coq or F*~\cite{AENEAS,
AENEAS_PART_2}. This approach builds on a long tradition of encoding language semantics in proof assistants~\cite{MackayMehnertPotaninGrovesCameron2012}. For a large class of Rust programs, this
translation eliminates the need to reason about pointers and heaps directly:
proof engineers can work with a pure functional model, while the design of CHARON/AENEAS
guarantee that proved properties carry back to the original Rust code. Indeed later in this report, (Section~\ref{sec:proof_obligations}) we will rely on this guarantee to back our verification pipeline.

More recently, Ho et al.\ introduced \emph{symbolic semantics} for Rust and
showed that they form a sound abstraction of concrete heap-based execution for
LLBC programs~\cite{AENEAS_PART_2}. A key outcome is a verified symbolic
interpreter that can serve as a proven-correct borrow checker for LLBC. This
provides strong assurance that tools built on CHARON/AENEAS rest on a solid
semantic foundation and, crucially, the have established a bridge between Rust's informal ownership
intuition and a mathematically precise model.

\vspace*{-2.5mm}
\subsubsection{Static Analysers and Model Checkers}
Beyond full functional verification, several tools provide lighter-weight
guarantees for Rust. \textit{Prusti} translates annotated Rust code into the
Viper verification language, using Rust's ownership and lifetime information to
encode permissions and prove user-specified pre- and post-conditions and
invariants~\cite{prust_in_practice,prusti_project}. \textit{RefinedRust} extends
this idea with a refinement type system for Rust, allowing rich safety and
correctness properties to be expressed as types and discharged in Coq, even for
some \texttt{unsafe} code~\cite{RefinedRust}.

On the model-checking side, \textit{Kani} is a bounded model checker for Rust
that operates on MIR and generates verification conditions for a SAT/SMT
solver~\cite{verifying_dynamic_trait_objects}. By working at the MIR level, Kani
can reason precisely about traits, generics, and dynamic trait objects~\cite{GarianoServettoPotaninArora2019}, and has
been applied to critical components such as cryptographic libraries. Dynamic
analysers like \textit{Miri} complement these tools by executing Rust programs
in an interpreter that detects undefined behaviour, providing strong testing-time
checks even when full verification is not attempted.

Together, RustBelt, CHARON/AENEAS, Prusti, RefinedRust, Kani, and Miri outline a
rich verification ecosystem. They demonstrate that Rust's ownership discipline is
not only a constraint but also a powerful abstraction layer that verification
tools can exploit.

\vspace*{-5mm}
\subsection{Integration with IDEs and Langauge Servers}
\label{sec:integration_with_ides}
\vspace{-1mm}
For refactoring and verification tools to be adopted in practice, they must
integrate smoothly into developers' workflows. For Rust, this has centred on
the compiler and the Language Server Protocol (LSP). The original Rust Language
Server (RLS) attempted to repurpose \texttt{rustc}'s internals for IDE support,
but struggled with performance and keeping pace with language evolution. The
community has since converged on \textit{Rust-Analyzer} (RA), a language
server that maintains its own syntax trees and performs incremental type
inference to answer editor queries quickly~\cite{Schiedt_2022}. RA
exposes refactorings and code transformations as "assists" that operate on its
internal representation; and these already include many common refactorings.

Industrial IDEs have also hosted Rust refactoring research. The original REM
prototype was implemented as an extension to the IntelliJ Rust plugin, showing
that research-grade refactorings can be prototyped on top of existing editor
infrastructure~\cite{AdventureOfALifetime,BorrowingWithoutSorrowing}. In
practice, maintaining such integrations is challenging: IDE APIs and Rust
tooling both evolve rapidly. To mitigate this, some tools deliberately avoid
\texttt{rustc} internals and instead rely on stable crates like
\texttt{syn}/\texttt{quote} for parsing and code generation, as in Lancet's
design~\cite{RustLancet}.

At the compiler level, projects like \textit{rustfix} show another style of
integration: \texttt{rustc} emits machine-applicable suggestions for certain
errors and deprecations, which rustfix can apply automatically across a
codebase. While not a general refactoring engine, this mechanism enables
small-scale, behaviour-preserving edits (for example, updating deprecated syntax
or inserting missing lifetime specifiers) driven directly by the compiler. As
Rust's stable analysis APIs mature and RA's assist framework
expands, there is increasing scope for external tools—such as the
\texttt{rem-server} daemon described in this report—to plug into these
ecosystems and offer more advanced, semantics-aware refactorings without being
tied to a single IDE.

\vspace*{-5mm}
\subsection{Related Work Beyond Rust}
\label{sec:related_work_beyond_rust}
Many of the challenges and opportunities in Rust refactoring echo earlier work
in other languages. Java and C\# have long enjoyed mature IDE support for
refactorings such as \emph{Rename}, \emph{Extract Method}, and \emph{Extract
Class}, and this ecosystem has inspired a large body of research. Beyond classic
examples such as generic-aware refactoring~\cite{GenericRefactoringJAVA}, more
recent work has focused on systematic migration to new language features.
Zhang et al.'s \textit{ReFuture} tool, for example, automatically refactors Java
asynchronous code to use the modern \icodeverb{CompletableFuture} API by combining
static analyses (visitor patterns, alias analysis) with transformation rules, and
has been evaluated on large projects such as Hadoop and ActiveMQ~\cite{AutomaticRefactoringAsyncJAVA}.
In the mobile domain, Lin et al.'s \textit{ASYNCDROID} refactors problematic
uses of Android's \icodeverb{AsyncTask} into safer constructs such as
\icodeverb{IntentService}, addressing common sources of leaks and UI bugs and
showing that domain-specific refactoring rules can be packaged as an IDE
plugin~\cite{AndroidAsncRefactoring}.

Refactoring has also been studied in other paradigms and ecosystems. In Haskell,
the \textit{HaRe} tool brought structured refactoring to a lazy, purely functional
language, handling challenges such as type classes and non-strict evaluation~\cite{HaRe}.
In dynamic languages like JavaScript and Python, the lack of static types has led
to refactoring approaches that rely on heuristics, runtime analysis, or
developer-guided transformations. Meanwhile, tools such as Facebook's \textit{JSCodeshift}
and Python's \textit{Bowler} allow developers to script custom codemods over
ASTs\cite{Bowler}.

Finally, refactoring is closely linked to adjacent areas such as code smell
detection, automated program repair, and large-scale migrations. Some
repair systems use refactoring-like transformations as behaviour-preserving
steps to simplify buggy code before or after applying a fix, and language
servers increasingly act as a common backbone for these capabilities
across many editors~\cite{AdventureOfALifetime}. The Rust work surveyed in this
section fits into this broader landscape: it adapts ideas from mature ecosystems
(Java, Android, \& Haskell) to a language whose ownership-based
semantics demand specialised analyses, but whose strong guarantees can in turn
be exploited by refactoring and verification tools.

%% file: chapter2.tex
\section{Expanding the Capabilities of REM}
\label{sec:expanding_rem}

Automated refactoring in Rust has long been constrained by the languages strict
ownership, borrowing and lifetime rules. Early prototypes of the Rusty Extraction
Maestro demonstrated that \emph{Extract Method} refactorings could be made
effective, but only within relatively narrow limits. Complex language features,
such as generic types and asynchronous code, were not supported. Moreover, the
entire toolchain was reliant on a fragile integration with IntelliJ, which
quickly became outdated. As a result, REM remained an academic proof-of-concept
rather than a tool developers could realistically adopt.

This section details how REM has evolved from a fragile prototype into a robust, modular, and extensible system. At its core, REM2.0 now operates through a streamlined, multi-component pipeline designed for speed, flexibility, and integration with modern Rust workflows.

The REM server orchestrates the entire process through a JSON-RPC interface, coordinating extraction, repair, and equivalence checking processes (the latter discussed in Section \ref{sec:verification}). The extraction tool, built directly on top of Rust Analyzer's incremental analysis engine, performs lightning-fast semantic extraction while integrating the capabilities of REM's original controller and borrower modules. The repairer tool, adapted from the original REM toolchain, continues to play a key role in resolving lifetimes and other semantic inconsistencies that may arise during refactoring.

Finally, a prototype Visual Studio Code (VSCode) extension connects with the REM server to deliver near-instant feedback—typically around 200 ms end-to-end (see Section \ref{sec:evaluation}). Together, these components transform REM into a practical refactoring system capable of integrating into everyday Rust development while preserving its foundation as a platform for continued research.

\vspace{-5mm}
\subsection{Motivation: Why REM needed to change}
\label{sec:expanding_motivation}

The original REM prototype demonstrated that automated extract-method refactoring for Rust was possible, but its design also exposed several fundamental limitations. At its core, REM relied on repeated invocations of \texttt{cargo check} and a set of highly unstable \texttt{rustc} internals, as well as the IntelliJ IDEA Rust plugin, which has since been superseded by RustRover and was never particularly stable during the lifetime of this project. As a consequence, the toolchain was slow, fragile, and extremely sensitive to minor changes in both the Rust ecosystem and the underlying editor integration. Its viability as anything beyond a research prototype was effectively non-existent: before any new work could begin, it took over a month of effort just to restore the original implementation to a usable state.

These architectural choices also severely constrained the kinds of Rust programs that REM could handle. Even small, modern Rust fragments routinely use language features that REM was simply unable to process. For example, REM could not extract from even minimal asynchronous functions:
\begin{lstlisting}[basicstyle=\footnotesize\ttfamily,language=Rust]
async fn fetch() {
    let v = client.get("/").await; // REM failed to recognise the `await`
}
\end{lstlisting}
nor could it handle functions involving generic parameters or bounds:
\begin{lstlisting}[basicstyle=\footnotesize\ttfamily,language=Rust]
fn min<T: Ord>(a: T, b: T) -> T { /* ... */ }
\end{lstlisting}
and it struggled with non-local control flow such as \texttt{break} or \texttt{return} inside loops:
\begin{lstlisting}[basicstyle=\footnotesize\ttfamily,language=Rust]
for x in xs {
    if x < 0 { break; } // difficult for REM to lift cleanly into a new function
}
\end{lstlisting}

Language features that are now commonplace in production codebases---such as \texttt{async}/\texttt{await}, const evaluation, generics with non-trivial bounds, trait objects, higher-ranked trait bounds, and non-local control flow\footnote{Non-local control flow \emph{is} supported by REM, but its implementation relies on one-off custom generics rather than the modern \icodeverb{std::ops::ControlFlow} abstraction.}---lay entirely outside REM's reliably supported fragment. Although the original codebase contained many scattered TODOs mentioning future support for generics or async, there was no coherent plan for how to integrate such features into the pipeline. In practice, this meant that REM could sometimes operate successfully on specific large projects such as \texttt{gitoxide}, but in our testing, it would frequently fail or time out when faced with more complex codebases like Deno, Tokio, or Vaultwarden, where these advanced features are pervasive.

Perhaps most importantly, REM provided no formal safety net. Its extraction decisions were driven largely by syntactic AST rewriting rather than a deep semantic understanding of the program, and there was no verification pipeline to check that transformations preserved behaviour. A refactoring could silently change program semantics while still compiling, and the tool had no mechanism to detect or prevent such regressions. Together, these limitations motivated a complete redesign of the system: if extraction was to be useful for everyday Rust development, REM had to become faster, more robust, feature-complete, and semantically trustworthy.

\subsection{High Level Vision: What REM2.0 Needed to Achieve}
\label{sec:high_level_vision}

REM2.0 was designed in response to these shortcomings with a clear set of high-level goals. The first was performance: the roughly one-second turnaround time of the original REM \emph{after} IntelliJ performed the initial extraction was unacceptable for an interactive refactoring workflow. To feel natural inside an IDE, extraction had to complete in well under a second, ideally within the sub-500\,ms range. Achieving this required abandoning the ``re-run the compiler'' model in favour of a persistent semantic engine. At the same time, REM2.0 needed to move beyond the restricted language subset of its predecessor and support essentially any piece of real Rust code. This meant handling async functions, const evaluation, generics (including complex bounds), trait objects, higher-ranked trait bounds, and non-local control flow as real language features rather than TODOs in the source tree.

A second, equally important, goal was accessibility. The original REM existed primarily as a research artefact and demanded significant effort to build, configure, and operate; very few everyday developers could realistically benefit from it. REM2.0, by contrast, is intended as a tool that can be adopted with minimal friction: it should integrate cleanly with common development workflows, expose its functionality through a familiar VSCode interface, and have no interaction with unstable compiler internals\footnote{The version of these internals available to the installed program is entirely dependent on the compiler being used by the developer, and thus the original REM could be completely incompatible with a codebase.}. Behind the straightforward interface, the system must still uphold strong correctness guarantees. Refactorings are deterministic and built on a stable representation designed for reliable transformation and verification. With help from a correctness backend, the tool ensures not only safety but behavioural equivalence. In short, REM 2.0 aims to be fast, compatible with modern Rust, and never silently break code.

\vspace{-5mm}
\subsection{REM2.0 At a Glance: Core Contributions}
\label{sec:at_a_glance}
Before delving into architecture and implementation details, it is useful to summarise the key contributions of REM2.0 at a high level. Collectively, the below changes / upgrades transform REM from a fragile research prototype into a practical, extensible refactoring toolchain for modern Rust.

\begin{itemize}[noitemsep, topsep=2pt,leftmargin=1em]

  \item \textbf{Rust-Analyzer-backed extraction engine.} REM~2.0 replaces direct dependence on \texttt{rustc} internals and editor-specific plugins with a new extraction engine that uses Rust-Analyzer as its backend. A custom ``single-file workspace'' abstraction allows us to run extract-method relative to the current file while still benefiting from Rust-Analyzer's full semantic analysis capabilities.

  \item \textbf{Full language-feature coverage for extraction.} Building on Rust-Analyzer's capabilities, REM2.0 supports extraction in the presence of async functions, const evaluation, generics (including complex bounds), trait objects, higher-ranked trait bounds, and non-local control flow, rather than restricting itself to a narrow, hand-picked fragment of Rust.

  \item \textbf{Preservation and extension of REM's lifetime repair.} The new system maintains all core capabilities of the original REM, in particular its ability to repair lifetime annotations and placements after extraction, while integrating these repairs into a far more useful and usable pipeline. REM2.0 gives the developer the ability to pick and choose between maximum correctness, maximum responsiveness, or a combination of the two.

  \item \textbf{Verification backend for behavioural equivalence.} REM2.0 introduces an end-to-end verification pipeline that translates extracted code through Charon and LLBC to Aeneas and Coq, where equivalence between the original and refactored functions can be proved (or refuted) automatically. See Section~\ref{sec:ref_to_verification} and Section~\ref{sec:verification} for a full explanation.

  \item \textbf{VSCode integration for interactive refactoring.} A new VSCode extension exposes REM~2.0 as an interactive refactoring tool, providing extract-method commands, preview and undo functionality, and surfacing verifier results directly to the developer.

\end{itemize}

Together, these contributions establish REM~2.0 as a fast, feature-complete, and semantically aware extract-method toolchain that is suitable both for research and for everyday Rust development.

\vspace{-5mm}
\subsection{Overview of the REM2.0 Architecture}
\label{sec:architecture_overview}

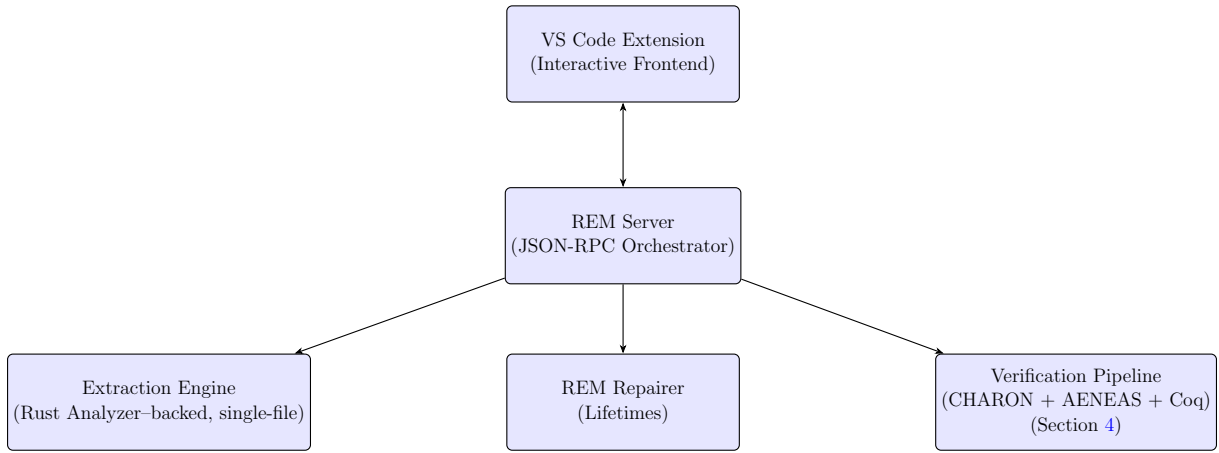
\begin{figure}[ht]
  \centering
  \large
  \resizebox{\textwidth}{!}{%
  \begin{tikzpicture}[node distance=2.2cm and 2.6cm, >=Stealth, thick, font=\Large]
    \tikzstyle{block}=[draw, rounded corners, fill=blue!10,
      minimum width=6cm, minimum height=2.5cm, align=center]

    \node[block] (vscode) {VS Code Extension\\(Interactive Frontend)};

    \node[block, below=of vscode] (server) {REM Server\\(JSON-RPC Orchestrator)};

    \node[block, below left=1.8cm and 5.0cm of server] (extractor)
      {Extraction Engine\\(Rust Analyzer–backed, single-file)};
    \node[block, below=1.8cm of server] (repairer)
      {REM Repairer\\(Lifetimes)};
    \node[block, below right=1.8cm and 5.0cm of server] (verify)
      {Verification Pipeline\\(CHARON + AENEAS + Coq) \\ (Section \ref{sec:verification})};

    \draw[<->] (vscode) -- (server);
    \draw[->]  (server) -- (extractor);
    \draw[->]  (server) -- (repairer);
    \draw[->]  (server) -- (verify);
  \end{tikzpicture}%
  }
  \captionsetup{justification=centering}
  \caption{High-level overview of the expanded REM architecture: VS Code communicates bidirectionally with the REM Server, which dispatches to the Extraction Engine, the REM Repairer, and the REM Verification pipeline.}
  \label{fig:rem_architecture_overview}
\end{figure}

The expanded REM architecture is organised around a modular, service-oriented design that is able to decouple the interactive editing from semantic analysis and its surrounding logic. Figure \ref{fig:rem_architecture_overview} illustrates the system's structure at the highest level.

At the centre of the system is the REM server, a lightweight orchestrator that exposes a JSON-RPC (Remote Procedure Call) interface to the VSCode extension. The server can also be rexported as a CLI to be integrated into local programs \footnote{https://crates.io/crates/rem-command-line}. The server is responsible for coordinating all major operations - \emph{extraction}, \emph{repair}, and \emph{verification}, for handling errors and for providing immediate feedback as to whether an operation is possible. Crucially, the interface is completely generic, and has been designed to be very easily adaptable to any other IDE.

The extraction engine (REM Extract) forms the analytical core of the new toolchain. Built directly atop of Rust Analyzer, its key innovation is being able to perform semantic extraction over a \textit{single source file} rather than an entire crate. This adaptation is key to REM's speed: whereas Rust Analyzer typically analyses entire projects to populate its HIR (High-Level Intermediate Representation), our engine is able to leverage internal features and interfaces to perform analysis on a single file without needing to consider the wider module / crate context. Given an extraction request (\verb|file path, selection range, function name|), the engine constructs a minimal semantic model of the file, determines the precise boundaries of the selection, and synthesises the extracted function and corresponding call-site edits. The engine is also capable of returning extract semantic information up the chain depending on the mode in which it is called and the requirements of the downstream elements.

Once extraction is complete`, the resulting transformation is then (optionally) passed into REM Repairer. The repairer is a direct descendant of the original REM prototype but re-engineered for performance and modularity. Its task is to reify and correct any missing lifetimes (as guided by \verb|cargo check|), before applying a series of lifetime elision rules to attempt to bring the resultant mess into a form that is (mostly) human readable. The main algorithm is still the same as was discussed in Adventure of a Lifetime \cite{AdventureOfALifetime}, so we have omitted a large discussion of it from this report.

Finally, the VSCode extension provides the user-facing layer of the system. Acting as a thin client, it serialises extraction requests to the REM server and applies the returned edits to the open buffer. Bidirectional communication over JSON-RPC enables sub-200 ms round-trip times from user action to visual feedback. From the user's perspective, REM behaves like a native IDE feature rather than an external tool.

\subsection{Single-File Workspaces over Rust-Analyzer}
\label{sec:single_file_workspace}

\vspace*{-2.5mm}
A central constraint for REM2.0's speed is the ability to run extract-method on demand, without re-analysing an entire project, or even really having access to the entire project. Rust-Analyzer (RA), however, is designed around crate-level workspaces: it expects a \texttt{Cargo.toml}, a dependency graph, and a complete module tree before it will perform semantic analysis. To bridge this mismatch, REM2.0 constructs a lightweight, synthetic ``single-file workspace'' in which the file currently being refactored is treated as an isolated crate. This workspace contains only the minimum metadata necessary for RA to parse, index, and type-check the file.

This abstraction gives us fast, incremental semantics while avoiding the cost of repeatedly analysing a full repository. However, as we discovered, a single-file workspace lacks any knowledge of the standard library: types available by default such as \icodeverb{Vec<T>}, \icodeverb{Option<T>}, \icodeverb{String}, or even \icodeverb{Result<T,E>} are not available unless for some reason they have been explicitly imported by the user. Rust-Analyser therefore assigns placeholder types (internally represented as ``\_'' or incomplete inference variables) whenever an extracted function must reference a std-provided type.

Figure~\ref{fig:std_extraction} highlights just how significant this limitation is. Extracting Listing~\ref{lst:std-original} should produce the fully inferred function in Listing~\ref{lst:std-fix}, but in a single-file workspace (without \icodeverb{std}/\icodeverb{core} loaded), RA instead generates the incomplete signature shown in Listing~\ref{lst:std-fail}. Because this placeholder type propagates into the call site and return position, the resulting code no longer compiles.

\begin{figure}[H]
\centering

\begin{subfigure}[t]{0.32\textwidth}
    \centering
\begin{lstlisting}[
        basicstyle=\scriptsize\ttfamily,
        numbers=left,
        frame=lines,
        language=Rust
    ]
fn main() {
//Rust-Analyzer(single-file) can't
//know what `Vec` is here.
//Even if `Vec` was strong typed
    let mut v = Vec::<i32>::new();
    for i in 0..10 {
        v.push(i);
    }

    println!("{:?}", v);
}
//
//
\end{lstlisting}
    \captionsetup{justification=centering}
    \sublistingcaption{Original Code}
    \label{lst:std-original}
\end{subfigure}
\hfill
\begin{subfigure}[t]{0.32\textwidth}
    \centering
\begin{lstlisting}[
        basicstyle=\scriptsize\ttfamily,
        numbers=left,
        frame=lines,
        language=Rust
    ]
fn main() {
    let v = build_vec();
    println!("{:?}", v);
}

//Rust-Analyzer's *generated* 
//signature uses a placeholder
//as it can't find `Vec`
fn build_vec() -> _ {
    let mut v = Vec::<i32>::new();
    for i in 0..10 { v.push(i); }
    v
}
\end{lstlisting}
    \captionsetup{justification=centering}
    \sublistingcaption{Failed Extraction \\ As \texttt{std} is out of scope}
    \label{lst:std-fail}
\end{subfigure}
\hfill
\begin{subfigure}[t]{0.32\textwidth}
    \centering
\begin{lstlisting}[
        basicstyle=\scriptsize\ttfamily,
        numbers=left,
        frame=lines,
        language=Rust
    ]
fn main() {
    let v = build_vec();
    println!("{:?}", v);
}

//With std in scope,
//This return type can be
//Correctly inferred
fn build_vec() -> Vec<i32> {
    let mut v = Vec::<i32>::new();
    for i in 0..10 { v.push(i); }
    v
}
\end{lstlisting}
    \captionsetup{justification=centering}
    \sublistingcaption{Correct Extraction \\ Now that \texttt{std} is in scope}
    \label{lst:std-fix}
\end{subfigure}

\captionsetup{justification=centering}
\caption{Comparison of an extraction prior to the \texttt{std} and \texttt{core} fixes and after. Before we return a `\_' as a placeholder, as Rust-Analyzer is unable to infer the type without the semantic information provided by \texttt{std}.}
\label{fig:std_extraction}
\end{figure}

\vspace*{-5mm}
To address this problem, REM~2.0 constructs a second, persistent workspace containing the complete set of standard crates and their dependency graph. This workspace is initialised once when the VSCode extension starts (a one-time expensive step) and is assumed to remain stable across refactorings. During extraction, REM merges the semantic graphs of the \icodeverb{std} workspace and the single-file workspace, producing an ``effective'' workspace with full knowledge of both user code and the standard library. With this view, RA regains the ability to infer types such as \texttt{Vec<T>} correctly, restoring the fully inferred extraction shown in Listing~\ref{lst:std-fix}. The complete set of standard crates imported into this persistent workspace is shown alongside.

\vspace{-2.5mm}
\begin{figure}[H]
\centering
\begin{minipage}[t]{0.5\textwidth}
To address this problem, REM2.0 constructs a second, persistent workspace containing the complete set of standard crates and their dependency graph. This workspace is initialised once when the VSCode extension starts (a one-time expensive step) and is assumed to remain stable across refactorings. During extraction, REM merges the semantic graphs of the \icodeverb{std} workspace and the single-file workspace, producing an ``effective'' workspace with full knowledge of both user code and the standard library. With this view, RA regains the ability to infer types such as \icodeverb{Vec<T>} correctly, restoring the fully inferred extraction shown in Listing~\ref{lst:std-fix}. The complete set of standard crates imported into this persistent workspace is shown alongside.
\end{minipage}\hfill%
\begin{minipage}[t]{0.48\textwidth}
\small
\footnotesize
\begin{verbatim}
crate Idx::<CData>(0) -> core
crate Idx::<CData>(1) -> compiler_builtins
crate Idx::<CData>(2) -> alloc
crate Idx::<CData>(3) -> panic_abort
crate Idx::<CData>(4) -> libc
crate Idx::<CData>(5) -> unwind
crate Idx::<CData>(6) -> panic_unwind
crate Idx::<CData>(7) -> std_detect
crate Idx::<CData>(8) -> cfg_if
crate Idx::<CData>(9) -> hashbrown
crate Idx::<CData>(10) -> rand_core
crate Idx::<CData>(11) -> rand
crate Idx::<CData>(12) -> rand_xorshift
crate Idx::<CData>(13) -> rustc_demangle
crate Idx::<CData>(14) -> std
crate Idx::<CData>(15) -> rustc_literal_escaper
crate Idx::<CData>(16) -> proc_macro
crate Idx::<CData>(17) -> getopts
crate Idx::<CData>(18) -> test
\end{verbatim}
\captionsetup{justification=centering}
\captionof{listing}{Crates in the std workspace}
\label{fig:std_crate_list_fallback}
\end{minipage}
\end{figure}

\vspace{-5mm}
\subsubsection{Leveraging Rust-Analyzer's Extract Method}
\label{sec:ra_extract_integration}

With the effective workspace in place, REM2.0 relies directly on RA's built-in \emph{Extract Function} assist. Instead of re-implementing extraction logic, REM2.0 issues the same internal request that a client editor would send to RA: the file contents, the selected range, and a desired function name. RA then returns a structured edit describing (1) a new function with fully inferred types, generics, lifetimes, and where-clauses, and (2) a rewritten call site at the extraction point.

This delegation is critical. RA already contains sophisticated logic for name resolution, trait and impl lookup, async/await de-sugaring, const evaluation, higher-ranked trait bounds, and non-local control flow analysis. By using its extract-method implementation as the semantic source of truth, REM2.0 inherits support for the full Rust language surface—including features that would be prohibitively difficult or brittle to handle manually. It also allows REM2.0 to fully benefit from years of active development in this space, allowing us to focus on expanding the capability of the whole REM2.0 toolchain rather than attempting to rewrite something that already exists. We are also able to query the semantic analysis returned by RA to improve the repair and verification stages that come next.

REM~2.0 treats the edits returned by RA as a semantic blueprint, adapting them only as needed for lifetime repair, call-site updates, and optional verification. Rather than maintaining its own parallel program state, REM is able to query RA directly for any additional type or lifetime information, allowing it to build on RA's mature transformations while adding its own layers of structure and correctness checking.

\vspace{-5mm}
\subsection{VSCode Integration and the REM Daemon}
\label{sec:vscode_and_daemon}
\begin{figure}[H]
    \centering
    \includegraphics[width=0.7\linewidth]{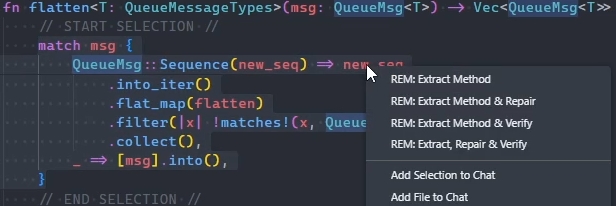}
    \captionsetup{justification=centering}
    \caption{Options presented to the user when they have selected a region to extract.}
    \label{fig:vscode_options}
\end{figure}

\vspace*{-2.5mm}
To make extraction accessible to everyday Rust developers, REM2.0 is exposed through a dedicated VSCode extension backed by a persistent ``REM daemon.'' The daemon runs as a long-lived process and communicates with the extension over a lightweight JSON-RPC protocol via \texttt{stdin}/\texttt{stdout}, enabling the lightning fast refactorings without the overhead of repeated process launches. On startup, the extension performs an environment check to ensure that the Daemon, CHARON, AENEAS and Coqc are available; if not, it presents targeted installation instructions directly in the editor. From the user's perspective, extraction appears as a standard VSCode code action: selecting a region triggers a panel, options for naming the new function, and (when equivalence checking is enabled) clear feedback on whether the refactoring preserves semantics. This architecture provides a seamless and familiar developer experience while hiding the complexity of the underlying pipeline. Figure~\ref{fig:vscode_options} below is the list of options presented to the developer, and they can also be given keyboard shortcuts for ease of use.

\vspace{-7.5mm}
\subsubsection{Preparing Refactorings for Verification}
\label{sec:ref_to_verification}

Although the verification pipeline is described in detail in Section~\ref{sec:verification}, with specific implementation details in Section~\ref{sec:implementation_verification}, REM2.0 performs a small amount of preparation at extraction time to ensure that refactorings can be checked efficiently. For each extraction, REM2.0 captures both the original function body and the newly generated function in isolated ``virtual crates'', preserving the exact syntax and inferred types returned by RA. These crates form the minimal input required by CHARON and allow the verifier to compare only the code relevant to the transformation, rather than the surrounding project.

%% file: chapter3.tex
\section{Automatic Equivalence Proofs for Refactored Code}
\label{sec:verification}


In the strictest sense, refactoring is defined as a behaviour-preserving code
transformation. In practice, refactoring is only valuable if it preserves the
original program's behaviour. In Rust, where ownership and lifetimes enforce
strict invariants, the mere fact that refactored code compiles does not
guarantee equivalence. Additionally, because REM2.0 performs complex,
compiler-guided repairs as part of its extraction process
\cite{AdventureOfALifetime}, there is an increased risk of introducing subtle
changes in program behaviour. Subtle shifts in aliasing or lifetime structure
can produce programs that pass the compiler yet diverge semantically from their
originals~\cite{YoungYangSergeyPotanin2024}. Additionally, as Section~\ref{subsec:extraction_failures} shows, automated extract method tooling can produce incorrect extractions that still compile. For high-assurance domains, this risk is unacceptable: automated
tools must not only generate compiling code but also provide evidence that
transformations are correct.

This section introduces a novel equivalence proof pipeline that extends REM2.0 with
automatic, annotation-free proofs of equivalence between original and refactored
code. Our approach combines the existing formal methods toolchains - CHARON
\footnote{More information accessible from their GitHub, \url{https://github.com/AeneasVerif/charon}}, which
translates Rust into an ownership-explicit intermediate from, and AENEAS
\footnote{More information accessible from their GitHub, \url{https://github.com/AeneasVerif/aeneas}}, which
then generates Coq code equivalent to the original Rust source through a pure
$\lambda$-calculus based intermediary. We then use this Coq code to formally prove that the refactored code is equivalent to the original.
The result is an end-to-end system in which the Extract $\rightarrow$ Repair
cycle is followed by a Verify phase, discharging proofs in seconds with no
additional burden on the developer. By embedding automated equivalence proofs directly into the
refactoring process, we provide developers with the option to move beyond compilation success to true semantic assurance, whilst also bridging the gap between theory and practical developer tools.

Throughout this section (and previously) we do occasionally refer to the equivalence checker as a ``verifier''. We acknowledge that the work it performs isn't verification in the strictest sense of the word, but it suits the language much better.

\subsection{What if Extraction Goes Wrong?}
\label{sec:extraction_goes_wrong}
\vspace{-2.5mm}


Automated extraction is not infallible: incorrect handling of ownership or mutation
can silently alter program semantics. To demonstrate the reason why we have built the verifier, and its role in REM2.0, we consider a trivial but illustrative example where the extractor mishandles a mutable
binding\footnote{Our extraction pipeline does not perform this extraction incorrectly, but previous tools, especially the ones evaluated against in Adventure of a Lifetime could make similar mistakes. This example is a demonstration of such a failed extraction.}. The verifier correctly identifies the non-equivalence and reports the
failure to the developer through the VSCode extension interface, offering an option
to automatically revert to the previous state. Figure~\ref{fig:extract_wrong_rust} shows the case we are working with, along with a correct extraction (passing the mutable reference to x, \icodeverb{\&mut x}), and an incorrect extraction (passing a cloned `x' instead \icodeverb{x.clone()}). This error sidesteps Rusts safety guarantees as the compiler cannot prove what \icodeverb{x} is at compile time, and thus we get a runtime crash as shown by Listing\ref{lst:incorrect_extraction_crash}.

\begin{figure}[H]
\centering

\begin{subfigure}[t]{0.32\textwidth}
    \centering
\begin{lstlisting}[
        basicstyle=\scriptsize\ttfamily,
        numbers=left,
        frame=lines,
        language=Rust
    ]
fn main() {
    let mut x = 0;
    x = 3;
    assert!(x == 3);
}
//
//
//
//
\end{lstlisting}
    \sublistingcaption{Original Code}
    \label{lst:extraction-original}
\end{subfigure}
\hfill
\begin{subfigure}[t]{0.32\textwidth}
    \centering
\begin{lstlisting}[
        basicstyle=\scriptsize\ttfamily,
        numbers=left,
        frame=lines,
        language=Rust
    ]
fn main() {
    let mut x = 0;
    mutate_x(&mut x);
    assert!(x == 3);
}

fn mutate_x(x: &mut i32) {
    *x = 3;
}
\end{lstlisting}
    \sublistingcaption{Correct Extraction}
    \label{lst:extraction-correct}
\end{subfigure}
\hfill
\begin{subfigure}[t]{0.32\textwidth}
    \centering
\begin{lstlisting}[
        basicstyle=\scriptsize\ttfamily,
        numbers=left,
        frame=lines,
        language=Rust
    ]
fn main() {
    let mut x = 0;
    mutate_x(x.clone());
    assert!(x == 3);
}

fn mutate_x(mut x: i32) {
    x = 3;
}
\end{lstlisting}
    \sublistingcaption{Incorrect Extraction}
    \label{lst:extraction-incorrect}
\end{subfigure}

\captionsetup{justification=centering}
\caption{Comparison of the original program, a correct extraction, and an incorrect extraction that clones rather than mutably borrows.}
\label{fig:extract_wrong_rust}
\end{figure}

\begin{lstlisting}[language=bash]
thread 'main' (543251) panicked at src/main.rs:6:5:
assertion failed: x == 3
note: run with `RUST_BACKTRACE=1` environment variable to display a backtrace
\end{lstlisting}
{
\captionsetup{justification=centering}
\captionof{listing}{Terminal output from running the incorrect extraction. The program crashes at runtime, not at compile time.}
\label{lst:incorrect_extraction_crash}
}

From here, we generate a semantically equivalent file, but in the Coq proof assistant language. In Figure~\ref{fig:extract_wrong_coq} below, we list the important parts of these translations. Other aspects (e.g. the complex definitions that translate Rust types to Coq types) have been ignored and will be covered later in Section~\ref{sec:implementation_verification}. Also note that rustc has performed some static optimisations on the code before translation, at this stage there is no way to disable them within CHARON so Listing~\ref{lst:coq-original} is visually different to the rust implementation.

\begin{figure}[H]
\centering

\begin{subfigure}[t]{0.32\textwidth}
    \centering
\begin{lstlisting}[
        basicstyle=\scriptsize\ttfamily,
        numbers=left,
        frame=lines,
        breaklines=true,
        language=Coq
    ]
(** [orig::main]:
    Source: 'src/main.rs', lines 1:0-5:1 *)
Definition main : result unit :=
  massert (3%i32 s= 3%i32).
(* *)
(* *)
(* *)
(* *)
(* *)
\end{lstlisting}
    \sublistingcaption{Original Code}
    \label{lst:coq-original}
\end{subfigure}
\hfill
\begin{subfigure}[t]{0.32\textwidth}
    \centering
\begin{lstlisting}[
        basicstyle=\scriptsize\ttfamily,
        numbers=left,
        frame=lines,
        breaklines=true,
        language=Coq
    ]
Definition mutate_x (x : i32) : result i32 :=
  Ok 3%i32.

Definition main : result unit :=
  x <- mutate_x 0%i32; massert (x s= 3%i32).
(* *)
(* *)
(* *)
\end{lstlisting}
    \sublistingcaption{Correct Extraction}
    \label{lst:coq-correct}
\end{subfigure}
\hfill
\begin{subfigure}[t]{0.32\textwidth}
    \centering
\begin{lstlisting}[
        basicstyle=\scriptsize\ttfamily,
        numbers=left,
        frame=lines,
        breaklines=true,
        language=Coq
    ]
Definition mutate_x (x : i32) : result unit :=
  Ok tt.

Definition main : result unit :=
  let i := core_clone_impls_CloneI32_clone 0%i32 in
  _ <- mutate_x i;
  massert (0%i32 s= 3%i32).
\end{lstlisting}
    \sublistingcaption{Incorrect Extraction}
    \label{lst:coq-incorrect}
\end{subfigure}

\captionsetup{justification=centering}
\caption{Comparison of the original program, a correct extraction, and an incorrect extraction that clones rather than mutably borrows.}
\label{fig:extract_wrong_coq}
\end{figure}
\vspace{-2.5mm}

Having obtained the AENEAS-generated Coq representations of both the correct and incorrect extractions, we now perform an automated equivalence check. This step constitutes a concrete implementation of the proof obligations described in Section~\ref{sec:proof_obligations}. In essence, the verifier constructs and discharges a proof showing that, for all possible inputs to the caller function, the original and refactored programs produce identical results. Because the verification operates within Coq's pure, functional semantics—free from hidden side effects or mutable state—the resulting proof provides strong assurance of behavioural equivalence.

\begin{figure}[H]
\centering
\begin{subfigure}[t]{0.48\textwidth}
    \centering
\begin{lstlisting}[basicstyle=\scriptsize\ttfamily, frame=lines, language=Coq]
(** EquivCheck.v Autogenerated by REM-verification **)
Require Import Primitives.
Import Primitives.
Require Import Orig.
Require Import Good.
Local Open Scope Primitives_scope.

Lemma main_equiv_check :
    Orig.main = Good.main.
    Proof.
    intros.
    unfold Orig.main.
    unfold Good.main.
    reflexivity.
    Qed.
\end{lstlisting}
    \sublistingcaption{Equivalence Check (Correct Extraction)}
    \label{lst:verify-equivalence-correct}
\end{subfigure}
\hfill
\begin{subfigure}[t]{0.48\textwidth}
    \centering
\begin{lstlisting}[basicstyle=\scriptsize\ttfamily, frame=lines, language=Coq]
(** EquivCheck.v Autogenerated by REM-verification **)
Require Import Primitives.
Import Primitives.
Require Import Orig.
Require Import Bad.
Local Open Scope Primitives_scope.

Lemma main_equiv_check :
Orig.main = Bad.main.
Proof.
intros.
unfold Orig.main.
unfold Bad.main.
reflexivity.
Qed.
\end{lstlisting}
    \sublistingcaption{Equivalence Check (Incorrect Extraction)}
    \label{lst:verify-equivalence-incorrect}
\end{subfigure}

\vspace{6pt}

\begin{subfigure}[t]{0.48\textwidth}
    \centering
\begin{lstlisting}[basicstyle=\scriptsize\ttfamily, frame=lines, language=bash]
COQ_PROJECT: cases/4.1_examples/good_verif/_CoqProject
EQUIVCHECK: cases/4.1_examples/good_verif/EquivCheck.v
PRIMITIVES: cases/4.1_examples/good_verif/Primitives.v
SUCCESS: true
OUTPUT_END
\end{lstlisting}
    \sublistingcaption{Verifier Output (Correct Extraction)}
    \label{lst:verify-output-correct}
\end{subfigure}
\hfill
\begin{subfigure}[t]{0.48\textwidth}
    \centering
\begin{lstlisting}[basicstyle=\scriptsize\ttfamily, frame=lines, language=bash]
COQ_PROJECT: cases/4.1_examples/bad_verif/_CoqProject
EQUIVCHECK: cases/4.1_examples/bad_verif/EquivCheck.v
PRIMITIVES: cases/4.1_examples/bad_verif/Primitives.v
SUCCESS: false
OUTPUT_END
\end{lstlisting}
    \sublistingcaption{Verifier Output (Incorrect Extraction)}
    \label{lst:verify-output-incorrect}
\end{subfigure}

\captionsetup{justification=centering}
\caption{Comparison of equivalence checking results for the correct and incorrect extractions. The verifier confirms semantic equivalence in the first case and correctly rejects the cloned variant as non-equivalent.}
\label{fig:verify-results}
\end{figure}
\vspace{-2.5mm}

In this example, the incorrectly extracted version introduces a cloned value (\icodeverb{x.clone()}) in place of a mutable borrow, thereby altering the program's semantics while still producing compilable Rust code. Through the equivalence checks shown in Figure~\ref{fig:verify-results}, the equivalence checker detects this discrepancy by symbolically comparing the Coq representations of both versions and rejecting the transformed function as non-equivalent. Within the REM2.0 workflow, this failure is brought directly to the developer via the VSCode extension as a clear error message, accompanied by an option to automatically revert to the pre-extraction state.

\vspace{-5mm}
\subsection{Demonstrating Non-Trivial Verification}
\label{sec:non_trivial_verification}
\vspace{-2.5mm}

The previous section showcased a deliberately trivial case to serve as an introduction to the concept, and to illustrate how the equivalence checker guards against obvious semantic breakages. In contrast, real refactorings can involve nested loops, const generics, ownership subtleties, and type-level invariants that interact in ways which are far less visible in the surface syntax. In this section we examine a more realistic example, inspired by an incorrect extraction found in the Vaultwarden code base, where the refactoring compiles, typechecks, and appears superficially reasonable, yet silently changes the program's behaviour. The aim here is to demonstrate that the verifier is not merely checking syntactic structure nor trivially returning success, but is able to detect subtle semantic divergences that Rust's type system alone cannot rule out.

\begin{figure}[H]
\centering

\begin{subfigure}[t]{0.48\textwidth}
    \centering
\begin{lstlisting}[
        basicstyle=\scriptsize\ttfamily,
        numbers=left,
        frame=lines,
        language=Rust
    ]
pub fn bump<const N: usize>(
    arr: &mut [i32; N],
    k: usize,
    mark: &mut bool,
) where [i32; N]: Copy, {
    let mut i = 0;
    let limit = if k < N { k } else { N };
    while i < limit {
        if arr[i] == 0 {
            *mark = true;
        }
        arr[i] = arr[i] + 1;
        i += 1;
    }
}
fn main() {
    let mut a = [0, 0, 5, 9];
    let mut touched_zero = false;
    bump(&mut a, 3, &mut touched_zero);
    assert_eq!(a, [1, 1, 6, 9]);
    assert!(touched_zero);
}
//
//
//
//
//
\end{lstlisting}
    \sublistingcaption{Original Code}
    \label{lst:complex-original}
\end{subfigure}
\hfill
\begin{subfigure}[t]{0.48\textwidth}
    \centering
\begin{lstlisting}[
        basicstyle=\scriptsize\ttfamily,
        numbers=left,
        frame=lines,
        language=Rust
    ]
fn bump_extracted<const N: usize>(
    mut arr: [i32; N], i: usize, mut mark: bool
) where  [i32; N]: Copy, {
    if arr[i] == 0 {
        mark = true;
    }
    arr[i] = arr[i] + 1;
}
pub fn bump<const N: usize>(
    arr: &mut [i32; N],
    k: usize,
    mark: &mut bool,
) where [i32; N]: Copy, {
    let mut i = 0;
    let limit = if k < N { k } else { N };
    while i < limit {
        bump_extracted::<N>((*arr).clone(), i, *mark);
        i += 1;
    }
}
fn main() {
    let mut a = [0, 0, 5, 9];
    let mut touched_zero = false;
    bump(&mut a, 3, &mut touched_zero);
    assert_eq!(a, [1, 1, 6, 9]); // fails: a: [0, 0, 5, 9]
    assert!(touched_zero); // fails
}
\end{lstlisting}
    \sublistingcaption{Incorrect Extraction}
    \label{lst:complex-wrong}
\end{subfigure}

\captionsetup{justification=centering}
\caption{Original Function and Incorrect Extraction produced by the Engine. Example inspired by the failed Vaultwarden extraction, see Section~\ref{subsec:extraction_failures}}
\label{fig:extract_wrong_rust_complex}
\end{figure}
\vspace{-2.5mm}

At a glance, the incorrect extraction appears harmless: the loop structure is preserved, the in-place update remains nested within an iterator, and Rust's type system accepts the transformation due to the array's \icodeverb{Copy} bound. However, because the extracted function takes its array \emph{by value} rather than by mutable reference, it mutates a temporary copy and discards it. This subtle deviation is difficult to detect manually and impossible for the compiler to reject. The equivalence checker, however, detects the mismatch by comparing the Coq translations of both versions and identifying that the caller's array is unchanged in the transformed program. As shown in Figure~\ref{fig:extract_wrong_rust_extended}, the equivalence check on the incorrect extraction fails\footnote{The assertions in listing \ref{lst:complex-wrong} also fail at runtime, they are used here to allow us to demonstrate the failure in Rust, and imitate a unit test.}, and REM2.0 is capable of passing this failure directly to the developer, preventing the silently incorrect refactoring from being applied.

\begin{figure}[H]
\centering

\begin{subfigure}[t]{0.48\textwidth}
    \centering
\begin{lstlisting}[
        basicstyle=\scriptsize\ttfamily,
        numbers=left,
        frame=lines,
        breaklines=true,
        language=Coq
    ]
Definition bump
  {N : usize} (coremarkerCopyArrayI32NInst : core_marker_Copy (array i32 N))
  (arr : array i32 N) (k : usize) (mark : bool) :
  result ((array i32 N) * bool)
  :=
  if k s< N then bump_loop0 k arr mark 0%usize else bump_loop1 arr mark 0%usize
.

(* *)
(* *)
(* *)
(* *)
(* *)
(* *)
(* *)
(* *)
(* *)
(* *)
(* *)
(* *)
(* *)
(* *)
(* *)
(* *)
(* *)
(* *)
(* *)
(* *)
\end{lstlisting}
    \sublistingcaption{Original Code}
    \label{lst:complex-original-coq}
\end{subfigure}
\hfill
\begin{subfigure}[t]{0.48\textwidth}
    \centering
\begin{lstlisting}[
        basicstyle=\scriptsize\ttfamily,
        numbers=left,
        frame=lines,
        breaklines=true,
        language=Coq
    ]
Definition bump
  {N : usize} (coremarkerCopyArrayI32NInst : core_marker_Copy (array i32 N))
  (arr : array i32 N) (k : usize) (mark : bool) :
  result ((array i32 N) * bool)
  :=
  if k s< N
  then (
    _ <- bump_loop0 coremarkerCopyArrayI32NInst arr k mark 0%usize;
    Ok (arr, mark))
  else (
    _ <- bump_loop1 coremarkerCopyArrayI32NInst arr mark 0%usize;
    Ok (arr, mark))
.

Definition bump_extracted
  {N : usize} (coremarkerCopyArrayI32NInst : core_marker_Copy (array i32 N))
  (arr : array i32 N) (i : usize) (mark : bool) :
  result unit
  :=
  i1 <- array_index_usize arr i;
  if i1 s= 0%i32
  then (_ <- i32_add i1 1%i32; _ <- array_index_mut_usize arr i; Ok tt)
  else (_ <- i32_add i1 1%i32; _ <- array_index_mut_usize arr i; Ok tt)
.
\end{lstlisting}
    \sublistingcaption{Incorrect Extraction}
    \label{lst:complex-wrong-coq}
\end{subfigure}

\vspace{0.5em}

\begin{subfigure}[t]{\textwidth}
    \centering
\begin{lstlisting}[
        basicstyle=\scriptsize\ttfamily,
        numbers=left,
        frame=lines,
        breaklines=true,
        language=bash
    ]
COQ_PROJECT: cases/4.3_examples/verif/_CoqProject
EQUIVCHECK: cases/4.3_examples/verif/EquivCheck.v
PRIMITIVES: cases/4.3_examples/verif/Primitives.v
SUCCESS: false
OUTPUT_END
\end{lstlisting}
    \sublistingcaption{Terminal output showing the verification failure}
    \label{lst:complex-output}
\end{subfigure}

\captionsetup{justification=centering}
\caption{Coq Translation and output of REM2.0's equivalence checker. // Note that some of Coq translation has been omitted for brevity. See Appendix~\ref{app:coq_translations} for the full translation.}
\label{fig:extract_wrong_rust_extended}
\end{figure}

\vspace{-2.5mm}
Together, these results demonstrate that the equivalence checker performs genuinely non-trivial reasoning: it succeeds only when two programs are behaviourally identical, and it rejects transformations that break behavioural equivalence even when the compiler accepts them and the syntactic structure appears valid. This establishes that the equivalence stage is a substantive correctness guarantee rather than a superficial or always-succeeding check.

\vspace*{-5mm}
\subsection{Handling of Likely Edge Cases}
\label{sec:handling_of_complex_cases}
\vspace{-2.5mm}

While the previous section(s) have demonstrated the need for the equivalence checker, going into examples in depth, this section aims to show how we handle likely boundary conditions that do arise as a result of using the extraction engine. Out goal is to show that: (i) invalid patterns are rejected early by rustc (and surfaced into REM2.0 as actionable diagnostics), (ii) well-typed but ``degenerate'' shapes (no params, unit return, locally inferred \icodeverb{\_}) translate cleanly through CHARON/AENEAS and verify as expected, and (iii) known limitations (e.g. \icodeverb{!}) are identified and cannot result in spurious ``success''.  The below table demonstrates each such edge case and what we expect to happen:

{
\setlength{\textwidth}{1.05\textwidth}   
\setlength{\LTleft}{-1cm}   
\setlength{\LTright}{0pt}     
\small

\begin{longtable}
{|
>{\raggedright\arraybackslash}p{0.15\textwidth}|
>{\raggedright\arraybackslash}p{0.23\textwidth}|
>{\raggedright\arraybackslash}p{0.10\textwidth}|
>{\raggedright\arraybackslash}p{0.1\textwidth}|
>{\raggedright\arraybackslash}p{0.34\textwidth}|
}
\hline
\textbf{Case} &
\textbf{Minimal Example} &
\textbf{Rustc / Cargo} &
\textbf{Translate} &
\textbf{Equivalence Outcome / Notes} \\
\hline
\endfirsthead

\hline
\textbf{Case} &
\textbf{Minimal Example} &
\textbf{Rustc / Cargo} &
\textbf{Translate} &
\textbf{Verifier Outcome / Notes} \\
\hline
\endhead

\hline
\multicolumn{5}{r}{Continued on next page} \\
\hline
\endfoot

\hline
\endlastfoot

No input arguments, Unit return or Both &
\texttt{fn ping() \{ /* ... */ \}} \newline
\texttt{fn reset(x: \&mut i32) \{ *x = 0; \}}  &
OK &
OK &
Proof engine capable of unfolding functions without a \icodeverb{forall} clause. Reflexivity guarantees equivalence. \\
\hline

Unit return (explicit or inferred) &
\texttt{fn touch(x:\&mut i32)\{ *x += 1; \}} \newline
\texttt{fn f()->() \{ \}} &
OK &
OK &
Equivalence checked via identical post-state; return value is \texttt{unit}. Side-effect modelling preserved. \\
\hline

\_ in parameter or return type &
\texttt{fn g(x: \_)-> i32 \{ x+1 \}} \newline
\texttt{fn h(x:i32)->\_ \{ x*2 \}} &
\textbf{Error} (invalid signatures) &
N/A &
Rejected upstream by the compiler. No translation or proof attempted. \\
\hline

\_ in local binding (within body) &
\texttt{let x: \_ = 42; let y = x+1;} &
OK &
OK &
Type is concretised before/at MIR; Coq has a concrete integer type. Verifier succeeds as usual. \\
\hline

\texttt{!} (never) return type &
\texttt{fn abort()->! \{ panic!("fail") \}} &
OK &
\textbf{Fails} &
Excluded from verification set. REM 2.0 reports an unsupported-feature diagnostic; extraction may proceed but is not verified. \\
\hline

Non-local control flow in extracted region &
\texttt{break}/\texttt{continue} inside loop body &
OK &
OK &
Control flow reified as \texttt{std::ops::ControlFlow}. Verifier checks pattern-match and proves behavioural equivalence. \\
\hline
\caption{Edge-case coverage for extraction and verification. "Translate" refers to CHARON/AENEAS; "Verifier" to the equivalence checker.}
\label{tab:edgecases} \\

\end{longtable}}

\vspace{-2.5mm}
Taken together, these cases show that the pipeline neither devolves into trivial proofs, nor silently accepts unsupported constructs. Invalid signatures involving \icodeverb{\_} are blocked by rustc and surfaced to the user; structurally simple but well-typed shapes (no params, unit return, local \icodeverb{\_}) translate cleanly and verify; and known limitations (e.g., \icodeverb{!}) are explicitly reported and excluded from proof. This provides broad coverage over edge conditions while keeping the verifier's guarantees meaningful and trustworthy. Whilst the design specifications and implementation are detailed in Section~\ref{sec:implementation_verification} and Appendix~\ref{app:implementation_verification}, what we have demonstrated here reflects a core philosophy of the equivalence checker: to the best of its capacity, it must never report that two pieces of code are equivalent when they are not.

\vspace*{-5mm}
\subsection{Proof Obligations}
\label{sec:proof_obligations}
\vspace{-2.5mm}

In the final stage of the verification pipeline, we discharge a formal
\textit{equivalence obligation} inside Coq. Intuitively, we must show that the
refactored program behaves identically to the original program for all valid
inputs. The central idea is functional equivalence: the observable outputs of
the two programs must match.

\begingroup
\setlength{\abovedisplayskip}{4pt}
\setlength{\belowdisplayskip}{4pt}
\setlength{\abovedisplayshortskip}{2pt}
\setlength{\belowdisplayshortskip}{2pt}

\vspace{-2.5mm}
\subsubsection{Simplified Obligation}
We write $\llbracket f \rrbracket$ for the semantic interpretation (meaning) of
the function $f$ as produced by the AENEAS translation into Coq. Simply put, the obligation can be stated as:
\begin{equation*}
  \forall x \in \mathrm{Dom}(f).\;
  \llbracket f \rrbracket(x) = \llbracket f' \rrbracket(x)
\end{equation*}
Where $f$ is the original function and $f'$ is its refactored counterpart. This
expresses that for every possible construct of inputs $x$, the translated functions return identical
results. In practice, this is the form we attempt to discharge automatically, as
shown in Figure~\ref{lst:verify-equivalence-correct}.

\vspace{-2.5mm}
\subsubsection{Whole-Program Context}
For a more complete characterisation, we must acknowledge that functions are
never executed in isolation, but as part of the enclosing program. Thus the
obligation can be strengthened to whole-program equivalence:
\begin{equation*}
    \forall x \in \mathrm{Dom}(f).\;
    \llbracket f \rrbracket(x) = \llbracket f' \rrbracket(x)
\end{equation*}
Where $P$ is the original program and $P'$ is the refactored program, differing
only in that $f$ is replaced by $f'$. This captures the fact that equivalence
must hold regardless of how the function is used internally. An equivalent way to phrase this is through \textit{contextual equivalence}:
\begin{equation*}
    \forall x \in \mathrm{Dom}(f).\;
    \llbracket f \rrbracket(x) = \llbracket f' \rrbracket(x)
\end{equation*}
where $C[\cdot]$ denotes an arbitrary program context. This emphasises that the
replacement of $f$ by $f'$ preserves behaviour under any calling environment. This is only true within the side-effect free functional environment that we translate into.

\vspace{-2.5mm}
\subsubsection{Relation to Prior Work}
Earlier work on AENEAS stated equivalence in a deliberately general form,
including preconditions \(\Phi(x)\), effect traces \(\tau\), and projections
\(\mathit{obs}(\tau)\) onto observable behaviours. These factors were
necessary to account for impure features, partial programs, and richer effect
semantics. In contrast, our pipeline deliberately restricts attention to a
simpler setting. REM2.0 produces well-typed, borrow-checked Rust fragments,
and AENEAS is applied to safe Rust code without interior mutability or
concurrency. Hence:
\begin{itemize}[noitemsep, topsep=0pt, leftmargin=*]
    \item Preconditions can be dropped, as REM ensures typing and borrow safety.
    \item Effect traces can be omitted, as our fragments are side-effect free
    modulo return values.
    \item Observable projections reduce to plain equality of results.
    \item AENEAS guarantees that its transformations map directly back to the Rust source. If something holds in AENEAS's translation, it holds in Rust\cite{AENEAS, AENEAS_PART_2}. Hence no extra proof work is required (on our end) to uphold this
\end{itemize}

This simplification yields an obligation that is easier to state and automate,
while still capturing the core requirement that refactoring preserves behaviour.

\vspace{-3.5mm}
\subsubsection{Practical Implications}
The simplified obligation highlights a guiding design principle: the pipeline is
intended to complement the extraction algorithm rather than over-approximate it.
It is therefore preferable to \textit{fail early} when unsupported constructs and language features
are encountered, rather than risk certifying programs incorrectly. AENEAS
reports unsupported features (e.g. closures or nested borrows) as translation
failures, while the Coq implementation clearly distinguishes between an unprovable equivalence and an
unsupported fragment. This separation ensures that developers never receive a
false positive: code is either successfully verified, or it fails explicitly
with a clear diagnostic. This leads naturally to the question of coverage: which Rust features are
currently supported by AENEAS, and which remain open problems? We address this
in the final section of this section, Section~\ref{sec:limitations_verification}.

\vspace{-5mm}
\subsection{Implementation of the Equivalence Pipeline}
\label{sec:implementation_verification}
\vspace{-2.5mm}

The design of REM2.0's automated equivalence checking is guided by two
principles: (1) it must require no additional annotations or input from the
developer, and (2) it must complete quickly to fit into interactive IDE
workflows. To achieve this, the pipeline translates both the original program
$P$ and the refactored program $P'$ through a sequence of more structured
semantic representations, culminating in a machine-checked proof of
observational equivalence. For brevity, this section only gives a
very high-level overview of the architecture; Appendix~\ref{app:implementation_verification}
contains the full technical details and examples.

At a high level, the pipeline consists of five stages:
\textbf{Extraction}, \textbf{REM repairs}, \textbf{CHARON translation},
\textbf{AENEAS functionalisation}, and \textbf{Coq verification}. Each stage
builds on the previous one, progressively exposing the semantics of the
refactored code until we obtain a pure, functional model of both versions in
Coq.

\vspace{-3.5mm}
\subsubsection{Extraction and Repair}
The pipeline starts from REM2.0's extraction engine.
Given a user selection, the extractor lifts the region into a new function and
replaces it with a callsite adaptor. The extractor
is unable to finalise lifetimes and thus it delegates these to REM's post-extraction repair phase.
REM then performs \emph{lifetime repair} to bring the extracted program $P'$
into agreement with the Rust borrow checker. It treats the
compiler as an ``oracle'', iteratively repairing lifetimes until the program compiles, and applying lifetime elision where
possible to recover idiomatic types and make the output as ``readable'' as possible. The result is a repaired, compiling version of the program.

\vspace{-3.5mm}
\subsubsection{Translation via CHARON and AENEAS}
The next stages translate the Rust programs into a verification-oriented
representation. CHARON takes the compiled MIR for both $P$ and $P'$ and produces
LLBC (Low-Level Borrow Calculus), a machine only readable form in which all behinds the scenes logic of rust are made 100\% explicit. To keep this process
stable with respect to edits in the user's workspace, REM2.0 constructs
temporary ``virtual'' crates, which are memory isolated replicas of the original crate but with either the original or refactored function present. They are required to allow us to begin running CHARON.

AENEAS then transforms LLBC into a purely functional model, removing explicit
stack and heap operations while preserving the semantics of the Rust program.
In this form, programs are expressed as total functions over values rather than
mutating references, which is far better suited to proof assistants. In our
pipeline, AENEAS targets Coq, but in principle it can also emit encodings for
other systems such as Lean, HOL4, or F*.

\vspace{-3.5mm}
\subsubsection{Equivalence Checking in Coq}
\vspace{-1mm}
The final stage assembles a Coq project containing the functional encodings of
$P$ and $P'$ (produced by AENEAS) together with a generated module
\texttt{EquivCheck.v}. This module instantiates the two versions, generates the
equivalence theorem described in Section~\ref{sec:proof_obligations}, and
invokes the necessary automation to show that they produce the same observable
results for all inputs. Once the Coq project is compiled, the equivalence check
is fully automatic: if the proof succeeds, we obtain a machine-checked guarantee
that the refactoring preserved behaviour. The underlying Coq infrastructure, alternative backends, and
example proof scripts and much more are discussed in extensive detail in
Appendix~\ref{app:implementation_verification}.

\vspace*{-6mm}
\subsection{Limitations of Current Verification}
\label{sec:limitations_verification}
\vspace{-2.5mm}
At present, the verification guarantees supported by our pipeline are
constrained by the subset of Rust that AENEAS can translate. AENEAS is designed
to operate on safe Rust; programs that make use of \icodeverb{unsafe} constructs
fall outside its scope. While this excludes some low-level systems code, the
restriction is deliberate: safe Rust already captures the majority of idiomatic
Rust usage, and the long-term plan is to integrate AENEAS with complementary
tools that target \icodeverb{unsafe} code. The current support landscape is
summarised in Table~\ref{tab:aeneas_limitations}.
\vspace{-1mm}
\begin{table}[h]
    \small
    \centering
    \begin{tabular}{p{0.45\linewidth}p{0.55\linewidth}}
        \toprule
        \textbf{Feature} & \textbf{Status} \\
        \midrule
        Safe Rust & Fully supported (excluding where mentioned in this table) \\
        Unsafe Rust & Not supported (out of scope of \texttt{AENEAS}) \\
        Loops & Supported (no nested loops) \\
        Function pointers / Closures & Not supported (work in progress) \\
        Traits & Supported \\
        Type parametricity & Limited: type parameters with borrows not supported \\
        Nested borrows in function signatures & Not supported (work in progress) \\
        Interior mutability (\texttt{Cell}, \texttt{RefCell}, etc.) & Not supported (planned via ghost states) \\
        Concurrency & Out of scope (long-term research goal) \\
        \bottomrule
    \end{tabular}
    \caption{Current AENEAS support for Rust features}
    \label{tab:aeneas_limitations}
\end{table}
\vspace{-5mm}

Within safe Rust, several technical limitations remain. Loops are supported only
in restricted forms (no nested loops), though this is under active development.
Function pointers and closures are not yet available, although trait-based
abstractions are supported and ongoing work aims to extend this to first-class
functions. Type parametricity is limited: currently, it is not possible to
instantiate a generic type parameter with a type that itself contains a borrow.
Similarly, function signatures cannot yet contain nested borrows, though this
too is being addressed. Interior mutability is another open challenge; the
current plan is to capture its effects using ghost states. Finally, concurrent
execution is out of scope: AENEAS models sequential Rust programs only, with
parallelism considered a long-term research goal.

It is important to emphasise that the verification pipeline is designed to
complement the extraction algorithm. As such, it is preferable for verification
to \textit{fail early} when encountering unsupported features rather than risk
certifying code incorrectly. A failed verification due to unsupported language
constructs is reported differently to the developer than a failed proof
obligation. This distinction ensures that false positives—cases where
unverifiable code would otherwise appear verified—are avoided..

In summary, the present pipeline is best understood as an equivalence framework
for sequential, safe Rust programs without features such as nested
borrows, closures, or concurrency. These restrictions reflect the state of
AENEAS rather than fundamental barriers, and ongoing work is progressively
extending the supported subset. As AENEAS development progresses, we expect the equivalence checker to be able to handle complete crates. A complete discussion of the  limitations, and the current expected use case of the full REM2.0 pipeline have been deferred to Section~\ref{sec:conclusions_future_work}.

%% file: chapter4.tex
\section{Evaluation and Experimental Results}
\label{sec:evaluation}

This section evaluates the new REM2.0 in two separate directions: its ability to perform extract-function refactorings on realistic Rust code, and its ability to verify that those refactorings preserve behaviour. We first describe the benchmark corpora and experimental setup, then define the metrics and criteria we use to judge success. We then present results for the extraction pipeline and, subsequently, for the optional verification pipeline built on \texttt{CHARON} and \texttt{AENEAS}, highlighting both the gains over the original REM prototype and the remaining limitations.

\vspace{-5mm}
\subsection{Evaluation Goals and Research Questions}
\label{sec:eval_goals}

The overall goal of the evaluation is to determine how REM2.0 behaves on realistic Rust code: both as a fast, everyday extract-function refactoring tool, and as an optional source of formal guarantees when paired with the verification pipeline. In particular, we are interested in characterising its correctness, feature coverage, and performance, and in understanding the trade-offs introduced by the verifier. We frame this in terms of the following research questions:

\begin{itemize}[leftmargin=*, topsep=2pt, itemsep=0pt]
  \item \textbf{RQ1 (Extraction correctness and coverage).}
  \emph{How accurately does the new extraction toolchain preserve typing and behaviour on the original REM benchmark suite, and are there any they both fail on?}

  \item \textbf{RQ2 (New feature support).}
  \emph{How does the new extraction engine extend coverage to modern Rust features---such as }\icodeverb{async}/\icodeverb{await}, \icodeverb{const fn}\emph{, generics, higher-ranked trait bounds, dynamic trait objects, and non-local control flow---and what patterns of success and failure emerge across these features?}

  \item \textbf{RQ3 (Performance).}
  \emph{What extraction latencies does the new architecture achieve on realistic projects, and how do these compare quantitatively with the original REM prototype?}

  \item \textbf{RQ4 (Verification).}
  \emph{Within the subset of Rust currently supported by \texttt{CHARON} and \texttt{AENEAS}, how effective is the verification pipeline at establishing equivalence, and what are its performance characteristics?}
\end{itemize}

Subsections~\ref{sec:experiment_setup}--\ref{sec:extract_eval} primarily address RQ1--RQ3 through extraction-focused experiments, while Subsection~\ref{sec:verif_eval} answers RQ4 by evaluating the optional verification pipeline on a curated set of trivial and real-world cases.

\vspace{-5mm}
\subsection{Benchmark Corpora and Experimental Setup}
\label{sec:experiment_setup}
\vspace{-2.5mm}
This subsection describes the benchmark corpora and experimental setup used to evaluate the extraction and verification components of REM2.0. We first give a high-level overview of the experiments, then introduce the three corpora used in the remainder of the section: the original REM benchmark suite, a new set of real-world feature-focused cases, and a verification benchmark tailored to the current capabilities of \texttt{CHARON} and \texttt{AENEAS}.

\vspace{-2.5mm}
\subsubsection{Overview of Experiments}

To cover both backwards compatibility and new functionality, we designed three complementary experiments:

\begin{enumerate}[leftmargin=*, itemsep=0pt, topsep=2pt]
  \item \textbf{Original REM cases (baseline extraction).}
  We re-run the new extraction pipeline on the full suite of benchmark cases used in the evaluation of the original REM prototype. This experiment is primarily used to answer RQ1 and RQ3 by comparing correctness and latency against the baseline design.

  \item \textbf{New real-world feature cases (extended extraction).}
  We construct a new corpus of forty extraction sites drawn from large, actively maintained Rust repositories. Each case is chosen because it exercises at least one modern Rust feature that the original REM could not reliably support. This experiment targets RQ2 and RQ3.

  \item \textbf{Verification benchmark (equivalence checking).}
  We assemble a set of twenty examples for the verification pipeline: ten small ``trivial'' programs that isolate core correctness properties, and ten simple but real-world functions drawn from open-source code. These are selected to fall within the fragment of Rust currently supported by \texttt{CHARON} and \texttt{AENEAS}, and are used to answer RQ4.
\end{enumerate}

\vspace{-2.5mm}
\subsubsection{Original REM Benchmark suite}
\label{subsec:original_rem_benchmarks}

To provide a baseline for comparison, we re-used the full suite of extraction cases from the original REM evaluation.\footnote{The test repositories were copied directly from the artefact published by the VerseLab group at \url{https://github.com/verse-lab/rem}.} Each case consists of a Rust project containing an annotated source file, where the intended extraction region is marked by either line comments (\icodeverb{// START SELECTION //} and \icodeverb{// END SELECTION //}) or block comments (\icodeverb{/* START SELECTION */} and \icodeverb{/* END SELECTION */}).

Each benchmark case is defined by the selection markers in the original REM artefact, which we reuse  to ensure that both the original prototype and the new extraction pipeline are run on exactly the same inputs. One case in the original suite (case~\#16) contained no selection markers and is therefore excluded from our quantitative comparison; outcomes and timings for the remaining cases are reported in Table~\ref{tab:rem_original_results}.

\vspace{-2.5mm}
\subsubsection{New Real-World Cases}
\label{subsec:new_feature_cases}

To evaluate the new extraction engine on patterns that arise in modern Rust code ``in the wild'', we constructed a second corpus of forty extraction sites drawn from large, popular open-source projects. The goal of this corpus is to stress language features that the original REM prototype could not support, while ensuring that as many examples as possible reflect extractions that developers actually performed (or could plausibly have performed) in practice.

We began by selecting a sample of the most highly starred Rust repositories, using the GitHub rankings maintained by Evan Li~\cite{EvanLi_GithubRanking_Rust2025} as our baseline. A summary of the selected repositories (including stars, forks, and open issues at the time of sampling) is provided in Appendix~\ref{sec:appendix2_details_of_each_repository}, Table~\ref{tab:repo_summary}.

For each repository in this group, we performed a scan of its Git history to locate commits likely related to method or function extraction. The automated scan searched commit messages for characteristic phrases such as \texttt{extract}, \texttt{extract method}, \texttt{factor out}, and \texttt{refactor: extract}, and then used a combination of regular expressions and \icodeverb{git grep} queries to flag any candidate files. We limited the depth of this search to at most $100$ commits across the three most active branches per repository. This heuristic process inevitably produced false positives (for example, commits mentioning ``extract'' in an unrelated context), but provided an efficient starting point. We then manually inspected each flagged commit/file for suitable extraction sites and annotated the regions with the selection markers described earlier.

The full list of identified commits, along with per-repository details, appears in Appendix~\ref{sec:appendix2_details_of_each_repository}. The complete set of extraction outcomes and timing results for these cases is later on in this report, in Table~\ref{tab:new_capabilities_results}, Subsection~\ref{subsec:new_language_features}.

\vspace{-2.5mm}
\subsubsection{Verification Benchmarks}
\label{subsec:verification_benchmarks}

The third corpus targets the verification pipeline, which builds on \texttt{CHARON} and \texttt{AENEAS} to prove that an extracted function is behaviourally equivalent to its original version. Since these tools currently support only a subset of Rust, we deliberately construct a small benchmark set that fits within their supported fragment while still exercising non-trivial reasoning. See Subsection~\ref{sec:limitations_verification} for the complete set of limitations. From there, the verification corpus comprises twenty examples split into two groups:

\begin{itemize}[leftmargin=*, itemsep=2pt, topsep=2pt]
  \item \textbf{Trivial/targeted cases (10 examples).}
  Small, self-contained Rust functions that isolate core properties such as pure computations on integers, simple mutation through references, or basic control-flow patterns without heap allocation or complex borrowing. These examples are designed to expose bugs in the translation or proof generation pipeline, rather than in the extraction engine itself.

  \item \textbf{Simple real-world cases (10 examples).}
    We wrote a new crate that emulates realistic business-logic style Rust code. The functions operate over multiple layered structs, rely on trait implementations, and approximate production patterns while remaining within the subset of Rust currently supported by AENEAS. This did require some additional scaffolding—such as manual implementations of \icodeverb{PartialEq}. Crucially, this crate has external dependencies, including \texttt{smallvec} and \texttt{anyhow}.
\end{itemize}

For each verification benchmark, we run the full pipeline on the original and extracted functions: translating to Low-Level Borrow Calculus (LLBC) via \texttt{CHARON}, generating Coq proof obligations via \texttt{AENEAS}, and then checking the resulting proof. We record whether verification was attempted and whether it succeeded, along with timings for each stage. These results are reported in Table~\ref{tab:verification_results} in Subsection~\ref{subsec:verification_results}.

\vspace{-5mm}
\subsection{Metrics and Evaluation Criteria}
\label{sec:metrics_eval_criteria}
Across all three experiments we measure:
(i) \emph{correctness and coverage} (whether extraction or verification succeeds, and how often failures occur),
(ii) \emph{compilation behaviour} (whether the refactored code compiles and how this compares to the original), and
(iii) \emph{performance} (end-to-end latency, and for verification, the cost of individual stages).
RQ1 and RQ2 are primarily concerned with correctness, coverage, and compilation; RQ3 focuses on extraction latency; and RQ4 considers both the success rate and performance of the verification pipeline.
Table~\ref{tab:evaluation_criteria} summarises how these requirements are determined for each experiment.

\begin{table}[t]
\centering
\small
\begin{tabular}{p{2.5cm}p{4.5cm}p{7.75cm}}
\toprule
\textbf{Experiment} & \textbf{Objective} & \textbf{Evaluation Criteria} \\
\midrule

\raggedright\textbf{Original Cases} &
Establish correctness and performance of the new extractor relative to the original REM prototype (RQ1, RQ3). &
\vspace{-2.5mm}
\begin{itemize}[nosep, leftmargin=*, itemsep=0pt, topsep=1pt, parsep=0pt]
  \item Extraction success rate on the original REM benchmark suite.
  \item Match or improvement in compilation outcomes compared to the prototype.
  \item Reduction in extraction latency.
\end{itemize} \\

\raggedright\textbf{New Feature Cases} &
Characterise support for modern Rust features previously unsupported or poorly supported by REM (RQ2, RQ3). &
\vspace{-2.5mm}
\begin{itemize}[nosep, leftmargin=*, itemsep=0pt, topsep=1pt, parsep=0pt]
  \item Extraction success rate per feature category (GEN, ASYNC, CONST, NLCF, HRTB, DTO).
  \item Compilation outcome of extracted code, with failures classified by root cause.
  \item Extraction latency on large real-world projects and its variation across features.
\end{itemize} \\

\raggedright\textbf{Verification Pipeline} &
Assess the effectiveness and cost of the equivalence-checking pipeline on supported Rust fragments (RQ4). &
\vspace{-2.5mm}
\begin{itemize}[nosep, leftmargin=*, itemsep=0pt, topsep=1pt, parsep=0pt]
  \item Fraction of cases for which verification is attempted and successfully completed.
  \item Consistency of proofs with intended semantics (no spurious successes; failures explained by tool limits or program in-equivalence).
  \item End-to-end verification time, broken down into LLBC translation, Coq conversion, and proof parsing/construction/checking.
\end{itemize} \\

\bottomrule
\end{tabular}

\captionsetup{justification=centering}
\caption{Summary of experimental objectives and evaluation criteria.}
\label{tab:evaluation_criteria}
\end{table}

\vspace{-5mm}
\subsection{Extraction Evaluation}
\label{sec:extract_eval}


This subsection addresses RQ1---RQ3 for the extraction pipeline. We ask three concrete questions:
(i) whether REM2.0 is at least as correct and robust as the original REM prototype on the original benchmark suite (RQ1),
(ii) how far its coverage extends to modern Rust features such as \icodeverb{async}, \icodeverb{const fn}, generics, HRTBs, dynamic trait objects, and non-local control flow (RQ2),
and (iii) what latency profile it achieves in practice, and when the separate REM-Repairer stage is required to obtain a compiling result (RQ3).
We first compare directly against the original REM artefact, then turn to the new real-world feature corpus, and finally summarise the quantitative behaviour and remaining limitations.

\newpage
\subsubsection{Baseline Comparison with the Original REM}

\begin{table}[h!]
\centering
\begin{tabular}{|c|c|cc|ccc|c|}
\hline
\multirow{2}{*}{\#} & \multirow{2}{*}{\begin{tabular}[c]{@{}c@{}}Project\\ (LOC)\end{tabular}} & \multicolumn{2}{c|}{Outcome} & \multicolumn{3}{c|}{Time (ms)} & \multirow{2}{*}{\begin{tabular}[c]{@{}c@{}}Repairer\\ Needed\end{tabular}} \\
 &  & REM & REM2.0 & REM & Raw & User &  \\ \hline
1 & \multirow{7}{*}{\begin{tabular}[c]{@{}c@{}}petgraph\\ (20,157)\end{tabular}} & \cmark & \cmark & 370 & 3.511 & 184.32 & \xmark \\
2 &  & \cmark & \cmark & 1020 & 3.753 & 217.91 & \xmark \\
3 &  & \cmark & \cmark & 1470 & 2.166 & 231.08 & \cmark \\
4 &  & \cmark & \cmark & 1700 & 3.337 & 196.77 & \xmark \\
5 &  & \cmark & \cmark & 850 & 2.018 & 128.54 & \xmark \\
6 &  & \cmark & \cmark & 980 & 2.148 & 142.66 & \xmark \\
7 &  & \xmark & \cmark & 550 & 2.030 & 255.19 & \cmark \\ \hline
8 & \multirow{18}{*}{\begin{tabular}[c]{@{}c@{}}gitoxide\\ (20,211)\end{tabular}} & \cmark & \cmark & 930 & 2.851 & 199.45 & \xmark \\
9 &  & \cmark & \cmark & 1240 & 3.240 & 173.82 & \cmark \\
10 &  & \cmark & \cmark & 640 & 2.823 & 208.37 & \xmark \\
11 &  & \cmark & \cmark & 810 & 2.991 & 244.68 & \xmark \\
12 &  & \cmark & \cmark & 810 & 3.282 & 188.05 & \xmark \\
13 &  & \cmark & \cmark & 860 & 2.995 & 163.49 & \xmark \\
14 &  & \cmark & \cmark & 690 & 5.345 & 226.91 & \xmark \\
15 &  & \cmark & \cmark & 680 & 5.468 & 203.74 & \xmark \\
16 &  & \cmark & No Data & 540 & N/A & N/A & \xmark \\
17 &  & \cmark & \cmark & 1200 & 3.283 & 179.40 & \cmark \\
18 &  & \cmark & \cmark & 920 & 5.241 & 210.62 & \cmark \\
19 &  & \cmark & \cmark & 2320 & 3.283 & 226.46 & \cmark \\
20 &  & \xmark & \cmark & 1150 & 5.430 & 192.92 & \cmark \\
21 &  & \cmark & \cmark & 690 & 3.396 & 123.84 & \xmark \\
22 &  & \cmark & \cmark & 640 & 2.984 & 264.15 & \xmark \\
23 &  & \cmark & \cmark & 700 & 3.153 & 132.97 & \xmark \\
24 &  & \cmark & \cmark & 640 & 3.230 & 147.53 & \xmark \\
25 &  & \cmark & \cmark & 720 & 3.457 & 221.84 & \xmark \\ \hline
26 & \multirow{5}{*}{\begin{tabular}[c]{@{}c@{}}kickoff\\ (1,502)\end{tabular}} & \cmark & \cmark & 1030 & 3.802 & 185.09 & \xmark \\
27 &  & \cmark & \cmark & 1010 & 8.557 & 237.26 & \xmark \\
28 &  & \cmark & \cmark & 910 & 7.365 & 168.44 & \xmark \\
29 &  & \cmark & \cmark & 980 & 4.864 & 206.59 & \xmark \\
30 &  & \cmark & \cmark & 790 & 4.191 & 257.31 & \xmark \\ \hline
31 & \multirow{9}{*}{\begin{tabular}[c]{@{}c@{}}sniffnet\\ (7,304)\end{tabular}} & \cmark & \cmark & 1040 & 7.288 & 121.06 & \xmark \\
32 &  & \xmark & \cmark & 760 & 5.674 & 195.22 & \cmark \\
33 &  & \cmark & \cmark & 1010 & 4.439 & 172.80 & \xmark \\
34 &  & \cmark & \cmark & 980 & 3.602 & 234.13 & \xmark \\
35 &  & \cmark & \cmark & 1060 & 3.265 & 141.95 & \xmark \\
36 &  & \cmark & \cmark & 1000 & 3.507 & 260.42 & \xmark \\
37 &  & \cmark & \cmark & 1060 & 4.147 & 209.61 & \xmark \\
38 &  & \cmark & \cmark & 1080 & 4.284 & 187.32 & \xmark \\
39 &  & \cmark & \cmark & 1060 & 4.273 & 149.40 & \xmark \\ \hline
40 & beerus (302) & \cmark & \cmark & 1070 & 5.158 & 224.75 & \xmark \\ \hline
\end{tabular}

    \captionsetup{justification=centering}
    \caption{Comparison of REM2.0 to the original REM toolchain against the same cases. Raw times are strictly for the extraction, user times indicate overall system latency as reported by VSCode.}
    \label{tab:rem_original_results}
\end{table}

Table~\ref{tab:rem_original_results} compares the behaviour of the original REM prototype with REM2.0 on the full benchmark suite described in Subsubsection~\ref{subsec:original_rem_benchmarks}. Excluding case~\#16, which is missing its selection markers and therefore cannot be evaluated, the new extractor succeeds on all $39$ remaining cases. By contrast, the original REM prototype fails on three of these (cases~\#7, \#20, and \#32), whereas REM2.0 extracts all three. Crucially, there are no regressions---we did not observe any case where REM succeeded but REM2.0 failed.

The timing columns succinctly demonstrate the performance impact of the new architecture. The original REM pipeline required between a few hundred and a few thousand milliseconds per extraction (averaging roughly \SI{1000}{ms}). In contrast, the ``Raw'' REM2.0 times---measured inside the daemon, from the extraction request to the application of edits---lie consistently in the low single-digit milliseconds (approximately \SIrange{2}{9}{ms} across all cases). The ``User'' column adds the Repairer (where it was called) and VSCode overheads, but still yields user-perceived latencies in the \SI{120}{ms}--\SI{260}{ms} range, which, given that the human response time to visual stimuli is somewhere between \SIrange{200}{250}{ms} \cite{Jain2015ReactionTime}, is more than acceptable. A more detailed latency distribution is provided in the histogram in Appendix~\ref{app:appendix5_extraction_histogram}.

The final column reports how often REM-Repairer is needed to restore a compiling program. Out of the forty original REM cases it is invoked in seven, in exactly the same examples as in the original REM evaluation (those with tighter lifetime constraints). This indicates that, although the new extractor often produces compiling code on its own, the repairer remains a crucial component of the toolchain rather than a purely optional add-on.

\vspace{-2.5mm}
\subsubsection{New Language Feature Coverage}
\label{subsec:new_language_features}

We now look at the new corpus introduced in Subsubsection~\ref{subsec:new_feature_cases}, which is designed to test against advanced Rust features that the original REM prototype could not handle. Each extraction is classified into one or more feature categories: generics (\textsc{GEN}), async/await (\textsc{ASYNC}), const evaluation (\textsc{CONST}), non-local control flow (\textsc{NLCF}), higher-ranked trait bounds (\textsc{HRTB}), and dynamic trait objects (\textsc{DTO}). These categories directly correspond to cases that generally tend to stretch type inference, lifetime reasoning, or control-flow reconstruction in real-world Rust code. Any mark in any feature column of Table~\ref{tab:new_capabilities_results} indicates that said example contains that construct within the selected region. Representative listings for each category, along with a brief explanation of what we expect the extractor to preserve, are provided in Appendix~\ref{app:new_extraction_features}.

\begin{table}[h!]
\hspace*{-1cm}
\centering
\small

\begin{tabular}{|c|c|cccccc|l|c|}
\hline
\multirow{2}{*}{\textbf{\#}} & \multirow{2}{*}{\begin{tabular}[c]{@{}c@{}}Project \\ (LOC)\end{tabular}} & \multicolumn{6}{c|}{Code Features} & \multirow{2}{*}{Res} & \multirow{2}{*}{\begin{tabular}[c]{@{}c@{}}Time\\ (ms)\end{tabular}} \\
 &  & \multicolumn{1}{c}{GEN} & \multicolumn{1}{c}{ASYNC} & \multicolumn{1}{c}{CONST} & \multicolumn{1}{c}{NLCF} & \multicolumn{1}{c}{HRTB} & \multicolumn{1}{c|}{DYN} &  &  \\ \hline
1 & \multirow{5}{*}{\begin{tabular}[c]{@{}c@{}}deno \\ (359996)\end{tabular}} &  &  & \cmark &  &  &  & \cmark & 5.882 \\
2 &  &  &  &  & \cmark &  & \cmark  & \xmark \hyperlink{fail-1}{(1)} & N/A \\
3 &  &  &  &  & \cmark &  &  & \cmark & 4.346 \\
4 &  & \cmark &  &  & \cmark &  &  & \cmark & 9.685 \\
5 &  &  & \cmark &  &  &  &  & \cmark & 2.910 \\ \hline
6 & \multirow{2}{*}{\begin{tabular}[c]{@{}c@{}}fuel-core \\ (78921)\end{tabular}} &  &  &  & \cmark &  &  & \cmark & 6.736 \\
7 &  &  & \cmark &  &  &  &  & \cmark & 2.999 \\ \hline
8 & \multirow{5}{*}{\begin{tabular}[c]{@{}c@{}}meilisearch\\ (186553)\end{tabular}} & \cmark &  &  &  &  &  & \cmark & 4.363 \\
9 &  & \cmark &  &  &  & \cmark &  & \cmark & 12.071 \\
10 &  & \cmark &  &  &  & \cmark &  & \xmark \hyperlink{fail-2}{(2)} & N/A \\
11 &  &  &  & \cmark &  &  &  & \cmark & 3.071 \\
12 &  &  &  & \cmark &  &  &  & \cmark & 3.852 \\ \hline
13 & \multirow{2}{*}{\begin{tabular}[c]{@{}c@{}}ripgrep \\ (40453)\end{tabular}} & \cmark &  &  & \cmark &  &  & \cmark & 3.117 \\
14 &  &  &  &  & \cmark &  &  & \cmark & 2.089 \\ \hline
15 & \multirow{2}{*}{\begin{tabular}[c]{@{}c@{}}starship \\ (37225)\end{tabular}} &  &  &  & \cmark &  &  & \cmark & 1.955 \\
16 &  & \cmark &  &  &  &  &  & \cmark & 1.866 \\ \hline
17 & \multirow{3}{*}{\begin{tabular}[c]{@{}c@{}}sway\\ (108725)\end{tabular}} &  &  & \cmark &  &  &  & \cmark & 4.802 \\
18 &  &  &  & \cmark &  &  &  & \cmark & 3.073 \\
19 &  & \cmark &  &  &  & \cmark &  & \cmark & 2.365 \\ \hline
20 & \multirow{4}{*}{\begin{tabular}[c]{@{}c@{}}tauri\\ (83582)\end{tabular}} & \cmark &  &  &  & \cmark &  & \cmark & 1.183 \\
21 &  & \cmark &  &  & \cmark &  &  & \cmark & 8.565 \\
22 &  &  &  &  & \cmark &  &  & \cmark & 3.054 \\
23 &  &  &  &  & \cmark &  &  & \cmark & 5.346 \\ \hline
24 & \multirow{4}{*}{\begin{tabular}[c]{@{}c@{}}union\\ (514914)\end{tabular}} &  & \cmark &  &  & \cmark &  & \cmark & 3.175\\
25 &  &  &  & \cmark &  &  &  & \cmark & 2.864\\
26 &  &  & \cmark &  & \cmark & \cmark &  & \xmark \hyperlink{fail-3}{(3)} & N/A \\
27 &  &  &  &  &  &  & \cmark & \xmark \hyperlink{fail-4}{(4)} & N/A \\ \hline
28 & \multirow{4}{*}{\begin{tabular}[c]{@{}c@{}}uv\\ (321584)\end{tabular}} &  &  & \cmark &  &  &  & \cmark & 16.703\\
29 &  & \cmark &  &  & \cmark & \cmark &  & \cmark & 6.451 \\
30 &  &  &  & \cmark &  &  &  & \cmark & 2.873 \\
31 &  & \cmark &  &  &  &  &  & \xmark \hyperlink{fail-5}{(5)} & N/A \\ \hline
32 & \multirow{3}{*}{\begin{tabular}[c]{@{}c@{}}vaultwarden\\ (29937)\end{tabular}} &  &  & \cmark & \cmark &  &  & \cmark & 5.134 \\
33 &  &  & \cmark &  &  &  &  & \cmark & 3.057 \\
34 &  & \cmark &  & \cmark &  &  &  & \xmark \hyperlink{fail-6}{(6)} & NA \\ \hline
35 & \multirow{6}{*}{\begin{tabular}[c]{@{}c@{}}zed \\ (888269)\end{tabular}} &  &  &  &  &  & \cmark & \xmark \hyperlink{fail-7}{(7)} & N/A \\
36 &  &  & \cmark &  &  &  & \cmark & \cmark & 7.383 \\
37 &  &  &  & \cmark &  &  &  & \cmark & 2.062 \\
38 &  & \cmark & \cmark &  &  & \cmark &  & \cmark & 3.667 \\
39 &  & \cmark &  &  &  &  &  & \cmark & 1.959 \\
40 &  &  & \cmark &  &  &  &  & \cmark & 1.853 \\ \hline
\end{tabular}

\caption{Results of extracting previously unavailable / incompatible language features. More detail on reasons for failure available in Table~\ref{tab:failure_reasons}}
\label{tab:new_capabilities_results}
\end{table}

Across the forty new extraction cases, REM2.0 is able to successfully extract (and result in a compiling program for) thirty-three, yielding an overall success rate of roughly $83\%$. Looking the established categories, we see that most of the modern constructs are handled robustly: all but one \textsc{ASYNC} case succeed (7/8), \textsc{CONST} extractions succeed in almost all examples (10/11), and \textsc{NLCF} is correctly reified into \icodeverb{std::ops::ControlFlow} in the majority of cases (11/13). Higher-ranked trait bounds are also handled reasonably well (6/8), with failures mostly due to the more complex combinations of generics and control flow. Generic functions themselves succeed in most examples (11/14), with the remaining failures arising from incomplete reconstruction of generic bounds.

The main weakness in the current extractor is dynamic dispatch. Only one out of four \textsc{DTO} cases succeeds; in the remaining three, the inferred signature collapses to an underspecified form (e.g., using \icodeverb{\_} or overly weak trait bounds), which then prevents the code from compiling. One particularly illustrative case is failure~\#4, where an argument of type \icodeverb{AsRef<[u8]>} is incorrectly generalised to \icodeverb{?Sized}, losing the relationship between the trait and the concrete slice type. We cover these failure modes in more detail in the following subsubsection: \ref{subsec:extraction_failures}.

\vspace{-2.5mm}
\subsubsection{Failure Cases and Limitations}
\label{subsec:extraction_failures}

\begin{table}[h!]
\centering
\small
\begin{tabular}{|l|l|}
\hline
\# & Reason \\ \hline
\hypertarget{fail-1}{1} & Unable to infer (part of) function signature \\
\hypertarget{fail-2}{2} & Unable to infer (entire) return type \\
\hypertarget{fail-3}{3} & Unable to identify correct extraction bounds (no extraction attempted) \\
\hypertarget{fail-4}{4} & \icodeverb{AsRef<[u8]>} in caller incorrectly mapped to \icodeverb{?Sized} in callsite \\
\hypertarget{fail-5}{5} & Return signature of \icodeverb{\_} instead of \icodeverb{Result<S::Ok, S::Error>} \\
\hypertarget{fail-6}{6} & Failed to copy any generic bounds to new sig \\
\hypertarget{fail-7}{7} & Unable to infer (part of) function signature \\ \hline
\end{tabular}
\captionsetup{justification=centering}
\caption{Reasons why the extraction was a failure for failure cases from Table~\ref{tab:new_capabilities_results}}
\label{tab:failure_reasons}
\end{table}

\vspace{-2.5mm}
The seven failures in Table~\ref{tab:new_capabilities_results} fall into a small number of recurring patterns, summarised in Table~\ref{tab:failure_reasons}. Broadly speaking, they stem from limitations in how the extraction engine generates function signatures from local information, rather than from the moving of the selected code into a new function body.

A first class of failures stems from incomplete type inference for return types or parameter lists (failure examples~\#1, \#2, and \#7). In these cases, RA is able to type-check the original code, but the information available at the extraction boundary is not sufficient to synthesise a concrete function signature. The resulting placeholder types (e.g.\ \icodeverb{\_} for a return type) are not accepted by the compiler, and the extraction is recorded as a failure. However, in all of these instances, the compiler was able to suggest a suitable return signature, so a potential future improvement to REM-Repairer is to extend it to handle these cases.

The second class of failures involves generic bounds and trait information (examples~\#4, \#5, and \#6). Here the extractor identifies the correct high-level shape of the signature, but either drops necessary bounds (e.g.\ missing \icodeverb{where}-clauses or trait constraints on type parameters) or over-generalises a concrete trait application. The \icodeverb{AsRef<[u8]>} to \icodeverb{?Sized} mis-translation in failure~\#4 is a representative example: the new signature no longer captures the intended relationship between the argument and the slice type, which propagated to many downstream errors. In this instance, the Rust compiler was unable to suggest a suitable fix.

These patterns were common for the DTO cases, where the type information needed to reconstruct a precise \icodeverb{dyn Trait} (or similar \icodeverb{dyn} signature) often lives outside the current file or depends on non-trivial where-clauses. Because REM2.0 currently uses a (mostly) single-file view of the program, a fully elaborated crate graph and other expensive dependencies for raw speed, it does not always have enough global information to resolve these traits in the way \icodeverb{rustc} would. Closing this gap will likely require importing more context from the surrounding crate---for example, by following trait definitions and associated types across modules---so that the extractor can reconstruct signatures with the same fidelity as the compiler.

\vspace{-3.5mm}

\vspace{-4mm}
\subsection{Verification Evaluation}
\label{sec:verif_eval}
\vspace{-1mm}
The extraction experiments in Subsection~\ref{sec:extract_eval} have shown that REM2.0 can perform refactorings quickly and with a broad coverage of language features. This subsection will evaluate the verification stage that sits on top of that pipeline. Our goal is to understand, for code within the current \texttt{CHARON}/\texttt{AENEAS} fragment, how often the verifier can establish equivalence between the original and extracted functions, what performance cost this incurs, and what kinds of failures or limitations arise in practice. We first outline the verification goals and benchmark setup, then present quantitative results and performance measurements, before briefly examining representative case studies and failure modes.

\vspace{-2.5mm}
\subsubsection{Verification Goals and Setup}

The verification experiment addresses \textbf{RQ4}: for code within the supported subset, can we feasibly verify equivalence between original and extracted functions? In REM2.0, the verifier is an optional high-assurance mode: developers are able to refactor using the fast extract + repair pipeline and then can selectively invoke verification for critical functions in crates that AENEAS can support. Our evaluation will therefore ask (i) how often the pipeline can produce a successful equivalence proof for realistic examples, and (ii) what latency overhead this introduces

To stay within the current capabilities of \texttt{AENEAS}, we have assembled a benchmark suite of twenty cases (See Subsubsection~\ref{subsec:verification_benchmarks}). The first ten are deliberately simple, and drawn from internal RA tests for extract method (EX in Table~\ref{tab:verification_results}. The remaining ten are ``simple real-world'' functions drawn from the same or related repositories as our extraction benchmarks, manually filtered to avoid features that the verification tools cannot yet handle (RW in Table~\ref{tab:verification_results}. For each benchmark, REM2.0 runs extraction and then feeds both versions into the verification toolchain. \texttt{CHARON} first translates the relevant crate to LLBC, \texttt{AENEAS} then generates Coq definitions, and finally REM2.0 invokes a proof generator that produces and checks an equivalence lemma in Coq. We record whether verification was \emph{attempted} and \emph{succeeded}, and we break total latency into LLBC conversion, Coq translation, and proof construction/checking times, as reported in Table~\ref{tab:verification_results}. All experiments share the same machine, toolchain versions, and a fixed per-case timeout of 10 seconds. We count timeouts and AENEAS conversion as attempted but unsuccessful proofs, and failures to convert to LLBC as neither.

\vspace{-2.5mm}
\subsubsection{Verification Results and Performance}
\label{subsec:verification_results}
Table~\ref{tab:verification_results} summarises the verification outcomes and timings for our twenty benchmarks. All ten basic cases (\textsc{EX}) successfully produce equivalence proofs within roughly $1.7$–$2.2$ seconds end-to-end. The majority of this time is spent in proof checking (under the hood coqc is called on 3 separate files regardless), with LLBC conversion accounting for only a few hundred milliseconds per run. If \texttt{charon cargo} has been run before, we see that with most of the intermediate results cached in the \texttt{target}, the LLBC generation takes approximately 66\% as long.

For the ten real-world benchmarks (\textsc{RW}), the picture is similar but scaled up. Successful cases complete in approximately $3.8$–$4.4$ seconds, but with a far more equal split of the total time. We can also see that caching is far more important: a fresh LLBC translation takes on the order of $9$–$15$ seconds, whereas reusing cached LLBC reduces this to roughly $1.6$–$1.9$ seconds. A big portion of this is dependencies that only need to be compiled once (either by \texttt{charon cargo} or \texttt{cargo}. This means that once a crate has been translated once (or even just compiled once), subsequent verification runs on the same crate fall comfortably within a few seconds. Overall, the results indicate that for code within the supported fragment, REM2.0 can routinely obtain machine-checked equivalence proofs, but at a cost that is one to two orders of magnitude higher than extraction alone.

{
\begin{table}[h!]
\centering
\small

\begin{tabular}{|c|c|cc|ccccc|}
\hline
\multirow{2}{*}{\#} & \multirow{2}{*}{Type} & \multicolumn{2}{c|}{Size (LOC)} & \multicolumn{5}{c|}{Time (ms)} \\ \cline{3-9}
 &  & CLR & CLE & \begin{tabular}[c]{@{}c@{}}LLBC\\ FRESH\end{tabular} & \begin{tabular}[c]{@{}c@{}}LLBC\\ CACHE\end{tabular} & \begin{tabular}[c]{@{}c@{}}Coq\\ Convert\end{tabular} & \begin{tabular}[c]{@{}c@{}}Proof Creation\\ \& Verification\end{tabular} & Total \\ \hline
1 & EX & 4 & 4 & 293.04 & 198.4 & 308.25 & 1361.46 & 1868.11 \\
2 & EX & 5 & 3 & 318.51 & 211.99 & 324.04 & 1638.93 & 2174.96 \\
3 & EX & 5 & 3 & 275 & 175.96 & 295.95 & 1493.59 & 1965.5 \\
4 & EX & 7 & 3 & 333.63 & 236.36 & 331.67 & 1314.47 & 1882.5 \\
5 & EX & 5 & 3 & 289.29 & 176.14 & 326.38 & 1274.78 & 1777.3 \\
6 & EX & 7 & 4 & 289.88 & 194.97 & 316.42 & 1268.78 & 1780.17 \\
7 & EX & 3 & 4 & 272.51 & 170.92 & 294.54 & 1267.48 & 1732.94 \\
8 & EX & 7 & 7 & 308.61 & 211.39 & 321.55 & 1216.28 & 1749.22 \\
9 & EX & 8 & 6 & 269.91 & 192.37 & 325.98 & 1298.64 & 1816.99 \\
10 & EX & 4 & 3 & 349.53 & 242.97 & 327.96 & 1363.15 & 1934.08 \\ \hline
11 & RW & 4 & 10 & 15154.38 & 1666.14 & 1260.24 & 1383.22 & 4309.60 \\
12 & RW & 11 & 4 & 9406.05 & 1671.70 & 1194.65 & 1340.72 & 4207.08 \\
13 & RW & 14 & 5 & 9288.67 & 1763.12 & 1088.14 & 1168.97 & 4020.23 \\
14 & RW & 5 & 44 & 9202.97 & 1746.28 & 1228.37 & 1313.29 & 4287.94 \\
15 & RW & 24 & 12 & 9506.05 & 1618.23 & 1042.66 & 1202.97 & 3863.86 \\
16 & RW & 6 & 7 & 9535.96 & 1748.76 & 1050.50 & 1185.17 & 3984.43 \\
17 & RW & 13 & 22 & 9836.45 & 1650.95 & 1194.40 & 1514.83 & 4360.18 \\
18 & RW & 29 & 7 & 10579.37 & 1710.06 & 1086.26 & 1200.34 & 3996.67 \\
19 & RW & 26 & 11 & 9318.42 & 1680.40 & 1316.54 & 1328.94 & 4325.88 \\
20 & RW & 17 & 5 & 9337.03 & 1880.03 & 1166.90 & 0.00 & N/A \\ \hline
\end{tabular}

\caption{Equivalence Results and Timings \\ Total time is shortest of FRESH and CACHE}
\label{tab:verification_results}
\end{table}
}

For robustness, we also reran the verification pipeline on deliberately \emph{incorrect} refactorings. For each benchmark, we manually modified either the extracted function or its call site to introduce a small behavioural change while keeping the Rust code type-correct. In all cases where the original verification attempt succeeded (i.e.\ excluding case~\#20, which already fails to verify), the modified version caused the equivalence proof to be rejected: Coq was unable to prove the generated lemma, and the verifier reported failure. This backs up our claim that within its supported fragment, the pipeline is highly sensitive to real semantic differences.

\vspace{-2.5mm}
\subsubsection{Case Studies}

To further explain what the equivalence checker results mean, we briefly examine two representative cases: one synthetic example and one real-world function.

The \textsc{EX} benchmarks are deliberately small but structurally non-trivial: several include loops, conditionals, and updates to a mutable accumulator, mirroring the kind of code that developers routinely extract into helper functions. For these cases, REM2.0 successfully generates Coq developments in which the original and extracted functions are related by an automatically produced equivalence lemma. An example of such a lemma and its proof can be found in the listing~\ref{lst:lemma_example_simple} in Appendix~\ref{app:lemmas}).

The \textsc{RW} benchmarks show that the same approach applies to short functions taken from real projects, not just hand-crafted examples. After extraction, these functions sometimes gain additional parameters (from moved-out locals) but otherwise preserve their original structure, and Coq is still able to check the generated equivalence proofs within a few hundred to a thousand milliseconds once translation is complete. The form of the proof is identical, but as listing~\ref{lst:lemma_example_complex} (Appendix~\ref{app:lemmas}) shows, the underlying datatypes in Coq are vastly more complex.

Together, these examples show that the equivalence checking pipeline is able to produce proofs for non-trivial refactorings, with the same underlying proof structure being able to comfortably scale from very simplistic (or even unitless) function definitions to substantially richer datatypes and modules.

\vspace{-2.5mm}
\subsubsection{Failure Modes of the Verifier}

Despite the generally positive results, a few benchmarks highlight current limitations of the verification pipeline and of \texttt{CHARON}/\texttt{AENEAS}.

The most obvious failure is case~\#20 in Table~\ref{tab:verification_results}, a ``real-world'' example that computes the total backoff delay. The original (Rust)function maintains a counter \icodeverb{attempts}, a mutable \icodeverb{delay}, and a running \icodeverb{total} inside a \icodeverb{while attempts < cfg.max\_attempts} loop. However, after the translation, \texttt{AENEAS} produced a recursive Coq \icodeverb{Fixpoint} where termination is expressed in terms of an \icodeverb{attempts} parameter. \texttt{coqc} then erorred with \emph{``Cannot guess decreasing argument of fix''}. Intuitively, we know that the Rust code terminates because the loop counter is incremented and compared against a fixed bound, but the way that AENEAS generated a recursive definition of this function resulted in this decrease not being syntactically obvious to Coq's termination checker. As a result, the entire Coq development for this benchmark fails to compile.

More broadly, several potential benchmarks had to be excluded earlier in the pipeline because they rely on language features that \texttt{AENEAS} does not yet support. These include certain core intrinsics, derived \icodeverb{Eq}/\icodeverb{PartialEq} implementations, and macro expansions such as \icodeverb{matches!}. When such constructs appear, the translation step fails before Coq code is generated, and no proof attempt is possible. In our current experiment design, we designed the real-world benchmarks to avoid these constructs, but this filtering necessarily hides some of the friction that developers would encounter in unrestricted code.

In the short term, the most realistic mitigation is to give developers better guidance about when they can expect verification to succeed. As a follow-on to this work, we plan to extend REM2.0 with a lightweight ``AENEAS-able'' checker that can quickly test whether a given function lies inside the current verified fragment and report obvious blockers (unsupported intrinsics, traits, or control-flow patterns). This checker is not available at the time of writing, but is planned for shortly after the completion of this project.

\vspace{-5mm}
\subsection{Discussion and Threats to Validity}
\label{sec:discussion_threats}
\vspace{-3mm}
\subsubsection{Discussion}
The extraction experiments show that embedding REM2.0 into RA's infrastructure meets the main design goals from Section~\ref{sec:expanding_rem}. On the original REM benchmarks, it is strictly better than the prototype---it covers all 39 available cases, and does so with a latency that would be undetectable to developers. On the new set of cases, it handles most examples involving \icodeverb{async}/\icodeverb{await}, \icodeverb{const fn}, NLCF, generics, and HRTBs, but also exposes a set of failures around DTOs and some of the more complex generic bounds. Taken together, these results support RQ1 and RQ2: REM2.0 is at least as capable as the original tool and substantially more expressive in the Rust features it can refactor.

For performance (RQ3), the contrast is highly impressive. The original REM pipeline had extraction times in the hundreds to thousands of milliseconds, likely not including the work done by IntelliJ. By reusing RA's analysis in a persistent daemon, REM2.0 reduces the raw extraction costs to a few milliseconds and user-experienced latency to well under a quarter of a second.

The verification results offer a new perspective (RQ4). For code within the subset supported by \texttt{CHARON} and \texttt{AENEAS}, the verifier is capable of producing end-to-end equivalence proofs between original and extracted functions. This is orders of magnitude slower than extraction and restricted to a subset of Rust, but provides a much stronger guarantee. In practice, this suggests that there is going to be a two-tier workflow: developers use REM2.0's fast extraction by default and selectively invoke the repairer or the verification pipeline for safety-critical code, tricky lifetime patterns, or APIs with strong behavioural contracts that need to be upheld.

Finally, the usage profile of REM-Repairer confirms its role as a targeted component of the toolchain. It is still essential for cases with tight lifetime constraints, but most extractions can succeed without it. However, as shown above, the large latency introduced with VSCode's overhead more than outweighs the call to the Repairer, and thus we have made the Extract + Repair pathway the default option inside our VSCode extension.

\vspace{-3.5mm}
\subsubsection{Threats to Validity}
As with any empirical study, our evaluation is subject to several threats to validity:

\textbf{Internal validity.}
Timing measurements for RQ3 and RQ4 are affected by system noise, caching, and warm-up behaviour in RA and the OS. Although we did our best to use a controlled environment and consistent configurations, individual runs may vary slightly, so the numbers should be read as relative trends rather than precise bounds.

\textbf{External validity.}
Our benchmarks, while drawn from real projects, are limited by necessity. The original REM suite captured a specific snapshot of Rust code and the new corpus focuses on large, popular GitHub repositories with manually curated extraction sites. This biases the evaluation toward more "mainstream'' code and as such cannot cover domains such as embedded \icodeverb{no\_std} crates or highly specialised libraries.

\textbf{Construct validity.}
We treat compilation success and, where applicable, successful \texttt{CHARON}/\texttt{AENEAS} proofs as proxies for refactoring correctness (RQ1, RQ2, RQ4). In practice, we can only have a superficial understanding of each example and largely rely on the Rust compiler and (where possible) the verification pipeline to detect problems. While the type system and borrow checker catch many errors, subtle behavioural changes (e.g.\ performance, logging, concurrency effects) may still go unnoticed.

\textbf{Tool ecosystem stability.}
REM2.0 depends on internal APIs of RA and on external verification tools (\texttt{CHARON}, \texttt{AENEAS}) that are under active development\footnote{specifically the internal APIs of RA are designed entirely for use within RA, and do break often. Our toolchain is built on version \texttt{0.0.262}, but with weekly releases the API is already up to version \texttt{0.0.3XX}. Whilst there has been no update to the major components we rely on, a brief attempt to migrate was quickly disbanded}. Future changes may alter behaviour, performance, or supported features, so our results capture a particular point in the evolution of this tool stack.

\subsection{Summary}
\label{sec:eval_summary}
\vspace{-2.5mm}
This section has evaluated REM2.0 against the four research questions posed at the start of this section. The extraction experiments show that REM2.0 matches and extends the capabilities of the original REM prototype (RQ1, RQ2), while the new architecture delivers interactive, IDE-like latency by reusing RA's incremental analysis in a persistent daemon (RQ3). For code within the currently supported \texttt{CHARON}/\texttt{AENEAS} fragment, the verification pipeline can further certify behavioural equivalence between original and extracted functions, at substantially higher performance cost but with correspondingly stronger guarantees (RQ4).

Overall, the results indicate that REM2.0 provides a practical extract-function refactoring for modern Rust, with a fast default path and targeted support for both automated repair and optional formal verification when additional assurance is required.

%% file: conclusions.tex
\section{Conclusions and Future Work}
\label{sec:conclusions_future_work}

\subsection{Revisiting Aims and Research Questions}
\label{sec:revisiting_aims}
\vspace{-2.5mm}

The aim of this project was to design, implement, and evaluate a practical extract-function refactoring toolchain for Rust that can handle modern language features and, where possible, provide stronger correctness guarantees than ``it still compiles''. Concretely, the work set out to build a new version of REM that integrates with \texttt{rust-analyzer}, supports contemporary Rust language features such as \icodeverb{async}/\icodeverb{await} and \icodeverb{const fn}, and connects to a verification pipeline for equivalence checking.

Section~\ref{sec:evaluation} framed this aim in terms of four research questions:
\begin{itemize}[nosep, leftmargin=*]
  \item \textbf{RQ1} -- Does the new toolchain match or exceed the original REM on the existing benchmark suite?
  \item \textbf{RQ2} -- To what extent can the new extraction pipeline handle modern Rust features (\icodeverb{async}, \icodeverb{const fn}, generics, higher-ranked trait bounds, dynamic trait objects, and non-local control flow)?
  \item \textbf{RQ3} -- What extraction latencies are achievable when we reuse \texttt{rust-analyzer} as a persistent analysis daemon, and how does this impact the user experience?
  \item \textbf{RQ4} -- For code within the supported subset, can we feasibly verify equivalence between original and extracted functions using \texttt{CHARON} and \texttt{AENEAS}?
\end{itemize}

The evaluation results give us clear answers. For \textbf{RQ1}, REM2.0 achieves strict parity with the original REM on the original benchmark suite and additionally succeeds on several cases that previously failed. For \textbf{RQ2}, it handles the majority of feature-focused examples involving \icodeverb{async}, \icodeverb{const}, NLCF, generics, and HRTBs, with remaining weaknesses concentrated around DTOs and intricate generic bounds. For \textbf{RQ3}, the new architecture reduces extraction latency from hundreds or thousands of milliseconds to a few milliseconds internally and to well under a quarter of a second as observed in the editor. For \textbf{RQ4}, the verification pipeline can construct end-to-end equivalence proofs for a non-trivial fragment of Rust, although at substantially higher performance cost and with a restricted language subset.

\subsection{Summary of Contributions}
\label{sec:summary_of_contributions}

Against this backdrop, the paper makes several technical and empirical contributions.

\vspace*{-4mm}
\subsubsection{A RA based architecture for extraction.}
The first contribution is the design and implementation of REM2.0 as a long-lived daemon built on top of RA. Instead of repeatedly invoking the Rust compiler on whole crates, REM2.0 reuses RA's analysis and exposes a JSON-RPC interface that can be driven from a VSCode extension or command-line client. This extension is currently available on the VSCode marketplace.

\vspace*{-4mm}
\subsubsection{A feature-rich extraction engine.}
Second, the project develops an extraction engine that supports a much broader range of Rust features than the original REM prototype. It operates correctly in the presence of \icodeverb{async}/\icodeverb{await}, \icodeverb{const fn}, HRTBs, generics with multiple trait constraints, and NLCF, using \icodeverb{std::ops::ControlFlow} to standardise \icodeverb{return}s, \icodeverb{break}, and \icodeverb{continue}. It still struggles on DTO-heavy cases, which remain the most complex feature class and a clear area for continued research.

\vspace*{-4mm}
\subsubsection{A repairer for lifetimes and type signatures.}
Third, the work refines and reuses REM-Repairer as a post-processing step that automatically adjusts extracted function signatures and lifetime annotations when the original extraction creates borrow-checker errors. Empirically, the repairer is needed in the same kinds of cases as in the original evaluation---those with tight lifetime constraints---and thus remains a crucial component of the overall toolchain rather than a mere convenience.

\vspace*{-4mm}
\subsubsection{Integration with verification tooling.}
Fourth, we prototyped a verification pipeline that connects REM2.0 to \texttt{AENEAS}. For functions within the supported fragment, this pipeline is able to translate original and extracted code to Coq definitions and construct proofs that the two versions are behaviourally equivalent. This provides a path from ordinary refactoring to machine-checked correctness guarantees for selected cases, building on the broader vision of correct-by-construction programming~\cite{RungeBordisPotaninThumSchaefer2023}.

\vspace*{-4mm}
\subsubsection{Empirical evaluation on real Rust projects.}
Finally, the project contributes a large empirical study across three benchmark corpora. These experiments demonstrated that the architecture scales to realistic workloads, quantify the latency improvements, and characterise the strengths, direct improvements, and failure modes of REM2.0.

\subsection{Limitations and Open Challenges}
\label{sec:limitations_open_challenges}
\vspace{-2.5mm}

Despite these contributions, REM2.0 has clear limitations that draw a line in the sand as to where it can be relied upon today and where further work is needed.

On the language-feature side, DTOs and complex generic bounds are the biggest current weaknesses. Several failure cases arose when the extractor could not reconstruct a precise function signature, falling back to unspecified types or dropping bounds. This is closely related to the second limitation: REM2.0 mostly operates with a modified single-file view based on RA's crate graphs, which makes it difficult to discover constraints defined in modules we do not explicitly trace and import.

The verification pipeline is similarly constrained by the subset that is supported by \texttt{AENEAS}. Many real-world functions make use of features that lie outside this subset, and even when verification is possible, it incurs significantly higher cost than extraction. As a result, the verifier cannot yet be applied indiscriminately to arbitrary refactorings and is best seen as an optional, high-assurance mode.

Finally, the evaluation focuses on correctness and performance, not on user experience. We have not performed a study to assess how developers actually adopt the tool, how often they invoke verification, or how much they trust the results. Understanding these human factors is an important but out-of-scope challenge.


\subsection{Future Directions}
\label{sec:future_directions}
\vspace{-2.5mm}

Several areas of future work follow directly from the limitations and from our broader vision of provably correct refactorings.

A first direction is \emph{deeper integration with RA}. By drawing on more global crate-graph information, REM2.0 should be able to extract richer trait and where-clause information, closing many of the gaps we have seen with DTOs and complex generic bounds. This may involve extending internal APIs or contributing changes upstream. Although unlikely due to its reliance on cargo check, integrating the repairer into RA remains a long term engineering goal of this project.

A second direction is to \emph{extend the equivalence fragment}. As \texttt{CHARON} and \texttt{AENEAS} improve, there is scope to support a larger subset of Rust---with interior mutability being a major feature to look forward to. At the same time, a deeper understanding of the LLBC (and potential integration with the polonius borrow checker) could be used to substantially increase how much of the original rust we can directly reason about without needing to perform the functional translation.

Third, there is room for \emph{smarter repair and analysis heuristics}. REM-Repairer could be upgraded to infer missing bounds or function annotations from the compiler feedback it already collects, or by learning patterns from previously successful repairs. Static checks that predict whether a given extraction is likely to be verifiable could help developers decide when it is worth paying the verification cost.

Finally, \emph{complex ownership cases} remain an important open challenge. Developers routinely work around borrow-checker rejections by introducing selective \icodeverb{clone}s, passing individual fields instead of whole structs, or using interior mutability. These borrow checking issues are something that the Polonius project aims to rectify.  At present, REM2.0 treats such situations as hard failures as we rely on the Rust compiler which is unable assist us with. A natural direction for future work is to recognise and apply a small catalogue of semantics-preserving ``borrow repair'' patterns, so that more extractions can be automatically reshaped into forms that satisfy the borrow checker without changing the intended behaviour. A potential advancement from there will be utilising the Polonius borrow checker or reaching into the LLBC semantics to further improve the original extraction.

\subsection{Broader Outlook}
\label{sec:broader_outlook}
\vspace{-2.5mm}

Stepping back, this paper sits at an intersection of refactoring tools, modern language servers, and formal verification. Rust is a demanding testbed for safe automated refactoring: its ownership and lifetime system make entire classes of bugs unpresentable, but as we and many others have discovered, those guarantees make na\"ive extractions impossible and well intentioned ones incredibly difficult.

REM2.0 shows that a RA-based design can deliver extract-function refactorings that are both fast enough for everyday use and expressive enough to handle a wide range of contemporary Rust language features. The addition of a repair stage and a verification pipeline demonstrates that we can go beyond compilation as our only correctness source, and selectively obtain machine-checked evidence that a refactoring preserves behaviour.

If such tools become commonplace, they could change how developers think about the large-scale evolution of Rust code: many refactorings would remain cheap, interactive, and IDE-driven, while critical feature changes could be escalated to formally justified transformations with minimal disruption to the developer's workflow. More broadly, the work suggests a path toward refactoring tools that treat correctness as their main concern rather than an afterthought. In this sense, REM2.0 is not just a faster extract-function command for Rust, but an early step toward a generation of refactoring tools that integrate verification into everyday software engineering practice.

%% file: appendix_main.tex

\makeatletter
\@ifundefined{r@chap:expanding_rem}{\label{chap:expanding_rem}}{}
\makeatother

\input{appendix1} 
\newpage

\input{appendix2} 
\newpage

\input{appendix3} 
\newpage

\input{appendix4} 
\newpage

\input{appendix5}
\newpage

\input{appendix6}
\newpage

\input{appendix7} 
\newpage

\input{appendix8} 
\newpage

\input{appendix9} 

%% file: appendix1.tex
\section{Appendix 1: Passing by Value and Reference}
\label{sec:appendix1}

This appendix provides three short, self-contained examples of ``na\"ive''
function hoisting in Rust. In each example, we will define a function that is
getting refactored, and we will demonstrate the case where the arguments to the
function are passed by value and the case where they are passed by reference. To
achieve this, we are imagining an automated refactoring tool that initially
takes an inline block of code and moves it into a new function. The compilation
errors (if any) guide a subsequent ownership analysis, which deduces how each
variable should be passed: by value (\icodeverb{T}), by shared reference
(\icodeverb{\&T}), or by mutable reference (\icodeverb{\&mut T}).

\subsection{Example 1: Passing by Value}
\subsubsection*{Original Code (inline)}
\begin{lstlisting}[language=Rust]
fn main() {
    let x = 42;
    let y = x * 2;
    println!("y is {}", y);
}
\end{lstlisting}
{\captionsetup{justification=centering}\captionof{listing}{Passing by value: Original}}

\subsubsection*{Na\"ive Hoisted Code (before fixes)}
Suppose an automated refactoring tool decides to “hoist” the multiplication
logic into a new function: \newline
\begin{lstlisting}[language=Rust]
fn hoisted_block(x: i32) {
    let y = x * 2;
    println!("y is {}", y);
}

fn main() {
    let x = 42;
    hoisted_block(x);
}
\end{lstlisting}
{\captionsetup{justification=centering}\captionof{listing}{Passing by value: Na\"ive extraction}}
In this example, no errors occur because after calling
\icodeverb{hoisted\_block(x)}, we do not need \icodeverb{x} again in \icodeverb{main}, so
moving (passing ownership) is safe. The automated repair analysis detects that
\icodeverb{x} is no longer needed in \icodeverb{main} and decides it can be passed by
value.

\subsection{Example 2: Passing by Shared Reference}
\subsubsection*{Original Code (inline)}
\begin{lstlisting}[language=Rust]
fn main() {
    let text = String::from("Hello world!");
    let length = text.len();

    // Some inline block that only reads from `text`:
    println!("First word is: {}", get_first_word(&text));
    println!("Total length is: {}", length);
}

fn get_first_word(s: &str) -> &str {
    s.split_whitespace().next().unwrap_or("")
}
\end{lstlisting}
{\captionsetup{justification=centering}\captionof{listing}{Passing by shared reference: Original}}

\subsubsection*{Na\"ive Hoisted Code (before fixes)}
Let's say the refactoring tool decides to hoist the usage of \icodeverb{get\_first\_word} (or
a bit more logic around it) into a new helper function:
\begin{lstlisting}[language=Rust]
fn main() {
    let text = String::from("Hello world!");
    let length = text.len();

    // Some inline block that only reads from `text`:
    println!("First word is: {}", get_first_word(&text));
    println!("Total length is: {}", length);
}

fn get_first_word(s: &str) -> &str {
    s.split_whitespace().next().unwrap_or("")
}
\end{lstlisting}
{\captionsetup{justification=centering}\captionof{listing}{Passing by shared reference: Na\"ive extraction}}
This version won't compile as expected if the function
\icodeverb{print\_first\_word\_block} takes \icodeverb{text} by value. Once \icodeverb{text}
is moved into \icodeverb{print\_first\_word\_block}, we can no longer use
\icodeverb{text} afterward in \icodeverb{main}. The compiler will complain that
\icodeverb{text} is moved, thus invalidating the line \texttt{println!("Total
length is: {}", length);} if we needed \icodeverb{text} for anything else.

\subsubsection*{Automated Repair Yields}
\begin{lstlisting}[language=Rust]
fn print_first_word_block(text: &String) {
    // Only reads from `text`, so it only needs a shared reference.
    println!("First word is: {}", get_first_word(text));
}

fn main() {
    let text = String::from("Hello world!");
    let length = text.len();

    print_first_word_block(&text); // Pass a shared reference
    println!("Total length is: {}", length);
}
\end{lstlisting}
{\captionsetup{justification=centering}\captionof{listing}{Passing by shared reference: Automated Repair}}
\noindent What's going on?
\begin{itemize}
    \item \textbf{Read-only usage}: The function \icodeverb{print\_first\_word\_block}
    just needs to read the string (in order to print the first word).
    \item \textbf{Shared reference}: Because the string must remain valid
    afterward in \icodeverb{main}, the ownership analysis decides that passing by
    shared reference (\icodeverb{\&String}) is correct.
\end{itemize}

\subsection{Example 3: Passing by Mutable Reference}
\subsubsection*{Original Code (inline)}
\begin{lstlisting}[language=Rust]
fn main() {
    let mut values = vec![1, 2, 3];

    // Some inline block that mutates `values`:
    values.push(4);
    println!("Values are now: {:?}", values);
}
\end{lstlisting}
{\captionsetup{justification=centering}\captionof{listing}{Passing by mutable reference: Original}}

\subsubsection*{Na\"ive Hoisted Code (before fixes)}
Hoisting the push operation into a new function could look like this:
\begin{lstlisting}[language=Rust]
fn push_block(values: Vec<i32>) {
    let mut v = values;
    v.push(4);
    println!("Values are now: {:?}", v);
}

fn main() {
    let mut values = vec![1, 2, 3];
    push_block(values);
    println!("Final values: {:?}", values); // error: `values` was moved
}
\end{lstlisting}
{\captionsetup{justification=centering}\captionof{listing}{Passing by mutable reference: Na\"ive extraction}}
Here, we have a compilation error similar to Example 2: once \icodeverb{values} is
passed by value, ownership is transferred, and we cannot use \icodeverb{values}
again in \icodeverb{main}. However, in this case, we truly do need to mutate
\icodeverb{values}, and we want \icodeverb{main} to see the updated vector.

\subsubsection*{Automated Repair Yields}
\begin{lstlisting}[language=Rust]
fn push_block(values: &mut Vec<i32>) {
    values.push(4);
    println!("Values are now: {:?}", values);
}

fn main() {
    let mut values = vec![1, 2, 3];
    push_block(&mut values);
    println!("Final values: {:?}", values); // Now values is still in scope and updated
}
\end{lstlisting}
{\captionsetup{justification=centering}\captionof{listing}{Passing by mutable reference: Automated Repair}}
\noindent What's going on?
\begin{itemize}
    \item \textbf{Mutation required}: The code needs to modify the original
    vector.
    \item \textbf{Mutable reference}: By passing \icodeverb{\&mut Vec<i32>}, the
    function can mutate \icodeverb{values} in place, and \icodeverb{main} retains
    ownership of \icodeverb{values} to use afterward.
\end{itemize}

%% file: appendix2.tex
\section{Appendix 2: Details of each Repository}
\label{sec:appendix2_details_of_each_repository}

\begin{table}[h]
\centering
\small

\begin{tabular}{|c|c|c|c|c|}
\hline
\textbf{Ranking} & \textbf{Project Repo}        & \textbf{Stars} & \textbf{Forks} & \textbf{Open Issues} \\ \hline
2  & denoland/deno                & 105077 & 5768  & 2384 \\
3  & rustdesk/rustdesk            & 102158 & 14993 & 61   \\
4  & tauri-apps/tauri             & 98717  & 3155  & 1161 \\
5  & unionlabs/union              & 74500  & 3835  & 155  \\
6  & astral-sh/uv                 & 72353  & 2212  & 2136 \\
7  & zed-industries/zed           & 69398  & 5885  & 3054 \\
8  & FuelLabs/sway                & 62106  & 5428  & 862  \\
9  & alacritty/alacritty          & 60977  & 3227  & 323  \\
10 & rust-lang/rustlings          & 60569  & 10978 & 46   \\
11 & FuelLabs/fuel-core           & 57463  & 2856  & 149  \\
12 & BurntSushi/ripgrep           & 57159  & 2304  & 85   \\
13 & topjohnwu/Magisk             & 56856  & 15853 & 36   \\
14 & sharkdp/bat                  & 55755  & 1409  & 318  \\
15 & meilisearch/meilisearch      & 54438  & 2254  & 221  \\
16 & lencx/ChatGPT                & 54214  & 6192  & 856  \\
17 & rust-unofficial/awesome-rust & 53670  & 3064  & 10   \\
18 & starship/starship            & 52113  & 2284  & 762  \\
19 & dani-garcia/vaultwarden      & 50694  & 2374  & 17   \\
20 & openai/codex                 & 50131  & 6215  & 1164 \\
21 & types/typst                  & 47876  & 1305  & 1007 \\ \hline
\end{tabular}

\captionsetup{justification=centering}
\caption{Brief summary of the 20 repositories searched to find real-world examples. \#1 is the Rust project itself.}
\label{tab:repo_summary}
\end{table}

\newpage

\begin{table}[h!]
\centering
\begin{tabular}{|c|p{3cm}|p{11cm}|}
\hline
\# & Name & URL \\ \hline

1  & denoland/\newline deno 
   & \url{https://github.com/denoland/deno/commit/104514b06dfbbb899235fb1147aa1fa62477a6d5} \\

2  & FuelLabs/\newline fuel-core
   & \url{https://github.com/FuelLabs/fuel-core/commit/ed92738280e068475682e220b93921bbbfebfdd7} \\

3  & meilisearch/\newline meilisearch
   & \url{https://github.com/meilisearch/meilisearch/commit/31142b36635ba062184b7d42f2e86021b11e911c} \\

4  & BurntSushi/\newline ripgrep
   & \url{https://github.com/BurntSushi/ripgrep/commit/7534d5144f5c5e89d67c791e663cdc0bee069b0e} \\

5  & starship/\newline starship
   & \url{https://github.com/starship/starship/commit/9093891acbe2c86b1615c37386dadbb0cc632199} \\

6  & FuelLabs/\newline sway
   & \url{https://github.com/FuelLabs/sway/commit/cdfc56c700ab32eefb84f75ed3048bd2a0ea22d2} \\

7  & tauri-apps/\newline tauri
   & \url{https://github.com/tauri-apps/tauri/commit/658e5f5d1dc1bd970ae572a42447448d064a7fee} \\

8  & unionlabs/\newline union
   & \url{https://github.com/unionlabs/union/commit/9b01ce6d75b84f049c1ede75750d33ce26ae73d9} \\

9  & astral-sh/\newline uv
   & \url{https://github.com/astral-sh/uv/commit/9af989e30c86391767830aa588c32fcb7c9a76b5} \\

10 & dani-garcia/\newline vaultwarden
   & \url{https://github.com/dani-garcia/vaultwarden/commit/952992c85bdf3147ddb8e663538e29c748b71f94} \\

11 & zed-industries/zed
   & \url{https://github.com/zed-industries/zed/commit/d00ad02b05981c79942b3bdff957e2384e68b89f} \\

\hline
\end{tabular}

\captionsetup{justification=centering}
\caption{Details of each repository in the evaluation, with the specific commit hash that we found examples from / evaluated against}
\label{tab:appendix2_details_of_each_repository}
\end{table}

%% file: appendix3.tex
\section{Appendix 3: GitOxide Example in Detail}
\label{sec:appendix_3}

The following diff shows how the `if add\_path` block was refactored into the
new helper method \texttt{extract\_include\_path}.

\begin{lstlisting}[breaklines=true,basicstyle=\small\ttfamily,language=diff]
-        if add_path {
-            if let Some(body) = target_config.sections.get(&id) {
-                let paths = body.values(&Key::from("path"));
-                let paths = paths.iter().map(|path| values::Path::from(path.clone()));
-                include_paths.extend(paths);
-            }
-        }
         // Some specifics ommited for simplicity
+        if add_path {
+            extract_include_path(target_config, &mut include_paths, id)
+        }
\end{lstlisting}

And here is the new method in full:

\begin{lstlisting}[numbers=left,breaklines=true,basicstyle=\small\ttfamily,language=Rust]
fn extract_include_path<'a>(
    target_config: &mut File<'a>,
    include_paths: &mut Vec<values::Path<'a>>,
    id: SectionId
) {
    if let Some(body) = target_config.sections.get(&id) {
        let paths = body.values(&Key::from("path"));
        let paths = paths.iter().map(|path| values::Path::from(path.clone()));
        include_paths.extend(paths);
    }
}
\end{lstlisting}
{\captionsetup{justification=centering}\captionof{listing}{Full gitoxide method}}

%% file: appendix4.tex
\section{Appendix 4: Complete MIR from \ref{subsec:charon}}

\begin{lstlisting}[basicstyle=\footnotesize\ttfamily,breaklines=true,language=bash]
// WARNING: This output format is intended for human consumers only
// and is subject to change without notice. Knock yourself out.
fn main() -> () {
    let mut _0: ();
    let _1: std::vec::Vec<i32>;
    let mut _2: std::boxed::Box<[i32]>;
    let mut _3: usize;
    let mut _4: usize;
    let mut _5: *mut u8;
    let mut _6: std::boxed::Box<[i32; 3]>;
    let _8: &i32;
    let mut _9: &std::vec::Vec<i32>;
    let _10: ();
    let mut _11: std::fmt::Arguments<'_>;
    let _12: &[&str; 2];
    let _13: &[core::fmt::rt::Argument<'_>; 1];
    let _14: [core::fmt::rt::Argument<'_>; 1];
    let mut _15: core::fmt::rt::Argument<'_>;
    let _16: &&i32;
    let mut _17: *const [i32; 3];
    let mut _18: *const ();
    let mut _19: usize;
    let mut _20: usize;
    let mut _21: usize;
    let mut _22: usize;
    let mut _23: bool;
    scope 1 {
        debug v => _1;
        let _7: &i32;
        scope 2 {
            debug x => _7;
        }
    }

    bb0: {
        _3 = SizeOf([i32; 3]);
        _4 = AlignOf([i32; 3]);
        _5 = alloc::alloc::exchange_malloc(move _3, move _4) -> [return: bb1, unwind continue];
    }

    bb1: {
        _6 = ShallowInitBox(move _5, [i32; 3]);
        _17 = copy (((_6.0: std::ptr::Unique<[i32; 3]>).0: std::ptr::NonNull<[i32; 3]>).0: *const [i32; 3]);
        _18 = copy _17 as *const () (PtrToPtr);
        _19 = copy _18 as usize (Transmute);
        _20 = AlignOf([i32; 3]);
        _21 = Sub(copy _20, const 1_usize);
        _22 = BitAnd(copy _19, copy _21);
        _23 = Eq(copy _22, const 0_usize);
        assert(copy _23, "misaligned pointer dereference: address must be a multiple of {} but is {}", copy _20, copy _19) -> [success: bb10, unwind unreachable];
    }

    bb2: {
        _9 = &_1;
        _8 = <Vec<i32> as Index<usize>>::index(move _9, const 0_usize) -> [return: bb3, unwind: bb8];
    }

    bb3: {
        _7 = copy _8;
        _12 = const main::promoted[0];
        _16 = &_7;
        _15 = core::fmt::rt::Argument::<'_>::new_display::<&i32>(copy _16) -> [return: bb4, unwind: bb8];
    }

    bb4: {
        _14 = [move _15];
        _13 = &_14;
        _11 = Arguments::<'_>::new_v1::<2, 1>(copy _12, copy _13) -> [return: bb5, unwind: bb8];
    }

    bb5: {
        _10 = _print(move _11) -> [return: bb6, unwind: bb8];
    }

    bb6: {
        drop(_1) -> [return: bb7, unwind continue];
    }

    bb7: {
        return;
    }

    bb8 (cleanup): {
        drop(_1) -> [return: bb9, unwind terminate(cleanup)];
    }

    bb9 (cleanup): {
        resume;
    }

    bb10: {
        (*_17) = [const 1_i32, const 2_i32, const 3_i32];
        _2 = move _6 as std::boxed::Box<[i32]> (PointerCoercion(Unsize, Implicit));
        _1 = slice::<impl [i32]>::into_vec::<std::alloc::Global>(move _2) -> [return: bb2, unwind continue];
    }
}

const main::promoted[0]: &[&str; 2] = {
    let mut _0: &[&str; 2];
    let mut _1: [&str; 2];

    bb0: {
        _1 = [const "", const "\n"];
        _0 = &_1;
        return;
    }
}
\end{lstlisting}
{\captionsetup{justification=centering}\captionof{listing}{Complete MIR from \ref{subsec:charon}}

%% file: appendix5.tex
\section{Appendix 5: Plotted Extraction time data}
\label{app:appendix5_extraction_histogram}

\begin{figure}[h!]
    \centering
    \includegraphics[width=\linewidth]{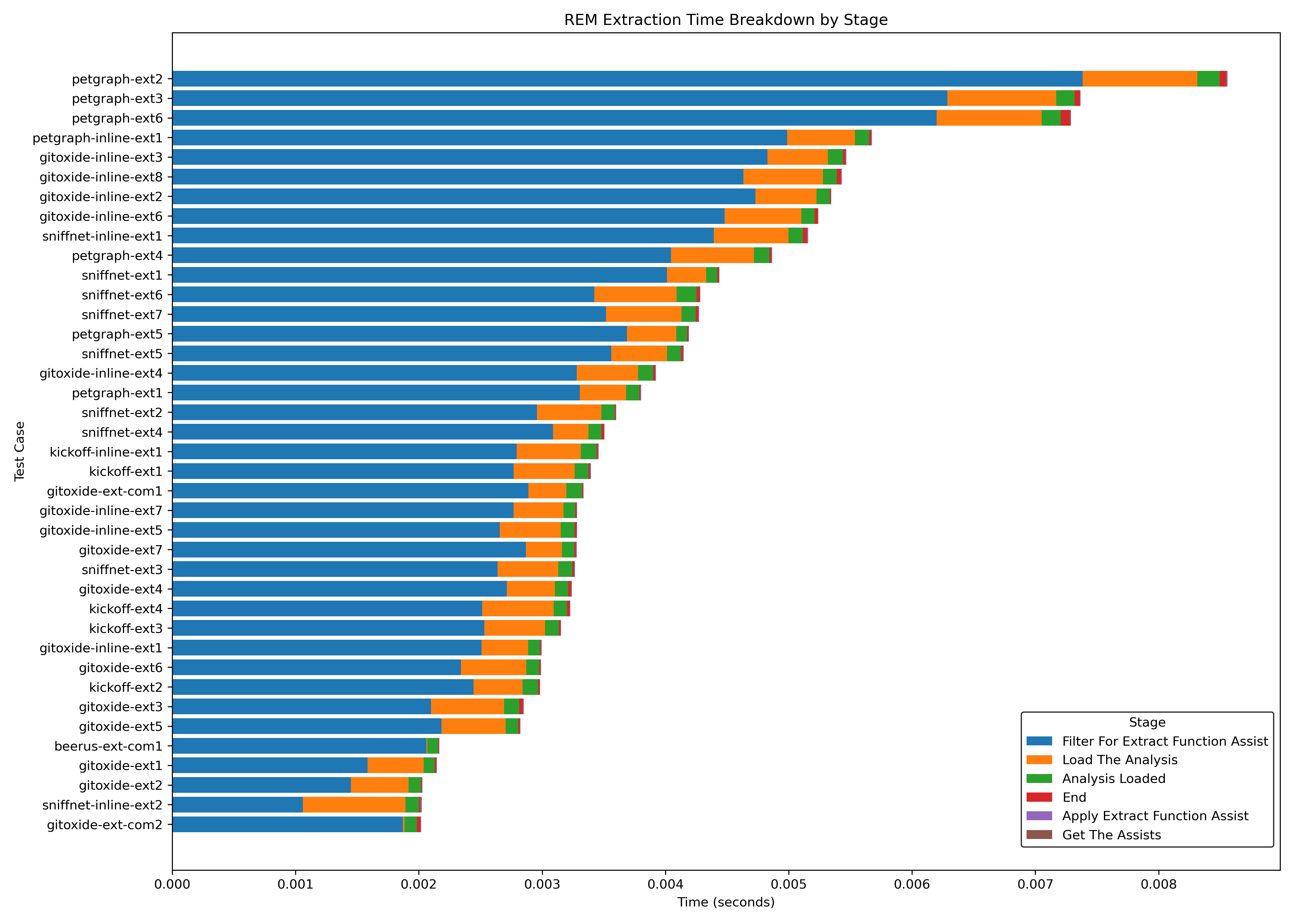}
    \caption{Execution time breakdown by pipeline stage for each test case.
    Times are reported in seconds (derived from nanosecond timing data).}
    \label{fig:rem_extraction_breakdown}
\end{figure}

\newpage

\begin{table}[h!]
\centering
\small
\begin{adjustwidth}{-1cm}{}
\begin{tabular}{|l|l|l|l|l|l|l|l|}
\hline
\multicolumn{1}{|c|}{Case} & \multicolumn{1}{c|}{\begin{tabular}[c]{@{}c@{}}Load the \\ Analysis\end{tabular}} & \multicolumn{1}{c|}{\begin{tabular}[c]{@{}c@{}}Analysis \\ Loaded\end{tabular}} & \multicolumn{1}{c|}{\begin{tabular}[c]{@{}c@{}}Get the \\ Assists\end{tabular}} & \multicolumn{1}{c|}{\begin{tabular}[c]{@{}c@{}}Filter For \\ EF Assist\end{tabular}} & \multicolumn{1}{c|}{\begin{tabular}[c]{@{}c@{}}Apply EF \\ Assist\end{tabular}} & \multicolumn{1}{c|}{Remaining} & \multicolumn{1}{c|}{Total} \\ \hline
beerus-ext-com1            & 9285                                                                              & 86226                                                                           & 2120                                                                            & 2060529                                                                                            & 2476                                                                                             & 5371                           & 2166007                    \\ \hline
gitoxide-ext-com1          & 307613                                                                            & 128739                                                                          & 2785                                                                            & 2887546                                                                                            & 2574                                                                                             & 7746                           & 3337003                    \\ \hline
gitoxide-ext-com2          & 10238                                                                             & 100897                                                                          & 2468                                                                            & 1872872                                                                                            & 2692                                                                                             & 28872                          & 2018039                    \\ \hline
gitoxide-ext1              & 454708                                                                            & 90926                                                                           & 2210                                                                            & 1583716                                                                                            & 3171                                                                                             & 13176                          & 2147907                    \\ \hline
gitoxide-ext2              & 468576                                                                            & 99266                                                                           & 2948                                                                            & 1449459                                                                                            & 3634                                                                                             & 6355                           & 2030238                    \\ \hline
gitoxide-ext3              & 593338                                                                            & 121557                                                                          & 2593                                                                            & 2098630                                                                                            & 4041                                                                                             & 30428                          & 2850587                    \\ \hline
gitoxide-ext4              & 387921                                                                            & 109024                                                                          & 2798                                                                            & 2715010                                                                                            & 3668                                                                                             & 21512                          & 3239933                    \\ \hline
gitoxide-ext5              & 520525                                                                            & 101651                                                                          & 2508                                                                            & 2183887                                                                                            & 2948                                                                                             & 11685                          & 2823204                    \\ \hline
gitoxide-ext6              & 529156                                                                            & 104946                                                                          & 2568                                                                            & 2342050                                                                                            & 2513                                                                                             & 10251                          & 2991484                    \\ \hline
gitoxide-ext7              & 290983                                                                            & 101383                                                                          & 2307                                                                            & 2869735                                                                                            & 5587                                                                                             & 11670                          & 3281665                    \\ \hline
gitoxide-inline-ext1       & 376787                                                                            & 96027                                                                           & 2954                                                                            & 2508828                                                                                            & 3532                                                                                             & 6906                           & 2995034                    \\ \hline
gitoxide-inline-ext2       & 496737                                                                            & 105101                                                                          & 2427                                                                            & 4729361                                                                                            & 3486                                                                                             & 8189                           & 5345301                    \\ \hline
gitoxide-inline-ext3       & 488481                                                                            & 122647                                                                          & 3248                                                                            & 4827718                                                                                            & 4000                                                                                             & 21855                          & 5467949                    \\ \hline
gitoxide-inline-ext4       & 495756                                                                            & 121965                                                                          & 3684                                                                            & 3281795                                                                                            & 5483                                                                                             & 15325                          & 3924008                    \\ \hline
gitoxide-inline-ext5       & 495051                                                                            & 112329                                                                          & 2474                                                                            & 2657133                                                                                            & 2991                                                                                             & 12808                          & 3282786                    \\ \hline
gitoxide-inline-ext6       & 622550                                                                            & 107609                                                                          & 2873                                                                            & 4480736                                                                                            & 3848                                                                                             & 22912                          & 5240528                    \\ \hline
gitoxide-inline-ext7       & 403207                                                                            & 96122                                                                           & 2356                                                                            & 2769977                                                                                            & 2617                                                                                             & 8734                           & 3283013                    \\ \hline
kickoff-ext1               & 494546                                                                            & 112377                                                                          & 3619                                                                            & 2768148                                                                                            & 2671                                                                                             & 14533                          & 3395894                    \\ \hline
kickoff-ext2               & 398752                                                                            & 124576                                                                          & 2850                                                                            & 2443569                                                                                            & 2733                                                                                             & 11091                          & 2983571                    \\ \hline
kickoff-ext3               & 492629                                                                            & 110374                                                                          & 2586                                                                            & 2531962                                                                                            & 2069                                                                                             & 13516                          & 3153136                    \\ \hline
kickoff-ext4               & 578567                                                                            & 108555                                                                          & 2736                                                                            & 2514134                                                                                            & 3164                                                                                             & 22551                          & 3229707                    \\ \hline
kickoff-inline-ext1        & 520277                                                                            & 126225                                                                          & 2583                                                                            & 2793580                                                                                            & 2920                                                                                             & 11691                          & 3457276                    \\ \hline
petgraph-ext1              & 374755                                                                            & 105993                                                                          & 2342                                                                            & 3306952                                                                                            & 3347                                                                                             & 9064                           & 3802453                    \\ \hline
petgraph-ext2              & 929776                                                                            & 181024                                                                          & 3519                                                                            & 7384406                                                                                            & 4936                                                                                             & 53650                          & 8557311                    \\ \hline
petgraph-ext3              & 881791                                                                            & 149205                                                                          & 2483                                                                            & 6287406                                                                                            & 2447                                                                                             & 42136                          & 7365468                    \\ \hline
petgraph-ext4              & 671999                                                                            & 123294                                                                          & 3069                                                                            & 4046105                                                                                            & 3188                                                                                             & 16542                          & 4864197                    \\ \hline
petgraph-ext5              & 400859                                                                            & 86750                                                                           & 2658                                                                            & 3688382                                                                                            & 3107                                                                                             & 9421                           & 4191177                    \\ \hline
petgraph-ext6              & 853792                                                                            & 154358                                                                          & 3156                                                                            & 6198179                                                                                            & 2902                                                                                             & 75377                          & 7287764                    \\ \hline
petgraph-inline-ext1       & 550082                                                                            & 114773                                                                          & 3004                                                                            & 4988262                                                                                            & 3287                                                                                             & 14740                          & 5674148                    \\ \hline
sniffnet-ext1              & 318576                                                                            & 89782                                                                           & 3191                                                                            & 4012997                                                                                            & 2925                                                                                             & 11313                          & 4438784                    \\ \hline
sniffnet-ext2              & 521024                                                                            & 108808                                                                          & 2324                                                                            & 2959298                                                                                            & 2719                                                                                             & 7431                           & 3601604                    \\ \hline
sniffnet-ext3              & 491132                                                                            & 114474                                                                          & 2420                                                                            & 2639553                                                                                            & 2501                                                                                             & 14870                          & 3264950                    \\ \hline
sniffnet-ext4              & 289206                                                                            & 104857                                                                          & 2458                                                                            & 3087829                                                                                            & 3184                                                                                             & 19270                          & 3506804                    \\ \hline
sniffnet-ext5              & 451829                                                                            & 112413                                                                          & 2904                                                                            & 3560519                                                                                            & 2571                                                                                             & 16898                          & 4147134                    \\ \hline
sniffnet-ext6              & 665962                                                                            & 163596                                                                          & 5027                                                                            & 3423886                                                                                            & 2428                                                                                             & 22809                          & 4283708                    \\ \hline
sniffnet-ext7              & 610395                                                                            & 115393                                                                          & 3062                                                                            & 3518957                                                                                            & 6090                                                                                             & 19601                          & 4273498                    \\ \hline
sniffnet-inline-ext1       & 606027                                                                            & 117642                                                                          & 2553                                                                            & 4392070                                                                                            & 8793                                                                                             & 30967                          & 5158052                    \\ \hline
sniffnet-inline-ext2       & 830359                                                                            & 110345                                                                          & 3269                                                                            & 1060569                                                                                            & 7246                                                                                             & 13375                          & 2025163                    \\ \hline
\end{tabular}
\captionsetup{justification=centering}
\caption{Raw timing data for the various extractions. Times are all in nanoseconds}
\end{adjustwidth}
\end{table}

%% file: appendix6.tex
\section{Appendix 6: (Almost) Complete Coq Translations}
\label{app:coq_translations}
Some of the Coq boilerplate has been removed to make these inclusions sensible. Still it is clear how massive the translation can get for even a small example. 

\captionsetup{justification=centering,type=listing}
\begin{lstlisting}[
    basicstyle=\scriptsize\ttfamily,
    numbers=left,
    frame=lines,
    breaklines=true,
    language=Coq
]
Definition bump
  {N : usize} (coremarkerCopyArrayI32NInst : core_marker_Copy (array i32 N))
  (arr : array i32 N) (k : usize) (mark : bool) :
  result ((array i32 N) * bool)
  :=
  if k s< N then bump_loop0 k arr mark 0%usize else bump_loop1 arr mark 0%usize
.

(* *)
(* *)
(* *)
(* *)
(* *)
(* *)
(* *)
(* *)
(* *)
(* *)
(* *)
(* *)
(* *)
(* *)
(* *)
(* *)
(* *)
(* *)
(* *)
(* *)
\end{lstlisting}
\captionof{listing}{Original Coq translation, less some boilerplate}
\label{lst:orig-coq}

\newpage
\captionsetup{justification=centering,type=listing}
\begin{lstlisting}[
    basicstyle=\scriptsize\ttfamily,
    numbers=left,
    frame=lines,
    breaklines=true,
    language=Coq
]
Definition bump
  {N : usize} (coremarkerCopyArrayI32NInst : core_marker_Copy (array i32 N))
  (arr : array i32 N) (k : usize) (mark : bool) :
  result ((array i32 N) * bool)
  :=
  if k s< N
  then (
    _ <- bump_loop0 coremarkerCopyArrayI32NInst arr k mark 0%usize;
    Ok (arr, mark))
  else (
    _ <- bump_loop1 coremarkerCopyArrayI32NInst arr mark 0%usize;
    Ok (arr, mark))
.

Definition bump_extracted
  {N : usize} (coremarkerCopyArrayI32NInst : core_marker_Copy (array i32 N))
  (arr : array i32 N) (i : usize) (mark : bool) :
  result unit
  :=
  i1 <- array_index_usize arr i;
  if i1 s= 0%i32
  then (_ <- i32_add i1 1%i32; _ <- array_index_mut_usize arr i; Ok tt)
  else (_ <- i32_add i1 1%i32; _ <- array_index_mut_usize arr i; Ok tt)
.
\end{lstlisting}
\captionof{listing}{Coq translation of the Incorrect Extraction, less some boilerplate}
\label{lst:wrong-coq}

%% file: appendix7.tex
\section{Appendix 7: New Extraction Feature Categories in Detail}
\label{app:new_extraction_features}

In this appendix we give concrete (although somewhat small) examples for each of the feature categories introduced in Section~\ref{subsec:new_language_features}. The examples are drawn from the repositories listed in Section~\ref{subsec:new_feature_cases}, or are lightly simplified variants of patterns we observed there, and are designed to isolate the behaviour of the extractor in the presence of a particular construct. In addition to asynchronous functions, const evaluation, generics, dynamic trait objects, and higher-ranked trait bounds, we include several instances of non-local control flow. Although the original REM toolchain could already handle some non-local control-flow refactorings, the new extraction engine implements them via the (now) standard \icodeverb{std::ops::ControlFlow} variants rather than custom enums, and we therefore evaluate them separately. Together, these examples characterise the kinds of advanced language mechanics that the original REM could not support and that the new toolchain is explicitly designed to handle, allowing us to check that it preserves semantics, typing, and compilation validity in each case.

\subsubsection{Generics (GEN)}

This case targets a generic function with multiple trait bounds, verifying that these are copied exactly to the extracted function. The extractor must not monomorphise or omit generic parameters, as doing so would restrict the function’s polymorphism and likely fail type checking. Additionally, if only a subset of the caller functions trait bounds apply to the callee (e.g. it only takes one of many generic arguments), then we would expect just the applicable trait bounds to be copied). 

\begin{minipage}[t]{0.43\linewidth}
    \vspace{0pt}
    \captionsetup{type=listing} 
\begin{lstlisting}[basicstyle=\scriptsize\ttfamily, frame=lines, numbers=left, breaklines=true, language=Rust]
fn min_of<T: Ord + Copy>(a: T, b: T) -> T {
    // EXTRACT START
    if a < b { a } else { b }
    // EXTRACT END
}
\end{lstlisting}
    \captionof{listing}{Generics Example: Before}
    \label{lst:generics-before}
\end{minipage}
\hfill
\begin{minipage}[t]{0.43\linewidth}
    \vspace{0pt}
    \captionsetup{type=listing}
\begin{lstlisting}[basicstyle=\scriptsize\ttfamily, frame=lines, numbers=left, breaklines=true, language=Rust]
fn min_of<T: Ord + Copy>(a: T, b: T) -> T {
    choose_min(a, b)
}

fn choose_min<T: Ord + Copy>(a: T, b: T) -> T {
    if a < b { a } else { b }
}
\end{lstlisting}
    \captionof{listing}{Generics Example: After}
    \label{lst:generics-after}
\end{minipage}

\subsubsection{Async and Await (ASYNC)}

This case targets extraction from within an \icodeverb{async fn}, where the inserted boundaries must remain within the compiler-generated state machine \footnote{\texttt{rustc} automatically generates a struct that acts as a state machine, with an internal state variable and a poll method. Each await point becomes a place where the state machine can pause, return \icodeverb{Poll::Pending}, and be resumed later by the runtime without losing its state.}.The extractor must correctly preserve the asynchronous context and the \icodeverb{Result} return type implied by \icodeverb{?}, while ensuring that lifetimes and borrows are not widened across the await suspension point. If the lifetimes produced by the extraction engine are incorrect, we expect REM-Repairer to correct them. 

\begin{minipage}[t]{0.43\linewidth}
    \vspace{0pt}
    \captionsetup{type=listing} 
\begin{lstlisting}[basicstyle=\scriptsize\ttfamily, frame=lines, numbers=left, breaklines=true, language=Rust]
async fn fetch_and_tag(client: &Client, id: u64) -> Result<String, Error> {
    // EXTRACT START
    let body = client.get(id).await?;     // await inside block
    let tagged = format!("[#{}] {}", id, body);
    // EXTRACT END
    Ok(tagged)
}
\end{lstlisting}
    \captionof{listing}{Async Example: Before}
    \label{lst:async-before}
\end{minipage}
\hfill
\begin{minipage}[t]{0.43\linewidth}
    \vspace{0pt}
    \captionsetup{type=listing}
    \label{lst:async-after}
\begin{lstlisting}[basicstyle=\scriptsize\ttfamily, frame=lines, numbers=left, breaklines=true, language=Rust]
async fn fetch_and_tag(client: &Client, id: u64) -> Result<String, Error> {
    fetch_and_tag_inner(client, id).await
}

async fn fetch_and_tag_inner(client: &Client, id: u64) -> Result<String, Error> {
    let body = client.get(id).await?;
    let tagged = format!("[#{}] {}", id, body);
    Ok(tagged)
}
\end{lstlisting}
    \captionof{listing}{Async Example: After}
\end{minipage}

\subsubsection{Constant declarations (CONST)}

This case covers extraction within a \icodeverb{const fn}, which must remain evaluable at compile time under Rust's const-checking rules. Thus the extraction engine must identify const contexts, and where possible, propagate the \icodeverb{const} qualifier to the new function to maintain those guaruntees.

\begin{minipage}[t]{0.43\linewidth}
    \vspace{0pt}
    \captionsetup{type=listing} 
\begin{lstlisting}[basicstyle=\scriptsize\ttfamily, frame=lines, numbers=left, breaklines=true, language=Rust]
// A const function that can be evaluated at compile time
const PREFIX: u32 = 10;
const fn compute_total(a: u32, b: u32) -> u32 {
    // EXTRACT START
    PREFIX + a + b
    // EXTRACT END
}
\end{lstlisting}
    \captionof{listing}{CONST Example: Before}
    \label{lst:const-before}
\end{minipage}
\hfill
\begin{minipage}[t]{0.43\linewidth}
    \vspace{0pt}
    \captionsetup{type=listing}
\begin{lstlisting}[basicstyle=\scriptsize\ttfamily, frame=lines, numbers=left, breaklines=true, language=Rust]
// A const function that can be evaluated at compile time
const PREFIX: u32 = 10;
const fn compute_total(a: u32, b: u32) -> u32 {
    compute_total_inner(a, b)
}

// Extracted function still marked as const
const fn compute_total_inner(a: u32, b: u32) -> u32 {
    PREFIX + a + b
}
\end{lstlisting}
    \captionof{listing}{CONST Example: After}
    \label{lst:const-after}
\end{minipage}

\subsubsection{Non Local Control Flow (NLCF)}

This case covers all aspects of NLCF, being \icodeverb{return}, \icodeverb{break}, and \icodeverb{continue} in Rust. NLCF objects cannot cross function boundaries, and as such the extraction engine must reify control flow into a return value, using \icodeverb{std::ops::ControlFlow} variants to communicate termination or continuation back to the caller. As mentioned previously, this differs from REM's original insertion of custom enums to outline control flow handling, as we rely on the extraction engine to provide this information rather than REM-Controller. 

\begin{minipage}[t]{0.43\linewidth}
    \vspace{0pt}
    \captionsetup{type=listing} 
\begin{lstlisting}[basicstyle=\scriptsize\ttfamily, frame=lines, numbers=left, breaklines=true, language=Rust]
fn sum_until_negative(xs: &[i32]) -> i32 {
    let mut sum = 0;
    for &x in xs {
        // EXTRACT START
        if x < 0 { break; }   // non-local `break`
        sum += x;
        // EXTRACT END
    }
    sum
}
\end{lstlisting}
    \captionof{listing}{NLCF Example: Before}
    \label{lst:nlcf-before}
\end{minipage}
\hfill
\begin{minipage}[t]{0.43\linewidth}
    \vspace{0pt}
    \captionsetup{type=listing}
\begin{lstlisting}[basicstyle=\scriptsize\ttfamily, frame=lines, numbers=left, breaklines=true, language=Rust]
use std::ops::ControlFlow;
use std::ops::ControlFlow::{Break, Continue};
fn sum_until_negative(xs: &[i32]) -> i32 {
    let mut sum = 0;
    for &x in xs {
        match step_sum_cf(x) {
            Break(()) => break,
            Continue(delta) => sum += delta, } }
    sum
}
fn step_sum_cf(x: i32) -> ControlFlow<(), i32> {
    if x < 0 { Break(()) }
    else { Continue(x) }
}
\end{lstlisting}
    \captionof{listing}{NLCF Example: After}
    \label{lst:nlcf-after}
\end{minipage}

\subsubsection{Higher Ranked Trait Bounds (HRTB)}

This case examines extraction from functions parameterised by higher-ranked trait bounds such as \icodeverb{for<'a> Fn(\&'a T) -> \&'a T}. These bounds are not inferred implicitly and must be explicitly copied to the new function signature to maintain lifetime correctness. Again, REM-Repairer is expected to be called here to either a) verify that the lifetimes are correct, and / or b) fix the incorrect lifetime bounds.

\begin{minipage}[t]{0.43\linewidth}
    \vspace{0pt}
    \captionsetup{type=listing} 
\begin{lstlisting}[basicstyle=\scriptsize\ttfamily, frame=lines, numbers=left, breaklines=true, language=Rust]
fn apply_str<F>(f: F) -> &'static str
where
    F: for<'a> Fn(&'a str) -> &'a str,
{
    let s = "hello";
    // EXTRACT START
    f(s)
    // EXTRACT END
}
\end{lstlisting}
    \captionof{listing}{HRTB Example: Before}
    \label{lst:hrtb-before}
\end{minipage}
\hfill
\begin{minipage}[t]{0.43\linewidth}
    \vspace{0pt}
    \captionsetup{type=listing}
\begin{lstlisting}[basicstyle=\scriptsize\ttfamily, frame=lines, numbers=left, breaklines=true, language=Rust]
fn apply_str<F>(f: F) -> &'static str
where
    F: for<'a> Fn(&'a str) -> &'a str,
{
    call_higher(f)
}

fn call_higher<F>(f: F) -> &'static str
where
    F: for<'a> Fn(&'a str) -> &'a str,
{
    let s = "hello";
    f(s)
}
\end{lstlisting}
    \captionof{listing}{HRTB Example: After}
    \label{lst:hrtb-after}
\end{minipage}

\subsubsection{Dynamic Trait Objects (DTO)}

Here we isolate extraction involving a \icodeverb{\&dyn Trait} parameter to test the extraction engines preservation of dynamic dispatch and object safety. The extractor must retain the trait object type and avoid replacing it with a generic bound, which would alter runtime behaviour. The category also includes cases where a function's arguments implement a trait, e.g. \icodeverb{fn ident(s: impl Into<String> -> ...}. 

\begin{minipage}[t]{0.43\linewidth}
    \vspace{0pt}
    \captionsetup{type=listing} 
\begin{lstlisting}[basicstyle=\scriptsize\ttfamily, frame=lines, numbers=left, breaklines=true, language=Rust]
use std::fmt::Display;

fn show_twice(d: &dyn Display) -> String {
    // EXTRACT START
    format!("{} | {}", d, d)
    // EXTRACT END
}
\end{lstlisting}
    \captionof{listing}{DTO Example: Before}
    \label{lst:dto-before}
\end{minipage}
\hfill
\begin{minipage}[t]{0.43\linewidth}
    \vspace{0pt}
    \captionsetup{type=listing}
\begin{lstlisting}[basicstyle=\scriptsize\ttfamily, frame=lines, numbers=left, breaklines=true, language=Rust]
use std::fmt::Display;

fn show_twice(d: &dyn Display) -> String {
    show_twice_inner(d)
}

fn show_twice_inner(d: &dyn Display) -> String {
    format!("{} | {}", d, d)
}
\end{lstlisting}
    \captionof{listing}{DTO Example: After}
    \label{lst:dto-after}
\end{minipage}

%% file: appendix8.tex
\section{Appendix 8: Implementation of the Verification Pipeline}
\label{app:implementation_verification}

This appendix contains a much more thorough dive into all of the implementation details behind the equivalence proover. It is intended to be read alongside the source code, and will in places contain duplicate information from Section~\ref{sec:implementation_verification}.

\vspace{5mm}

\begin{minipage}[t]{0.55\textwidth}
\vspace{-90mm}
The design of REM2.0's equivalence checking is guided by two principles: (1)
verification must require no additional annotations or input from the developer,
and (2) it must complete quickly enough to fit into interactive IDE workflows.
To achieve this, the pipeline translates both the original program $P$ and the
refactored program $P'$ into progressively more structured semantic
representations, culminating in a machine-checked proof of observational
equivalence.

At a high level, the pipeline consists of five stages: \textbf{Extraction},
\textbf{REM repairs}, \textbf{CHARON translation}, \textbf{AENEAS
functionalisation}, and \textbf{Coq verification}. Each stage builds on the
previous one, gradually exposing the semantics of the refactored code, before we end up with the pure, functional model in Coq. Figure
\ref{fig:verification_pipeline} provides an overview of the entire process. We will only very briefly cover off on extraction and lifetime repairing in this section, please refer to Section \ref{sec:expanding_rem} for the full scope.
\end{minipage}
\hfill
\begin{minipage}[t]{0.35\textwidth}
\centering
\begin{tikzpicture}[node distance=2cm, >=stealth, thick]
  \node[draw, rounded corners, fill=blue!10, minimum width=6cm, minimum height=1cm] (extract) {1. Extraction};
  \node[draw, rounded corners, fill=blue!10, minimum width=6cm, minimum height=1cm, below of=extract] (rem) {2. REM Repairs};
  \node[draw, rounded corners, fill=blue!10, minimum width=6cm, minimum height=1cm, below of=rem] (charon) {3. CHARON (LLBC)};
  \node[draw, rounded corners, fill=blue!10, minimum width=6cm, minimum height=1cm, below of=charon] (aeneas) {4. AENEAS (Functionalisation)};
  \node[draw, rounded corners, fill=blue!10, minimum width=6cm, minimum height=1cm, below of=aeneas] (coq) {5. Coq Verification};

  \draw[->] (extract) -- (rem);
  \draw[->] (rem) -- (charon);
  \draw[->] (charon) -- (aeneas);
  \draw[->] (aeneas) -- (coq);
\end{tikzpicture}


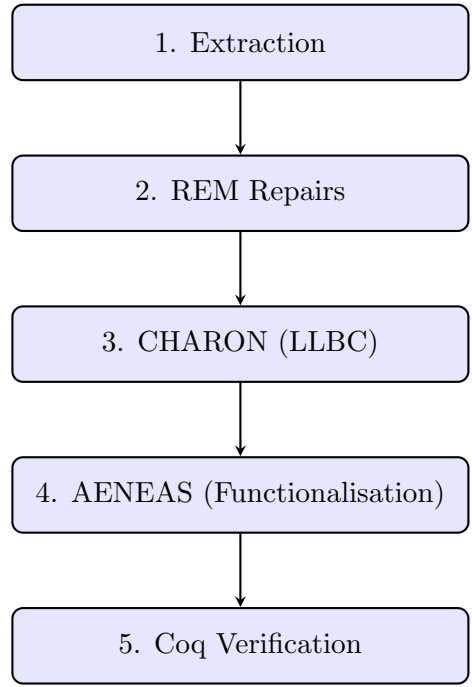
\captionof{figure}{Overview of the verification pipeline: from source code extraction to proof of observational equivalence in Coq.}
\label{fig:verification_pipeline}
\end{minipage}

\vspace{-2.5mm}
\subsection{Extraction}
Extraction lifts a user-selected region into a new function and replaces the selection with a call at the original site. As detailed in Section \ref{sec:single_file_workspace}, this step is powered by a Rust-Analyzer (RA) backed engine that performs single-file incremental analysis. It resolves the selection’s AST and symbol information in memory and synthesises a function skeleton together with the call-site edit.

The extraction result is returned as a deterministic set of edits (a patch) and a machine-readable summary for downstream stages. By design, the extractor does not attempt to finalise lifetimes or advanced ownership modes—that responsibility is delegated to REM’s repair phase. This separation keeps extraction millisecond fast and predictable, and completely independent of crate size, while ensuring a stable interface for the later verification.

\vspace{-2.5mm}
\subsection{REM}
REM performs \textit{post-extraction} semantic repair to attempt to bring the extracted code back into alignment with Rust's ownership and lifetime rules. Whilst much of the original REM toolchain's efforts have been integrated into the extraction process, we still rely heavily on the most complex part of REM's analysis: the Lifetime reification. We introduce, propagate, and constrain lifetime parameters to satisfy the borrow checker, generalising where necessary.

As detailed earlier, REM treats the compiler as an oracle: it iterates repairs until the program compiles, and then attempts to apply lifetime elision to produce a human-readable signature (where practical) and a function body that aligns as closely as possible to idiomatic Rust. The output of REM is thus the repaired program $P'$, which is then passed onto CHARON to begin the process of discharging equivalence obligations. 

\vspace{-2.5mm}
\subsection{CHARON}
\label{subsec:charon}

The first (new) stage of our verification pipeline is translation from Rust into a
lower-level, verification-friendly form. This role is performed by
\textbf{CHARON}, a Rust to LLBC (Low-Level Borrow Calculus) translator. CHARON
accepts Rust source code (or more precisely, its compiler intermediate
representation MIR) \footnote{Even more precisely, the tool itself accepts Rust
code, but its initial steps rely on a complex interaction with \texttt{rustc} to
acquire the MIR}. CHARON's main task is to extract complex semantic information
from \verb|rustc| and produce a machine readable output containing said
information. Up until this point, we have been working with / reasoning about a single file. Whilst this kept extraction lightning fast, CHARON's support for single file conversion is limited at best\footnote{CHARON can convert just a single file, but it relies on \texttt{rustc} to do so, and as such the single file must be self contained, which we cannot guarantee}. 

Thus at this stage, we construct two ``virtual'' crates within the users temporary directory (\icodeverb{std::env::temp()}). The advantages of this approach are:
\begin{enumerate}
    \item Any changes the user makes to their source code whilst we are performing the later stages of analysis will not get passed down to us (potentially introducing half written words or other bugs that would fail immediately)
    \item We do not require a file lock on the users code, so they can continue work
    \item We can safely abstract away the creation / deletion of the virtual crates, and have them live for only so long as the equivalence engine runs for, using a custom implementation of the \icodeverb{Drop} trait. 
    \item We can limit the data passing over the JSON-RPC bridge to just the original and extracted code for the file the user is working in. 
\end{enumerate}

CHARON then performs several non-trivial transformations:
\begin{itemize}
  \item \textbf{Structured control flow.}  
    MIR is a control-flow graph with arbitrary \icodeverb{goto}s; CHARON reconstructs
    this into structured \icodeverb{if}, \icodeverb{match}, and \icodeverb{loop} constructs,
    or preserves the raw form with the \texttt{--ullbc} option.

  \item \textbf{Trait and type resolution.}  
    Explicitly records how trait bounds are proven, normalises default methods,
    and can transform associated types into explicit parameters
    (\texttt{--remove-associated-types}).

  \item \textbf{Lifetime handling.}  
    Hides the distinction between early- and late-bound variables and makes
    implied bounds explicit, simplifying reasoning about generic functions.

  \item \textbf{Closures and vtables.}  
    Represented as ordinary structs implementing the relevant \texttt{Fn*}
    traits, making them accessible to later verification steps.

  \item \textbf{Noise reduction.}  
    Options such as \texttt{--hide-marker-traits} remove built-in traits
    (\icodeverb{Sized}, \icodeverb{Send}, etc.) that would otherwise clutter the output.
\end{itemize}

The result is a cleaned, semantically faithful program in LLBC form. Unlike raw
MIR, which is compiler-oriented, LLBC is designed for verification: ownership
and borrowing are represented directly, trait resolution is explicit, and
control flow is structured. A sample program and small portion of its MIR are
shown below. However, the output of CHARON is not designed for human
readability, and even the tiny program in Figure \ref{fig:charon_translation}
produces over 140,000 characters of LLBC. This (very verbose) explicitness is
what enables the subsequent translation into pure functional code for
verification.

\begin{figure}[ht]
  \small
  \centering
  \begin{minipage}[t]{0.45\linewidth}
\begin{lstlisting}[basicstyle=\footnotesize\ttfamily, language=Rust]
fn main() {
    let v = vec![1, 2, 3];
    let x = &v[0];
    println!("{}", x);
}
\end{lstlisting}
    \vfill
    \captionsetup{type=listing}
    \caption{Rust source}
  \end{minipage}
  \hfill
  \begin{minipage}[t]{0.45\linewidth}
\begin{lstlisting}[basicstyle=\footnotesize\ttfamily]
bb2: {
    _9 = &_1;
    _8 = <Vec<i32> as Index<usize>>::
        index(move _9, const 0_usize);
}

bb3: {
    _7 = copy _8;
    _16 = &_7;
    _15 = core::fmt::rt::Argument::<'_>::
        new_display::<&i32>(copy _16);
}
\end{lstlisting}
    \vfill
    \captionsetup{type=listing}
    \caption{Excerpt of MIR (\texttt{rustc})}
  \end{minipage}
  \captionsetup{justification=centering}
  \caption{Illustration of CHARON's role. While only Rust and MIR are shown here, CHARON translates the MIR into LLBC. The LLBC is only designed to be machine readable, and as such is ommitted}
  \label{fig:charon_translation}
\end{figure}

\vspace{-2.5mm}
\subsection{AENEAS}
With the program translated into LLBC by CHARON, the next stage of the pipeline
is handled by \textbf{AENEAS}. AENEAS takes the borrow- and lifetime–explicit
LLBC and translates it into a \emph{purely functional form}, stripping away
low-level memory operations while preserving the program's semantics. This
functionalisation step is what makes the program suitable for downstream
reasoning in Coq: instead of reasoning about references and loans, we can reason
about values and functions. Currently AENEAS supports translation into
\emph{Coq, Lean, HOL4}, and \emph{F*}.

In practice, AENEAS acts as the bridge between compiler-level detail and
proof-level abstraction. CHARON ensures that all the complexities of lifetimes,
loads, and trait resolution are made explicit; AENEAS then packages these into a
form where correctness properties can be stated and proved without touching the
borrow checker or the stack/heap. Within the context of our pipeline, this is the point where Rust
code ``stops being Rust'' and becomes a functional program whose equivalence can
be checked in Coq.

\vspace{-2.5mm}
\subsection{Coq}
AENEAS translates the LLBC into a purely functional form, targeting Coq. The
final stage of the pipeline then combines the two generated Coq files ($P_{coq}$
and $P'_{coq}$), together with all auxiliary information needed to
automatically verify the equivalence between the original and refactored Rust
programs. A key challenge at this stage is the verbosity of the generated Coq
code (as illustrated in Listing~\ref{lst:complex-wrong-coq}), which led us to
implement a custom Coq tokeniser and parser tailored to the format emitted by
AENEAS.

In parallel with these translations, the equivalence prover’s main task is to
produce the \texttt{EquivCheck.v} file. This module instantiates both the
original and refactored definitions and generates an equivalence theorem, which
is a Coq encoding of the obligation(s) described in Section~\ref{sec:proof_obligations}. In practice, this means the Coq project is set up so that
it automatically checks that the two versions produce the same results for all
inputs. We defer the details of the proof obligations to that section; the key
point here is that the generated Coq project is self-contained: once compiled,
the equivalence is verified without further user intervention.

\vspace{-2.5mm}
\subsubsection*{What is Coq?}
Coq is an interactive theorem prover and proof assistant built on a dependently
typed functional language. It allows developers to define mathematical objects,
programs, and logical propositions, and to construct machine-checked proofs
about them. Every proof is validated by Coq's small, trusted kernel, giving very
high assurance of correctness. Coq has a long history of use in large-scale
verification projects—such as CompCert (a verified C compiler) and formally
verified operating systems—making it a natural fit for reasoning about the
safety and correctness of Rust programs.

\vspace{-2.5mm}
\subsubsection*{Why Coq?}
Coq was selected over alternatives such as HOL4, Lean, or F* for three main reasons:
\begin{itemize}[leftmargin=*,itemsep=0pt,topsep=2pt]
  \item \textbf{Mature ecosystem}  
    Decades of development have produced robust tooling, extensive libraries,
    and a stable kernel.

  \item \textbf{Direct integration}  
    AENEAS already targets Coq, whereas HOL4 and Lean would require substantial
    new backends for compatibility.

  \item \textbf{Automation}  
    Coq's tactic language and SMT integrations allow many routine proofs—
    such as equivalence checks—to be discharged automatically.
\end{itemize}

Lean and HOL4 both provide highly expressive grammars, but at the cost of
significant engineering effort to integrate with AENEAS \footnote{They both require even more additional supporting architecture on top of the current pipeline}. Whilst F* would require the least additional infrastructure, the platform is not that mature mature and would also duplicate existing
functionality. Coq therefore offers the best combination of maturity,
automation, and direct compatibility with our pipeline.

%% file: appendix9.tex
\section{Appendix 9: EquivCheck.v Lemmas from the Extraction Evaluation}
\label{app:lemmas}

\begin{figure}[h!]
    \centering
\begin{lstlisting}[language=Coq]
(** EquivCheck.v Autogenerated by REM-verification **)
Require Import Primitives.
Import Primitives.
Require Import Input.
Require Import Out.
Local Open Scope Primitives_scope.

Lemma foo_equiv_check :
    Input.foo = Out.foo.
Proof.
    intros.
    unfold Input.foo.
    unfold Out.foo.
    reflexivity.
Qed.
\end{lstlisting}
    \captionof{listing}{EqivCheck.v file from one of the simple verification cases}
    \label{lst:lemma_example_simple}
\end{figure}

\begin{figure}[h!]
    \centering
\begin{lstlisting}[breaklines=true, language=Coq]
(** EquivCheck.v Autogenerated by REM-verification **)
Require Import Primitives.
Import Primitives.
Require Import BusinessLogic.
Require Import BusinessLogicRefactored.
Local Open Scope Primitives_scope.

Lemma example5_config_normalise_config_equiv_check : forall (raw : example5_config_RawConfig_t),
    BusinessLogic.example5_config_normalise_config raw = BusinessLogicRefactored.example5_config_normalise_config raw.
Proof.
    intros raw.
    unfold BusinessLogic.example5_config_normalise_config.
    unfold BusinessLogicRefactored.example5_config_normalise_config.
    reflexivity.
Qed.
\end{lstlisting}
    \captionof{listing}{EqivCheck.v file from one of the complex verification cases}
    \label{lst:lemma_example_complex}
\end{figure}